\documentclass[superscriptaddress,prx,twocolumn,amsmath,amssymb,nofootinbib,]{revtex4-1}
\usepackage[T1]{fontenc}
\usepackage{amsthm}
\usepackage{graphicx}
\usepackage{times}
\usepackage{hyperref}
\hypersetup{
colorlinks=true,
linkcolor=red,
citecolor=red,}
\usepackage{textcomp}
\usepackage{enumerate}
\usepackage{multirow}
\usepackage{tabularx}
\usepackage[mathscr]{euscript}
\usepackage{mathtools}
\usepackage{mhchem}
\usepackage[mathscr]{euscript}
\usepackage{dsfont}
\usepackage{amsfonts}
\usepackage[normalem]{ulem}
\usepackage{centernot}
\usepackage{stmaryrd}
\usepackage{longtable}
\usepackage{enumitem,kantlipsum}
\usepackage{array}
\usepackage{wrapfig}
\usepackage{tabu}
\usepackage{graphicx}
\usepackage{placeins}
\usepackage{enumitem}
\usepackage{dutchcal}
\usepackage{tabularx}
\usepackage[export]{adjustbox}
\usepackage{color,soul}
\usepackage{longtable}
\usepackage[dvipsnames]{xcolor}

\usepackage[caption=false]{subfig}
\usepackage{adjustbox}

\usepackage{tikz}
\usetikzlibrary{quantikz}

\newcommand\s[1]{_{\rm #1}}

\newcommand\us[1]{^{\rm #1}}

\newcommand{\Bell}{{\rm{Bell}}}

\newcommand{\ketbra}[1]{ | #1 \rangle\!\langle #1 |}

\newcommand{\ii} {\textbf{i}}

\newcommand{\expec}[1]{\left\langle #1 \right\rangle}
\newcommand{\one}{\leavevmode\hbox{\small1\normalsize\kern-.33em1}}

\newcommand{\oprod}[1] {^{\otimes #1}}

\newcommand{\PP}{{\cal P}}

 \def\ee{\mathord{\rm e}}
 
 \def\ii{\mathord{\rm i}}

\def\half{\textstyle\frac{1}{2}}

\def\fourth{\textstyle\frac{1}{4}}

\renewcommand{\ii}{{\rm i}}
\renewcommand{\ee}{{\rm e}}
\def\beq{\begin{equation}}
\def\eeq{\end{equation}}

\def\barray{\begin{eqnarray}}
\def\earray{\end{eqnarray}}

\newcommand{\FS}[1]{\textcolor{red}{#1}}

\begin{document}

\title{Witnessing entanglement in  trapped-ion quantum error correction under realistic   noise}

\author{Andrea Rodriguez-Blanco}.
\affiliation{Departamento de F\'{i}sica Teorica, Universidad Complutense de Madrid, 28040 Madrid, Spain}
\email{Current affiliation: University of California Berkeley, rodriguezblanco@berkeley.edu.}
\author{Farid Shahandeh}
\affiliation{Department of Computer Science, Royal Holloway, University of London, Egham TW20 0EX, United Kingdom}

\author{Alejandro Berm{u}dez}
\affiliation{Instituto de F\'{i}sica Te\'{o}rica, UAM-CSIC, Universidad Aut\'{o}noma de Madrid, Cantoblanco, 28049 Madrid, Spain}


\begin{abstract}
Quantum Error Correction (QEC) exploits redundancy  by encoding logical information into multiple physical qubits. In current implementations of QEC, sequences of non-perfect two-qubit entangling gates are used to codify the information redundantly into multipartite entangled states. Also, to extract the error syndrome, a series of two-qubit gates are used to build parity-check readout circuits. In the case of noisy gates, both steps cannot be performed perfectly, and an error model needs to be provided to assess the performance of QEC. We present a detailed microscopic error model to estimate the average gate infidelity of  two-qubit light-shift gates used in trapped-ion platforms. We analytically derive  leading-error contributions in terms of microscopic parameters and present effective error models that connect the error rates typically used in phenomenological accounts to the microscopic gate infidelities hereby derived. We then apply this  realistic error model to quantify the multipartite entanglement generated by circuits that act as QEC building blocks. We do so by using entanglement witnesses,  complementing in this way the recent studies in~\cite{PRXQuantum.2.020304, PhysRevX.12.011032} by exploring the effects of a more realistic microscopic noise. 
\end{abstract}

\maketitle

\setcounter{tocdepth}{2}
\begingroup
\hypersetup{linkcolor=black}
\tableofcontents
\endgroup

\section{\textbf{Introduction}}\label{sec:intro_text}

Fault-tolerant  quantum computation promises to solve hard computational problems by harnessing   the power inherent in the laws of quantum physics~\cite{nielsen00}. In this way, quantum computers  have the potential to outperform their classical counterparts for  certain computations~\cite{Montanaro2016}, and  great efforts are being devoted to their development using various experimental platforms~\cite{Ladd2010}. 
However, the  sensitivity of  these quantum systems to the interactions with the environment, as well as the unavoidable imperfections of the quantum operations used to control them,  lead to processing errors that pose a  hurdle for  bringing these  devices to the  large scales required for  useful  quantum computations.  At present, specific tasks have been designed where  such noisy intermediate-scale quantum (NISQ)~\cite{Preskill2018quantumcomputingin} computers can display quantum advantage with respect to classical devices, even in the presence of errors ~\cite{Arute2019,doi:10.1126/science.abe8770}. In order to benefit from  such advantages in more generic algorithms, a possible solution is to  fight against  error accumulation by active quantum error correction (QEC)~\cite{PhysRevA.54.1098,  PhysRevA.54.1098, PhysRevLett.77.793, RevModPhys.87.307}. Using this strategy,  the quantum information is encoded redundantly in logical qubits composed of multiple physical qubits using  specific QEC codes.
Such logical encodings allow  for the detection and correction of the errors that occur during the computation without gaining any knowledge of the encoded information (and thus, without perturbing  the quantum state by  measurement) and the fed-forward correction operation. Since the operations used to infer and correct errors and those used to manipulate  ancillary qubits will also be faulty, all operations on the quantum register must obey a fault-tolerant (FT) circuit design~\cite{FTQEC}, ensuring that errors do not spread uncontrollably through the quantum register during the computation. These FT considerations usually entail an increase in the circuit complexity and, importantly, set stringent conditions for the accuracy of each gate~\cite{https://doi.org/10.48550/arxiv.quant-ph/0508176,https://doi.org/10.48550/arxiv.0711.1556}.

Among the architectures for large-scale FT quantum computers, trapped ions provide one of the most promising platforms~\cite{doi:10.1063/1.5088164}.
We note that there are different ion species, various possible qubit encodings, different ion-trap hardware, and a variety of quantum-information processing schemes. Although the performance and scaling capabilities of each of these approaches can vary, we can draw the following picture of current capabilities.  In general, state preparation  is achieved by illuminating the ions with laser radiation and exploiting  optical-pumping schemes, which yields low error rates on the order of $10^{-3}$  to $10^{-4}$~\cite{PhysRevLett.113.220501}. The accuracy of state detection, which relies on collecting photons from a state-dependent fluorescence, depends on the specific qubit encoding as well as the scheme for collecting the scattered photons and the collection time. In general, measurements are performed with low error rates also on the order of $ 10^{-3}$, while the 10$^{-4}$ level can be achieved in specific cases~\cite{PhysRevLett.113.220501}.  
Trapped-ion qubits  also provide excellent realizations of quantum memory. When encoded in the ground state manifold, relaxation $T_1$ times are essentially infinite, and decoherence $T_2$ times can reach seconds  by using mu-metal magnetic-field shielding to minimize the dephasing noise~\cite{Ruster2016}. 
By exploiting qubit states that are first-order insensitive to Zeeman shifts, as used in trapped-ion clocks, the coherence time can be even further increased~\cite{PhysRevLett.113.220501}.

Quantum information can also be processed in qubit registers corresponding to trapped-ion crystals with remarkable levels of performance. High-fidelity single-qubit rotations have been achieved for all different qubit encodings, by shining either laser or microwave radiation onto the ions, achieving errors in the $10^{-2}$-$10^{-6}$ range~\cite{doi:10.1063/1.5088164}. Laser-driven single-qubit gates for Zeeman qubits which exploit two-photon transitions (discussed below in more details) have achieved error rates of $10^{-2}$~\cite{Ruster2016} which can be lowered to $10^{-3}$ by using RF radiation~\cite{Keselman_2011} at the expense of individual qubit addressing. Note that these numbers also include state-preparation and measurement errors and \FS{the} subsequent randomized benchmarking~\cite{Emerson_2005} experiments have singled out the gate contribution to the error to be on the order of $10^{-4}$.

Together with a  two-qubit entangling gate, these operations lead to a universal gate set for quantum information processing and, combined with state preparations and measurements, allow for flexible trapped-ion computations which have enabled the very first implementations of several quantum algorithms such as a scalable Shor's factorization algorithm, quantum chemistry algorithms, and portfolio optimization.~\cite{doi:10.1126/science.aad9480,PhysRevX.8.031022,https://doi.org/10.48550/arxiv.2111.14970}. One should note that the  two-qubit entangling gates typically display a higher error in comparison to other quantum operations which lies in the $10^{-2}$-$10^{-3}$ range and depends on the specific ion encoding and the gate scheme~\cite{doi:10.1063/1.5088164}. M\o lmer-S\o rensen gates~\cite{PhysRevLett.82.1835,PhysRevA.62.022311} have reached an error on the order of  $10^{-3}$ level using hyperfine~\cite{PhysRevLett.117.060505} and optical~\cite{Erhard2019} qubits. Among other  schemes~\cite{PhysRevA.85.040302,Lemmer_2013} that have stimulated  high-fidelity implementations of entangling gates~\cite{PhysRevLett.110.263002,PhysRevLett.117.140501}, the so-called light-shift gates~\cite{Leibfried2003} have reached  errors as low as $10^{-3}$ for hyperfine qubits~\cite{PhysRevLett.117.060504,PhysRevA.103.012603}. We also note that one of the advantages of trapped ions with respect to other solid-state platforms is the programmable connectivity of the two-qubit gates. Using accurate and fast ion-crystal re-configurations, typically referred to as ion shuttling~\cite{Kielpinski2002,doi:10.1116/1.5126186}, it is possible to perform an entangling gate on any desired qubit pair~\cite{doi:10.1126/science.1177077,PhysRevLett.119.150503, Pino2021,PhysRevX.11.041058,PhysRevX.12.011032}. Alternatively, in  static ion crystals, any desired pair of qubits is individually addressed by laser beams and interactions mediated by radial phonons to achieve programmable entangling gates~\cite{Debnath2016,Figgatt2019,PRXQuantum.2.020343}.

For Zeeman qubits,  current light-shift gates have errors  on the order of $10^{-2}$ and thus, fall short of the topological codes' fault-tolerance requirements~\cite{ RevModPhys.87.307}.

Regardless of the qubit encoding, and the specific thresholds of the QEC strategy, one can see that current entangling-gate errors are a major bottleneck for increasing the depth of trapped-ion quantum circuits.  It is also important to improve the two-qubit gates for the demonstration of practical advantages of QEC encodings where the encoded logical operations are expected to improve upon those performed on bare physical qubits.  In this context, it is crucial to have realistic models of the microscopic errors that go beyond a single number and  can be incorporated in the simulations of the corresponding circuits. The resulting models are then used to assess the performance and the required improvements of the current hardware/software, guiding near-term developments. There has been recent progress in this direction in  the context of optical trapped-ion qubits~\cite{PhysRevX.7.041061, PhysRevA.99.022330,PhysRevA.100.062307,ParradoRodriguez2021crosstalk}, as well as  hyperfine qubits~\cite{Trout_2018, Debroy_2020,https://doi.org/10.48550/arxiv.2004.04706,9407237,Tinkey_2021}. The  native set of operations, as in the case of optical qubits~\cite{Schindler2013}, includes the entangling M\o lmer-S\o rensen gate and ion-crystal reconfigurations~\cite{doi:10.1116/1.5126186}.
In contrast, this type of study has not been carried out for Zeeman qubits with other entangling gates.

In this manuscript, we fill this gap by presenting a detailed microscopic error model for  two-qubit light-shift gates. 
Various sources of state infidelities for light-shift gates have previously been discussed in the literature~\cite{ballance_thesis,PhysRevA.105.022437} in parallel with  M\o{}lmer-S\o{}rensen gates~\cite{PhysRevA.62.022311}.
Such microscopic derivations are very useful, as they provide insight into different contributions to errors (i.e. error budget) and experimental  optimizations to minimize them. Note, however, that these estimations focus on a single target state that would be produced by the ideal gate acting on  a single initial state. Nonetheless, the same entangling gates in a larger circuit will typically act on a variety of initial states and could  exhibit a different error. Hence, it is desirable to obtain estimations for the  gate infidelity which averages over the initial states. Moreover, it is desirable to relate these estimates to an accurate microscopic model of the noisy quantum dynamics as addressed for M\o{}lmer-S\o{}rensen gates~\cite{PhysRevX.7.041061,PhysRevA.100.062307}. This model is not characterized by a single quantum number, but rather by a dynamical quantum map that faithfully describes the noisy gate. The effective noise channels can then be used to study the performance of QEC primitives, going beyond the oversimplified phenomenological and circuit noise models.

As noted above, QEC exploits redundancy to protect logical information from external noise and correct errors. The encoded logical qubits typically display  entanglement~\cite{preskill_notes}, and there is indeed a very interesting interplay between entanglement and QEC codes~\cite{ PhysRevA.54.3824, entanglementassistQECC,Almheiri2015,PRXQuantum.2.020304}. Following this connection, the  generation and verification of  entangled states can be used as a  benchmarking protocol for QEC primitives, as advocated  recently in~\cite{PRXQuantum.2.020304,PhysRevX.12.011032}.
In these works, {key} aspects of QEC circuits are  characterized in terms of their power to generate maximally entangled output states, which can be efficiently and robustly witnessed  when the QEC codes are described under the stabilizer formalism~\cite{nielsen00,stabilisers}. To model the noise at the circuit level, these works use an idealized depolarizing channel. In this manuscript, we take further steps in this direction, evaluating and comparing the performance of the witnessing methods  under a microscopic trapped-ion noise model for Zeeman qubits and light-shift gates. We  focus on entanglement witnesses that can be efficiently inferred from experimental data, those in which the number of measurements scales linearly with the number of qubits. \\

The manuscript is organized as follows. In Sec.~\ref{sec:FT_colorcode}, we present the primitives of the topological QEC code we wish to characterize. We show the compilation of those QEC circuits into the trapped-ion native gates set and introduce the entanglement tests we will use to characterize the aforementioned circuits in terms of their ability to generate maximally entangled output states. In Sec.~\ref{Sec:non_ideal_conditions}, we revise the construction of the entangling light-shift gate in terms of laser-ion  interaction Hamiltonians. We then derive a microscopic noise model for this unitary under non-ideal conditions and present effective error models for the entangling light-shift gate, which now connect the error probability of each model to the microscopic gate infidelities .  Finally, in Sec.~\ref{Sec:results}, we  analyze the results of  witnessing maximally entangled states for these trapped-ion QEC circuits with realistic noise.

\section{\textbf{Quantum error correction primitives, fault tolerance and entanglement}}\label{sec:FT_colorcode}

Trapped-ion quantum computers are one of the leading experimental platforms in experimental QEC~\cite{Chiaverini2004,Schindler1059,Nigg302,Linkee1701074,Stricker2020, Erhard2021, Egan2021, PhysRevLett.127.240501, PhysRevLett.127.240501}. Recent experiments with trapped ions~\cite{PhysRevX.12.011032,PhysRevX.11.041058, Postler2022,https://doi.org/10.48550/arxiv.2208.01863} have employed color codes~\cite{PhysRevLett.97.180501}, a family of topological QEC codes~\cite{RevModPhys.87.307}, in order  to detect and correct errors in a fault-tolerant manner. 
In particular, these works have employed FT constructions via the so-called  flag ancilla qubits~\cite{PhysRevLett.121.050502,Chamberland2018flagfaulttolerant,Chao2020} which have been used for an FT measurement of the stabilizers~\cite{PhysRevX.12.011032}, a demonstration of repetitive rounds of fault-tolerant QEC cycles interleaved with logical operations ~\cite{PhysRevX.11.041058,https://doi.org/10.48550/arxiv.2208.01863}, and  implementing a universal set of logical single and two-qubit  operations~\cite{Postler2022}. 
The target of these studies has mainly been the smallest-possible color code which corresponds to the 7-qubit Steane code~\cite{doi:10.1098/rspa.1996.0136}. In this work, we choose to work with this distance-3 topological color code and focus on the specific stabilizer measurements. 
 \begin{figure}[t!]
  \includegraphics[width=1\columnwidth]{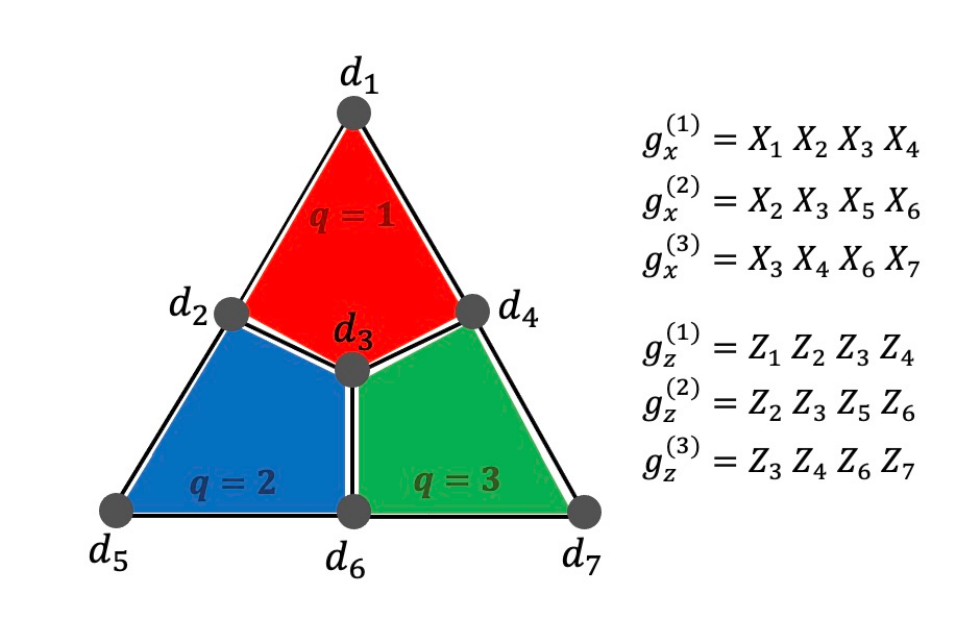}
  \caption{{ Schematics of the standard $d=3$ topological color code. The $n_c=7$ physical data qubits that  encode the logical qubit lay on the vertices of the three colorable plaquettes. There are two parity-check per colorable plaquette $q$: $g_{x}^{(q)}$ for detecting single qubit phase-flip errors, and $g_{z}^{(q)}$ for detecting bit-flip errors. That makes a total of $G=6$ parity checks, leading to $k=n_c-G=1$ logical qubits.}}
  \label{fig:standard_color_code}
\end{figure}

\subsection{Stabilizer codes}

A stabilizer QEC code of type $[[n_c,k,d]]$ 
encodes $k$ logical qubits into $n_c$ physical qubits with a code distance $d$, and  can correct up to $t=\lfloor{(d-1)/2}\rfloor$ errors. 
The stabilizers $S$ are operators that act on the Hilbert space of physical qubits $\mathcal{H}=\mathbb{C}^{2n_c}$, and  form an Abelian subgroup $\mathcal{S}_G$ 
of  the $n$-qubit Pauli group $\mathcal{P}_{n_c}=\times\{c_0I,c_1X,c_2Y,c_3Z: \hspace{1ex}c_i\in\{\pm 1, \pm \ii\}\}^{\otimes n_c}$, where $I,X,Y,Z$ are the identity and the single-qubit Pauli matrices. This group is  obtained by taking $n$-fold tensor products of these Pauli matrices, and can also include  the  multiplicative factors $\pm 1, \pm i$. 
The stabiliser subgroup, which must fulfil $-I^{\otimes n_c}\notin \mathcal{S}_G$,  has $|\mathcal{S}_G|=2^G$ elements  that can be  generated   by  a  subset 
of $G$ linearly-independent and mutually-commuting Pauli operators, $\mathcal{g}=\{g_{i}\}_{i=1}^G$. These generators, commonly known as 
\textit{parity checks},  are   Hermitian and involutory operators within  the Pauli group  $g_i^\dagger=g_i\in\mathcal{P}_{n_c}: g_i^2=I^{\otimes n_c}$. 
Accordingly, any element of the stabilizer  $S\in \mathcal{S}_G=\langle g_{1},...,g_{G}\rangle$ can be obtained
as a  certain product of  generators specified by the indexing set $\mathcal{I}$
\begin{equation}
    S=\prod_{i\in \mathcal{I}} g_{i} \quad \text{with}\quad g_i\in\mathcal{g}\,\forall i\in \mathcal{I}.
\end{equation}
In light of the above constraints,  the parity checks  correspond to a set of compatible observables with $\pm 1$ eigenvalues, such that 
the measurement  of one of them   projects the state onto   a $2^{n_c-1}$-dimensional subspace of  the $n_c$-qubit Hilbert space $\mathcal{H}$.
Since the parity checks commute, it is possible to unambiguously specify a $2^{n_c-G}$-dimensional subspace through their common $+1$ eigenspace, namely,
\begin{equation}
   \mathcal{L}= \{\ket{\psi}\in\mathbb{C}^{2n_c}:\hspace{2ex}S\ket{\psi}=\ket{\psi},\hspace{1ex} \forall S\in\mathcal{S}_G\}.
\end{equation}
The subspace $\mathcal{L}\subset\mathcal{H}$ is known as the \textit{code space} and can be used to embed or codify  $k=n_c-G$ logical qubits.
The logical operators
$\{X^{L}_\ell,Z^L_\ell\}_{\ell=1}^k\notin\mathcal{S}_G$ correspond to Hermitian elements of the Pauli group $\PP_n$ that commute  with any stabilizer $[X^{L}_\ell,S]=[Z^{L}_\ell,S]=0$, $\forall S\in\mathcal{S}_G$ and, thus, act non-trivially within the code space $X^{L}_\ell,Z^L_\ell:\mathcal{L} \to \mathcal{L}$. Moreover, they are required  to fulfil the corresponding algebra for $k$ independent logical qubits $(X^{L}_\ell)^2=(Z^{L}_\ell)^2=I^{\otimes n_c}$, and $X^{L}_\ell Z^L_{\ell'}=(1-2\delta_{\ell,\ell'}) Z^L_{\ell'}X^{L}_\ell$, $\forall\ell,\ell'\in\{1,\cdots,k\}$.\\
Topological QEC codes are a specific type of stabilizer code in which the parity checks have local support on a lattice where the physical qubits reside, whereas the logical qubits  are encoded in global topological properties. For instance, in topological color codes~\cite{PhysRevLett.97.180501}, the physical qubits are arranged on the vertices of a trivalent three-colorable planar lattice (see, e.g. Fig.~\ref{fig:standard_color_code} for the smallest-distance color code). 
The parity check operators are the following pair of $X$- and $Z$-type operators  per plaquette $q$, and have local support, acting only on those qubits located at the plaquette vertices 
\begin{equation}\label{eq:plaquette_gen}
    g_x^{(q)}=\bigotimes_{i\in v(q)}X_i,\qquad g_z^{(q)}=\bigotimes_{i\in v(q)}Z_i,
\end{equation}
where $v(q)$ is the set of vertices that belong to plaquette $q$.
As depicted in Figure~\ref{fig:standard_color_code}, this color code has $n_c=7$ physical qubits and
 $G=6$ parity checks, each of which acts locally
on only 4 qubits  (i.e. weight-4 parity checks),  so that the code space is two-dimensional ${\rm dim}\,\mathcal{L}=2^{7-6}=2$, encoding $k=n_c-G=1$ logical qubit. Up to deformations by application of parity checks, the logical operators correspond to strings of Pauli operators along the sides of the triangle $X_L=X_5X_6X_7$, $Z_L=Z_1Z_4Z_7$, which clearly show that the distance of the code is $d=3$, such that it can correct $t=1$  error. These errors are typically referred to as single-qubit bit- and phase-flip errors $E\in\{X_i, Z_i\}_{i=1}^7$, and anti-commute with some of the parity checks. Accordingly, the measurement of parity checks provides an error syndrome $s=\{s^{(1)}_x,s^{(2)}_x,s^{(3)}_x,s^{(1)}_z,s^{(2)}_z,s^{(3)}_z\}$, where $s^{(q)}_\alpha=\pm1$, and $\alpha=x,z$, which allows inferring, and subsequently correcting, the most likely error $\ket{\Psi(0)}=c_0\ket{0}_L+c_1\ket{1}_L\mapsto E\ket{\Psi(0)}$ without obtaining any information about  $c_0$ and $c_1$.

\subsection{ Trapped-ion parity-check measurements}

Let us now discuss how the parity check measurements  are performed in practice, considering a specific implementation based on trapped ions. The  native trapped-ion entangling gates in current designs are not CNOT gates~\cite{nielsen00}. 
Instead, they are based on state-dependent dipole forces and effective phonon-mediated interactions   that dynamically generate entanglement  between distant qubits~\cite{PhysRevLett.82.1835,PhysRevA.62.022311,Sackett2000,Leibfried2003,Benhelm2008}. 
In this work, we shall be concerned with  $Z$-type state-dependent forces. These forces, which stem from  bi-chromatic laser beams in a $\Lambda$-scheme configuration (see Fig.~\ref{fig:2ndorder}) in  a specific parameter regime, will be discussed in the following section.  Under certain conditions, we show that these forces  give rise to the entangling light shift gates~\cite{Leibfried2003,PhysRevLett.119.150503}, namely,
\begin{equation}
\label{eq:zz}
    U^{\rm ZZ}_{{i,j}}(\theta)=e^{-\ii\frac{\theta}{2}{Z}_{i}{Z}_{j}},
\end{equation}
where  $\theta$ is the corresponding pulse area. This unitary leads to a fully entangling gate for $\theta=\pi/2$, i.e.~ they map product states onto GHZ-type entangled states. They  will be denoted by the symbol $   U^{\rm ZZ}_{{i,j}}(\pi/2)=\,\,\boldsymbol{|}^{\hspace{-0.82ex}\bullet\hspace{-1.62ex}\boldsymbol{-}\hspace{-0.82ex}\boldsymbol{-}i}_{\hspace{-0.8ex}\bullet\hspace{-1.62ex}\boldsymbol{-}\hspace{-0.82ex}\boldsymbol{-}j}$ in the  circuits below. In addition, we shall also consider single-qubit rotations. In particular, the unitaries 
\begin{equation}
\label{eq:z_rot}
    R^{Z}_{i}(\theta)=e^{-\ii\frac{\theta}{2}Z_i},
\end{equation}
are obtained by local ac-Stark shifts~\cite{Poschinger_2009}, and correspond to rotations about the $z$-axis of the Bloch's sphere of each qubit.
Rotations  about an axis contained in the equatorial plane of the Bloch sphere are obtained by  the
driving of the so-called carrier transition~\cite{RevModPhys.75.281}, which leads to
\begin{equation}
\label{eq:XYrotation}
    R^{\perp}_{\phi,i}(\theta)=e^{-\ii\frac{\theta}{2}\sum_i(\cos\phi X_i+\sin\phi Y_i)}.
\end{equation}
With this notation,  $R^{\perp}_{0,i}(\theta)$ and $ R^{\perp}_{\pi/2,i}(\theta)$ correspond to rotations around $x$ and $y$ axes of the Bloch sphere, respectively. 
We note that, although these rotations act on all illuminated ions, the equatorial rotations can be applied to a particular set of ions by using spin-echo-type refocusing pulses that interleave these rotations with the addressable $Z$-type rotations~\cite{Nebendahl2009}. Alternatively, one can also use laser addressing techniques for both single and two-qubit gates~\cite{SchmidtKaler2003,Debnath2016,Figgatt2019,PRXQuantum.2.020343}. 

\begin{figure}[t!]
  \includegraphics[width=1\columnwidth]{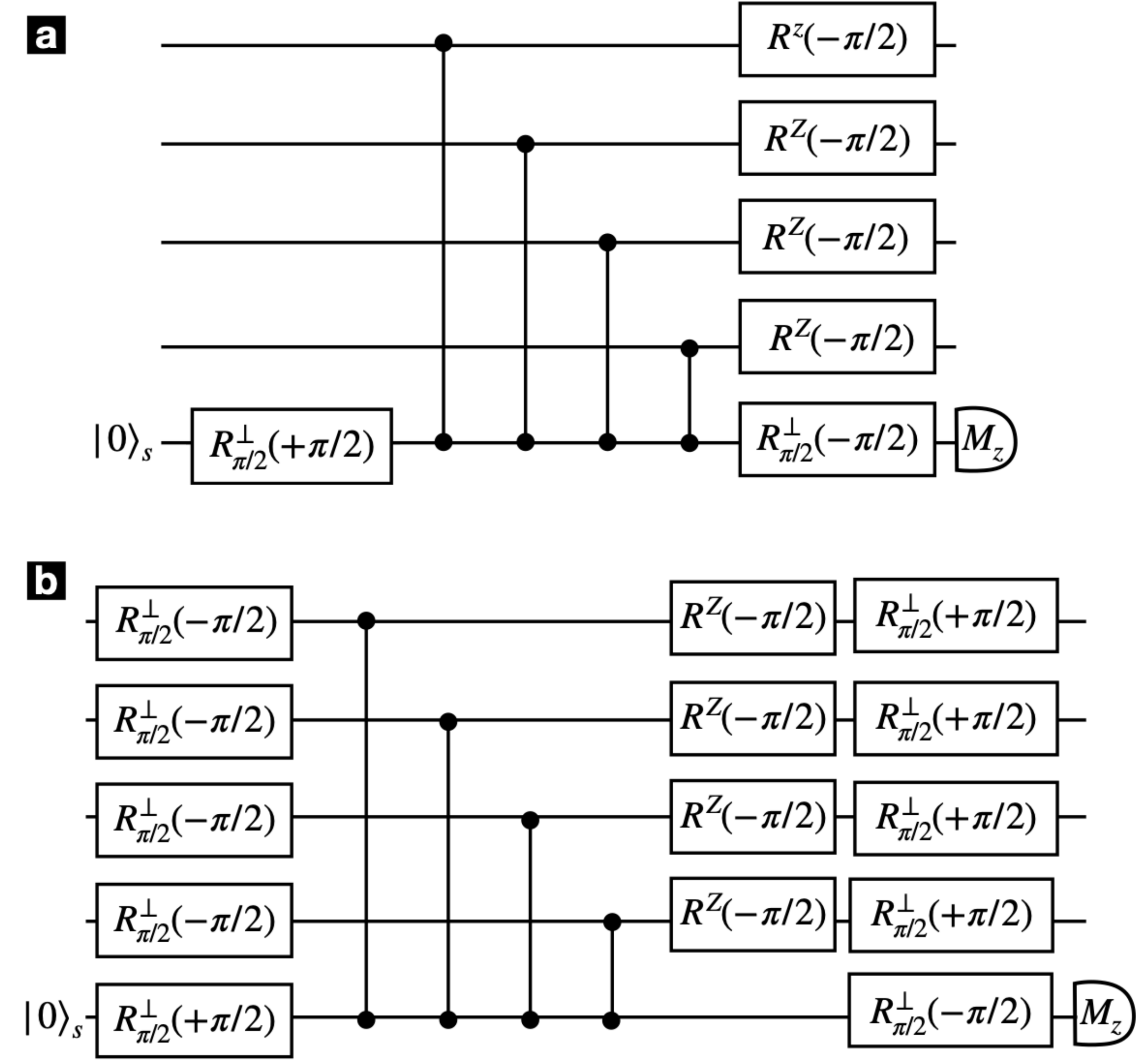}
  \caption{Parity-check measurement circuit using the native trapped-ion gates. (a) $Z$-type parity checks, (b) $X$-type parity checks.} 
  \label{fig:parity_check_measurement}
\end{figure}

Although this collection of gates is not the standard one in quantum computation~\cite{nielsen00}, it forms a universal gate set, so that any quantum algorithm can be decomposed into a sequence of such elementary operations~\cite {Nebendahl2009,Schindler2013}.
 Let us now discuss how this native gate set can be used for the measurement of the parity checks. Since we cannot measure the physical qubits directly, as this would project the state and thus perturb the encoded quantum information, the measurement needs to be performed on a set of ancillary qubits into which the parity-check information has been previously mapped. In principle, this would require an ancilla qubit per plaquette. In the following, we will focus on a single plaquette and thus restrict momentarily to a single ancilla qubit, which we initialize in state $\ket{+}_{\rm s}$, where the subscript refers to the syndrome readout and $\ket{\pm}_{\rm s}=(\ket{0}_{\rm s}\pm\ket{1}_{\rm s})/\sqrt{2}$. We note that the action of the maximally-entangling light-shift gate and a single rotation on the target qubit about the $z$-axis is equivalent, up to an irrelevant global phase,  to a controlled-$Z$ gate ${\rm CZ}_{c,t}=\ket{1_c}\bra{1_c}\otimes Z_t$,
\beq
\label{eq:cz_equivalence}
{\rm CZ}_{c,t}=R^Z_t(-\pi/2)\,\,\,\,\boldsymbol{|}^{\hspace{-0.82ex}\bullet\hspace{-1.62ex}\boldsymbol{-}\hspace{-0.82ex}\boldsymbol{-}c}_{\hspace{-0.8ex}\bullet\hspace{-1.62ex}\boldsymbol{-}\hspace{-0.82ex}\boldsymbol{-}t}=\,\,\,\,\boldsymbol{|}^{\hspace{-0.82ex}\bullet\hspace{-1.62ex}\boldsymbol{-}\hspace{-0.82ex}\boldsymbol{-}c}_{\hspace{-0.8ex}\bullet\hspace{-1.62ex}\boldsymbol{-}\hspace{-0.82ex}\boldsymbol{-}t}\,\,R^Z_t(-\pi/2).
\eeq
In addition, the Hadamard gate $H_i=\ket{+_i}\bra{0_i}+\ket{-_i}\bra{1_i}$ can be decomposed, again up to an irrelevant global phase, as 
\beq
\label{eq:had_equivalence}
H_i=R_i^Z(\pi)R^{\perp}_{\pi/2,i}\left(-{\pi}/{2}\right)=R^{\perp}_{\pi/2,i}\left({\pi}/{2}\right)R^Z_i(\pi).
\eeq
With these  identities, it is straightforward to construct the circuits for the measurement of  parity checks~\eqref{eq:plaquette_gen} starting from the generic circuit  for measuring  Hermitian involutory operators~\cite{nielsen00}. For the $Z$-type parity checks, the circuit is that of Fig.~\ref{fig:parity_check_measurement} {\bf (a)} which is obtained by using Eq.~\eqref{eq:cz_equivalence} for each of the target physical qubits together with Eq.~\eqref{eq:had_equivalence} for the syndrome ancillary qubit. The latter can be simplified by noting that the action of the rotations about the $z$-axis becomes trivial when acting on the initial state, or prior to the measurement $M_z$ in the $Z$-basis.  For the $X$-type parity checks, the circuit is that of Fig.~\ref{fig:parity_check_measurement} {\bf (b)}, and can be obtained by applying  the Hadamard gate~\eqref{eq:had_equivalence}  to all physical qubits twice,  right before and right after the gates of Fig.~\ref{fig:parity_check_measurement} {\bf (a)}, which rotates the controlled-$Z$ operations to the CNOTs required for the measurements of the $X$-type plaquette operators. For each of the Hadamards, one can choose  a specific ordering~\eqref{eq:had_equivalence} to simplify the circuits considering their commutation with the light-shift gates~\eqref{eq:zz}.

In this work, we are interested in assessing this QEC building block from the perspective of multipartite entanglement. We note that for an arbitrary state of the physical qubits  $\ket{0_s}\ket{\psi}\mapsto \ket{0_s}\half(I^{\otimes n}+g_{\alpha}^{(q)})\ket{\psi}+\ket{1_s}\half(I^{\otimes n}-g_{\alpha}^{(q)})\ket{\psi}$ , such that the   measurement $M_z$ on the 
ancilla qubit projects the state onto the respective $\pm 1$ eigenspace of the parity check, giving access to the  complete error syndrome $\{s_\alpha^{(q)}\}$ when performed over all plaquettes. In order to characterize this QEC primitive using its ability to generate multipartite entanglement, we will dispense with the ancilla using a rotation about the $y$-axis followed by a $Z$ measurement in the circuit of a single $X$-type plaquette (cf. Fig.~\ref{fig:parity_check_measurement} {\bf (b)}).  Considering the specific initial state $\ket{\psi_{+,z}}=\ket{0,0,0,0}$, one readily sees that a 5-qubit GHZ state is produced deterministically,
\beq
\label{eq:ghz_z}
\ket{\psi_{+,z}}\ket{0_s}\mapsto\frac{1}{\sqrt{2}}\left(\ket{0}^{\otimes 5}+\ket{1}^{\otimes 5}\right)=\ket{\textrm{GHZ}_{z,+}^5}.
\eeq
Analogously, in the circuit of a single $Z$-type plaquette (cf. Fig.~\ref{fig:parity_check_measurement} {\bf (a)}), we  dispense with the ancilla using a single rotation about the $z$-axis, $R^Z_s(\pi)$, prior to the last rotation of ancilla about the $y$-axis followed by a $Z$ measurement such that, altogether, they lead to the Hadamard gate~\eqref{eq:had_equivalence}.  If we now consider the initial state  $\ket{\psi_{+,x}}=\ket{+,+,+,+}$, one readily sees that a 5-qubit GHZ state is produced deterministically,
\beq
\label{eq:ghz_x}
\ket{\psi_{+,x}}\ket{0_s}\mapsto\frac{1}{\sqrt{2}}\left(\ket{+}^{\otimes 5}+\ket{-}^{\otimes 5}\right)=\ket{\textrm{GHZ}_{x,+}^5}.
\eeq

Both of these states are maximally entangled.
Let us note that none of the above schemes are fault-tolerant, as a single qubit error could propagate through the entangling gates,  which leads to a pair of errors.
Such errors could not be corrected by the small 7-qubit color code. In order to see that, recall the error-propagation relations \cite{PRXQuantum.2.020304}
\beq
\label{eq:loght_shift_gate_propagation}
\begin{split}
\,\,\,\,\boldsymbol{|}^{\hspace{-0.82ex}\bullet\hspace{-1.62ex}\boldsymbol{-}\hspace{-0.82ex}\boldsymbol{-}c}_{\hspace{-0.8ex}\bullet\hspace{-1.62ex}\boldsymbol{-}\hspace{-0.82ex}\boldsymbol{-}t}\,\,X_c&=Y_cZ_t\,\,\,\,\boldsymbol{|}^{\hspace{-0.82ex}\bullet\hspace{-1.62ex}\boldsymbol{-}\hspace{-0.82ex}\boldsymbol{-}c}_{\hspace{-0.8ex}\bullet\hspace{-1.62ex}\boldsymbol{-}\hspace{-0.82ex}\boldsymbol{-}t}, \hspace{2ex} \,\,\,\,\boldsymbol{|}^{\hspace{-0.82ex}\bullet\hspace{-1.62ex}\boldsymbol{-}\hspace{-0.82ex}\boldsymbol{-}c}_{\hspace{-0.8ex}\bullet\hspace{-1.62ex}\boldsymbol{-}\hspace{-0.82ex}\boldsymbol{-}t}\,\,X_t=Z_cY_t\,\,\,\,\boldsymbol{|}^{\hspace{-0.82ex}\bullet\hspace{-1.62ex}\boldsymbol{-}\hspace{-0.82ex}\boldsymbol{-}c}_{\hspace{-0.8ex}\bullet\hspace{-1.62ex}\boldsymbol{-}\hspace{-0.82ex}\boldsymbol{-}t},\\
\,\,\,\,\boldsymbol{|}^{\hspace{-0.82ex}\bullet\hspace{-1.62ex}\boldsymbol{-}\hspace{-0.82ex}\boldsymbol{-}c}_{\hspace{-0.8ex}\bullet\hspace{-1.62ex}\boldsymbol{-}\hspace{-0.82ex}\boldsymbol{-}t}\,\,Z_c&=Z_c\,\,\,\,\boldsymbol{|}^{\hspace{-0.82ex}\bullet\hspace{-1.62ex}\boldsymbol{-}\hspace{-0.82ex}\boldsymbol{-}c}_{\hspace{-0.8ex}\bullet\hspace{-1.62ex}\boldsymbol{-}\hspace{-0.82ex}\boldsymbol{-}t}, \hspace{4ex} \,\,\,\,\boldsymbol{|}^{\hspace{-0.82ex}\bullet\hspace{-1.62ex}\boldsymbol{-}\hspace{-0.82ex}\boldsymbol{-}c}_{\hspace{-0.8ex}\bullet\hspace{-1.62ex}\boldsymbol{-}\hspace{-0.82ex}\boldsymbol{-}t}\,\,Z_t=Z_t\,\,\,\,\boldsymbol{|}^{\hspace{-0.82ex}\bullet\hspace{-1.62ex}\boldsymbol{-}\hspace{-0.82ex}\boldsymbol{-}c}_{\hspace{-0.8ex}\bullet\hspace{-1.62ex}\boldsymbol{-}\hspace{-0.82ex}\boldsymbol{-}t}.
\end{split}
\eeq
Accordingly, phase flip errors commute through the gates and no further errors are triggered. On the other hand, the bit flip errors occurring on any of the two qubits are rotated into a $Y$- type error, and also triggers an additional phase-flip error in the other qubit. In light of this, recalling propagation rules for $Y$-type errors is also useful
\beq
\,\,\,\,\boldsymbol{|}^{\hspace{-0.82ex}\bullet\hspace{-1.62ex}\boldsymbol{-}\hspace{-0.82ex}\boldsymbol{-}c}_{\hspace{-0.8ex}\bullet\hspace{-1.62ex}\boldsymbol{-}\hspace{-0.82ex}\boldsymbol{-}t}\,\,Y_c=X_cZ_t\,\,\,\,\boldsymbol{|}^{\hspace{-0.82ex}\bullet\hspace{-1.62ex}\boldsymbol{-}\hspace{-0.82ex}\boldsymbol{-}c}_{\hspace{-0.8ex}\bullet\hspace{-1.62ex}\boldsymbol{-}\hspace{-0.82ex}\boldsymbol{-}t}, \hspace{2ex} \,\,\,\,\boldsymbol{|}^{\hspace{-0.82ex}\bullet\hspace{-1.62ex}\boldsymbol{-}\hspace{-0.82ex}\boldsymbol{-}c}_{\hspace{-0.8ex}\bullet\hspace{-1.62ex}\boldsymbol{-}\hspace{-0.82ex}\boldsymbol{-}t}\,\,Y_t=Z_cX_t\,\,\,\,\boldsymbol{|}^{\hspace{-0.82ex}\bullet\hspace{-1.62ex}\boldsymbol{-}\hspace{-0.82ex}\boldsymbol{-}c}_{\hspace{-0.8ex}\bullet\hspace{-1.62ex}\boldsymbol{-}\hspace{-0.82ex}\boldsymbol{-}t}.
\eeq
In both circuits {\bf (a)} and {\bf (b)} of Fig.~\ref{fig:parity_check_measurement}, we see that a bit-flip error between the second and the third light-shift gates results in a pair of phase-flip errors in the plaquette, which leads to an uncorrectable error in the 7-qubit code. On the contrary, phase flip errors do not lead to this cascading of errors and, as shall be discussed below, circuits based on light-shift gates can be beneficial for certain biased noise models.

\begin{figure}[t!]
  \includegraphics[width=1\columnwidth]{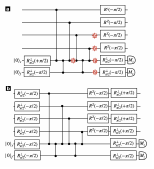}
  \caption{Parity-check measurement circuit using flag qubits. (a) $Z$-type parity checks, (b) $X$-type parity checks.} 
  \label{fig:parity_check_measurement_FT}
\end{figure}

A possibility to make parity-check circuits fault tolerant (FT) is to  add as many ancilla qubits as the weight of the parity checks ~\cite{PhysRevX.7.041061}.
This method requires preparing and verifying those ancillae in  a GHZ state~\cite{PhysRevLett.77.3260,PhysRevLett.98.020501}, which increases the complexity of the trapped-ion circuits considerably, and thus, requires further improvements before one can see the benefits of QEC~\cite{PhysRevX.7.041061}.
More recent FT schemes exploit the so-called flag qubits, which are ancillary qubits introduced to detect the dangerous cascading of errors.  
The flag-based technique reduces qubit overhead and circuit complexity~\cite{PhysRevLett.121.050502, Chamberland2018flagfaulttolerant, Chamberland2019faulttolerantmagic,PhysRevA.100.062307, Reichardt_2020, PhysRevA.101.012342, Chamberland_2020,PhysRevA.101.012342,PRXQuantum.1.010302}. 
For the distance-3 color code, this technique uses a single additional ancilla, the \textit{flag} qubit, which is coupled to the syndrome qubit by a pair of light-shift gates. These extra gates serve to detect the single-qubit errors that occur on the ancillary qubit and propagate into a pair of errors in the data qubits (see Fig.~\ref{fig:parity_check_measurement_FT} {\bf (a)}). 
These additional gates ensure that these weight-2 errors  also propagate onto the flag qubit, which would be signalled by a $-1$ outcome of the flag-qubit measurement.
The flag readout, when combined with the subsequent parity check measurements, can be used to unequivocally identify and deterministically correct  dangerous errors, achieving the desired fault tolerance~\cite{PhysRevLett.121.050502,PhysRevA.100.062307}.

In the present context, we will instead consider a post-selection scheme in which, after running the circuits from Fig.~\ref{fig:parity_check_measurement_FT}, one only keeps the events where the flag is not triggered, and the circuit generates the GHZ entangled states in Eqs.~\eqref{eq:ghz_z}-\eqref{eq:ghz_x} depending on the initial state {with} the corresponding $Z$- or $X$-type stabilizer.

\subsection{Entanglement witnesses for QEC primitives}

In this section, following the work presented in~\cite{PRXQuantum.2.020304,PhysRevX.12.011032}, we discuss how to quantify the capability  of the parity-check  circuits to generate maximally-entangled states in the presence of  noise. As argued in~\cite{PRXQuantum.2.020304}, verifying genuine multipartite entanglement  is not an easy task as the number of bipartitions in which   entanglement must be checked grows exponentially with the number of qubits. Moreover, in some cases, the required measurements designed theoretically are difficult to implement experimentally. An interesting strategy to overcome these limitations for entanglement detection is to use entanglement witnesses~\cite{HORODECKI19961,Terhal2000,Lewenstein2000,Sperling2013,Horodecki2009}, which are observables signalling the entanglement of the state  when a negative expectation value is obtained.  Interestingly, entanglement witnesses can be constructed from local and experimentally-friendly observables~\cite{Toth2005,PRXQuantum.2.020304}.
We will consider noisy parity-check  circuits using imperfect measurements  in two different ways:
{\it (i)} \textit{standard linear witnessing} (SL) with a single witness that only relies on  a linear number of measurements~\cite{Toth2005}, and
{\it (ii)}  \textit{conditional linear witnessing} (CL) that requires a linear number of witnesses and measurements~\cite{PRXQuantum.2.020304}.\\

Let us start by discussing how to construct these witness operators $W_{n}$ for the  parity-check circuits, following the design principles discussed in~\cite{Toth2005,PhysRevLett.94.060501} for stabilizer states. An entanglement witness $W$ is an observable whose expectation value is, by construction, non-negative $\text{tr}(W\rho_{s}) \geq 0$ for all separable states $\rho_s=\sum_kp_k\rho^{(k)}_1\otimes\rho^{(k)}_2\otimes\cdots\otimes\rho^{(k)}_n$~\cite{PhysRevA.40.4277}.  Therefore, if the witness is negative for a certain state, $\text{tr}(W\rho) < 0$, one can ascertain that $\rho$ is not separable and, thus, that it contains some sort of  entanglement. Note that, however,  with the above definition of the witness using fully separable states with $n$ parties, the witnessed entanglement may be present only within a specific bipartition of the system, whereas other bipartitions may be consistent with separability.

A stronger entanglement criteria is obtained by considering states that are not biseparable, i.e., those who do not admit a decomposition of the form $\rho_s=\sum_kp_k\rho^{(k)}_A\otimes\rho^{(k)}_B$ in any bipartition $AB$.
A state possesses the strongest form of entanglement, namely, genuine multipartite entanglement (GME), if it further cannot be written as a convex combination of biseparable states with respect to different bipartitions.

If a noisy imperfect evolution produces a state $\rho$ that is expected to be close to a target entangled $n$-qubit pure state $\ket{\psi^{(n)}}$, one can construct a projector-type   witness 
\begin{eqnarray}
\centering
W_{n} =l_{n}\mathbb{I}-\ketbra{ \psi^{(n)}},
\label{eq:witnessgeneral}
\end{eqnarray}
where $l_{n}$ is the smallest constant for which every biseparable state $\rho_{s}$ satisfies $\text{tr}(W_n\rho_{s}) \geq 0$ \cite{Toth2005,PhysRevLett.94.060501}. Accordingly, when the fidelity of a state with respect to the target entangled state fulfils $\mathcal{F}=\bra{\psi^{(n)}}\rho\ket{\psi^{(n)}}>l_n$, the state is sufficiently close to the target state so that $\text{tr}(W_n\rho) < 0$ and one concludes that $\rho$ is indeed entangled. For the specific target states in Eqs.~\eqref{eq:ghz_z}-\eqref{eq:ghz_x}, the  GME witness bound is $l_n=1/2$~\cite{PhysRevLett.94.060501}.

Depending on the specific target state, the above witnesses might  be  complicated observables with a large overhead in measurement complexity. In general, they require  $2^{n-1}-1$ different local Pauli measurements, which is exponential in the number of qubits $n$. This has motivated the development of alternative witnesses, such as the witnesses proposed by T\'oth and G\"uhne \cite{Toth2005,PhysRevLett.94.060501}, or the conditional witness method~\cite{PRXQuantum.2.020304}, both of which are efficient in the number of bipartitions in which entanglement must be checked. They reduce the overhead on the required number of measurements to $n$ rather than $2^{n-1}-1$. 

We describe below both protocols by focusing on a specific type of entangled state. These are stabilizer states fulfilling 
\beq
g_i\ket{\psi^{(n)}}=+\ket{\psi^{(n)}},\,\,\forall i\in\{1,\cdots,n\},
\eeq
where the $n$ Hermitian, involutory, and mutually-commuting operators of the Pauli group $g_i\in\mathcal{P}_n$ generate the stabilizer group $\mathcal{S}_{n}= \langle g_{1}, g_{2}..., g_{n} \rangle$, and uniquely determine the target entangled state $\ket{\psi^{(n)}}$. We note that the number of qubits of the multipartite state  $n$ should not be confused with the number of physical qubits in the code $n_c$ and that, likewise, the stabilizer group associated with the state  $\tilde{\mathcal{S}}_n$ should not be confused with that of the QEC codes $\mathcal{S}_G$ discussed above. In our case, the target state for the $X$-type~\eqref{eq:ghz_z} ($Z$-type~\eqref{eq:ghz_x}) circuit contains $n=5$ qubits, such that the respective generators $g_i^x$ ($g_i^z$) are not the two  parity checks~\eqref{eq:plaquette_gen} of the corresponding plaquette, rather the  two following sets
\beq
\begin{split}
g_i^x=Z_iZ_{i+1},\forall i\in\{1,\cdots,4\},
\hspace{2ex}g_5^x=\prod_{i=1}^5X_i,\\
g_i^z=X_iX_{i+1},\forall i\in\{1,\cdots,4\},
\hspace{2ex}g_5^z=\prod_{i=1}^5Z_i.
\end{split}\label{eq:CSS_stabilizers}
\eeq
Note that, in light of the GHZ-type target states in Eqs.~\eqref{eq:ghz_z} and~\eqref{eq:ghz_x}, one could also define the weight-two parity checks as being composed of other pairs of physical qubits. This will be important when considering the robustness of the witness against biased noise. 

The  projector that appears in the general entanglement witness for $n$-partite pure states~\eqref{eq:witnessgeneral} can be expressed in terms of these stabilizer generators. In particular, the projector used in the construction of the witness can be obtained  by the product of all projectors onto the $+1$ eigenspace of the generators 
$\ketbra{ \psi^{(n)}}=\prod_{i=1}^{n}\half\big(I^{\otimes n}+g_{i} \big)=L_{n}$, where $L_n$ will be  denoted as the entanglement test operator. This operator contains all products of the stabilizers and, since they belong to the Pauli group, can be inferred from local measurements. However, the corresponding product  increases the number of required  measurement basis. We now discuss two witnesses that minimize the measurement overhead.
\subsubsection{Standard linear witnessing}\label{sec:SL_witness}
As discussed in~\cite{Toth2005}, a GME  witness that only requires a linear number of measurements can be defined by
\begin{equation}
\centering
W_{n}^{\rm{SL}}=(n-1){I}^{\otimes n}-\sum_{i=1}^{n}g_{i},
\label{eq:optimizedW}
\end{equation}
where $l_n=(n-1)$ is the corresponding entanglement bound \cite{Toth2005}. In this case, the linear number of measurements is given by measuring the expectation value of the $n$ generators of the target state, rather than all stabilizers.
In this work,  for convenience, we choose  a normalized version
\begin{eqnarray}
\centering
\widetilde{W}_{n}^{\rm{SL}} = \tilde{l}_{n}^{\rm{SL}}{I}^{\otimes n}-L_{n}^{\rm{SL}},
\label{eq:normalizedwitness}
\end{eqnarray}
where the test operator $L_{n}^{\rm{SL}}$ is  the average of  the generators    
\beq
L_{n}^{\rm{SL}}=\frac{1}{n}\sum_{i}^{n} g_{i} ,
\eeq
and the witness bound is 
\beq
\tilde{l}_{n}^{\rm{SL}}=\frac{n-1}{n}.
\eeq
One can see that the expectation value of the test operator 
 becomes +1 for the ideal GME states in Eqs.~\eqref{eq:ghz_z}-\eqref{eq:ghz_x}.
If $\langle L_{n}^{\rm {SL}}\rangle$ surpasses the critical value of $\tilde{l}_{n}^{\rm{SL}}$, one would obtain $\langle \widetilde{W}^{\rm SL}_{n} \rangle < 0$, signalling a GME state of $n$ qubits.

In the SL witnessing, we need  to measure $n$ different stabilizer generators, which could, in principle, require up to $n$ different measurement settings. For the particular case of parity check circuits whose output states have CSS-type stabilizers \cite{Devitt_2013}, i.e. its generators $g_i$ are formed of either $X$- or $Z$-type Pauli operators only as in Eqs.~\eqref{eq:CSS_stabilizers}, it has been shown  that the required number of measurement settings reduces from $n$ to just 2 settings: measuring all qubits in  the $X$- and the $Z$- basis suffices to determine the expectation value of all operators contributing to the witness $\widetilde{W}_{n}^{\rm{SL}}$~\cite{PhysRevLett.94.060501}. For both, non-FT and FT  circuits that allow us to measure the $X$-type parity checks, as shown in Fig.~\ref{fig:parity_check_measurement} {\bf (b)} and in  Fig.~\ref{fig:parity_check_measurement_FT} {\bf (b)}, the target 5-qubit output state is $\ket{\textrm{GHZ}_{z,+}^5}$, as given in Eq.~\eqref{eq:ghz_z}. For the case of the FT-circuit, we obtain the target 5-qubit GME state after post-selecting on the measurement events in which  the flag-qubit is not triggered. This state is stabilized by the $Z$-type operators of Eq.~\eqref{eq:CSS_stabilizers} and, therefore, just two measurement settings are needed. Likewise, for the non-FT and FT $Z$-type circuits shown in Fig.~\ref{fig:parity_check_measurement} {\bf (a)} and  Fig.~\ref{fig:parity_check_measurement_FT} {\bf (a)}, the target output state is given by the 5-qubit GME state $\ket{\textrm{GHZ}_{x,+}^5}$  shown in  Eq.~\eqref{eq:ghz_x}. For the FT circuit, we again postselect on the flag qubit. In this case, the stabilizer generators are 
the $x$-type operators of Eq.~\eqref{eq:CSS_stabilizers}, showing that two measurement settings suffice.\\

\subsubsection{Conditional linear witnessing}\label{subsec:GMEwitness_flag}

With the same goal, namely  reducing the exponential number of bipartitions that need to be checked to a linear one, and hence, making the  entanglement detection efficient,  we have introduced conditional entanglement witnessing in~\cite{PRXQuantum.2.020304} as a robust and efficient technique to test multipartite entanglement (ME)  in  quantum systems. For the conditional witness, we will refer to the verification of ME instead of GME, following the definition as in Refs.~\cite{Guhne2009,Huber2010}. In this definition, GME detection terminology is explicitly used for states that are not within the convex hull of all biseparable states, while ME describes the states that are not separable within any bipartition. We proved that, given an $n$-partite system, certification of entanglement between the subsystems $s_1$ and $s_x$ for all $s_x\in\{s_1\}\us{c}=\{s_1,\cdots,s_n\}\setminus\{s_1\}$, conditioned on an outcome $i$ of suitable local measurements on the remaining $n-2$ subsystems, is actually sufficient for the certification of  ME. Moreover, 
the number of bipartitions needed to be checked gets reduced to $n-1$ ~\cite{PRXQuantum.2.020304}. In this case, the conditional entanglement witness has an analogous form as before in~\eqref{eq:normalizedwitness}, namely, 
\begin{eqnarray}
\centering
W^{\rm CL} =l^{\rm CL}\mathbb{I}- L^{\rm CL},
\label{eq:witnessconditional}
\end{eqnarray}
where $L^{\rm{CL}}$ is the conditional test operator, and $l^{\rm CL}$ is the corresponding witness bound.
Let us now give the explicit expressions of the conditional witness for the 5-qubit target multipartite entangled state in Eqs.~\eqref{eq:ghz_z} and~\eqref{eq:ghz_x}.     
In a standard  scenario to  certify  $n$-party entanglement in a $n=5$ partite system, we would need to test entanglement in $N^{b}=2^{4}-1=15$ bipartitions; a number that grows exponentially with the number of subsystems $n$. On the contrary, conditional entanglement achieves the verification of entanglement in a number of bipartitions that just grows linearly with the number of subsystems. The method works by localizing entanglement between two partitions of the 5-qubit multipartite system, by performing local measurements on the three remaining subsystems ~\cite{Verstraete2004,Popp2005}. This reduces the exponential complexity of the $n$-party entanglement detection  to  a  linear  number  of  tests. For the particular case of study, it would be then enough to test entanglement in $n-1=4$ bipartitions instead of in 15. We are going to focus merely on the four bipartitions $[s_5|s_1]$, $[s_5|s_2]$, $[s_5|s_3]$, and $[s_5|s_4]$. In this case, within each bipartition, we check if each subsystem data qubit is entangled with the syndrome ancilla subsystem. However, a different choice of bipartitions is also valid. Then, for the certification of multipartite entanglement in the 5-qubit states from Eqs.~\eqref{eq:ghz_z} and~\eqref{eq:ghz_x}, it suffices to test entanglement in each bipartition $[s_5|s_{x}]$ for $x\in\{1,2,3,4\}$, conditioned on the outcome of a projective measurement on the remaining three qubits $s_{x'}$ $\in\{s_5,s_x\}\us{c}$. The conditional entanglement test operators for each bipartition are
\begin{equation}\label{eq:condWitness_5_numeric}
\begin{split}
    & L_{[s_{5}|s_{x}]}^{\rm{CL,z}}=\ket{\Bell_{s_{5}|s_{x}}^{\rm{z}}}\bra{\Bell_{s_{5}|s_{x}}^{\rm{z}}} \otimes P^0_{\{s_{5},s_{x}\}^c}\\
    & L_{[s_{5}|s_{x}]}^{\rm{CL,x}}=\ket{\Bell_{s_{5}|s_{x}}^{\rm{x}}}\bra{\Bell_{s_{5}|s_{x}}^{\rm{x}}} \otimes P^{+}_{\{s_{5},s_{x}\}^c}\\
\end{split}    
\end{equation}
 where $L_{[s_{5}|s_{x}]}^{\rm{CL,z}}$ is the conditional test operator for the $Z$-type parity-check circuit output state, whereas $ L_{[s_{5}|s_{x}]}^{\rm{CL,x}}$ is the conditional test operator for the $X$-type. In this case, in addition to the  projector  of the qubit pair $s_{5}$ and $s_{x}$ onto the Bell state $\ket{\Bell^{\rm{z}}}=(\ket{++}+\ket{--})/\sqrt{2}$, we condition on the projection of the remaining qubits into  
\beq
P^0_{\{s_{5},s_{x}\}^{c}}=\bigotimes_{s_{x'}\in\{s_{5},s_{x}\}^{c}}\ketbra{0_{s_{x'}}}.
\eeq
The test operator in the X-type circuit for the bipartition $[s_{5}|s_{x}]$ is given by $ L_{[s_{5}|s_{x}]}^{\rm{CL,x}}$. The main difference with respect to the Z-type conditional test operator is that now the projector on the remaining three qubits is
\beq
P^+_{\{s_{5},s_{x}\}^{c}}=\bigotimes_{s_{x'}\in\{s_{5},s_{x}\}^{c}}\ketbra{+_{s_{x'}}},
\eeq
and the conditional Bell pair is  $\ket{\Bell^{\rm{x}}}=(\ket{00}+\ket{11})/\sqrt{2}$. 

The Bell pairs can be described in terms of the same weight-2 stabilizer generators $\{X_{s_x}X_{s_5}, Z_{s_x}Z_{s_5}\}$, so that the projector can be specified by the product of these two generators  and, thus, the conditional test operators can be expressed as a   linear combination of two-point correlations 
\begin{equation}\label{eq:CL_test_operator}
    L_{[s_{5}|s_{x}]}^{\rm{CL,x/z}}={I_{s_{x}}I_{s_{5}}+X_{s_{x}}X_{s_{5}}-Y_{s_{x}}Y_{s_{5}}+Z_{s_{x}}Z_{s_{5}} \over 4},
\end{equation}
 In addition, the entanglement bounds for  Bell states are $l_{[s_{5}|s_{x}]}^{\rm{CL,x/z}}=1/2$  ~\cite{PhysRevLett.94.060501, PRXQuantum.2.020304} for all the bipartitions. Thus, the conditional test operator must surpass this value, $\expec{L_{[s_{5}|s_{x}]}^{\rm{CL,x/z}}}>1/2$, in order to obtain a negative value for the conditional witness, that is, $\langle W^{\rm CL, x/z}_{[s_{5}|s_{x}]} \rangle < 0$.
 
 In this occasion, we need three measurement settings $\{XX, ZZ, YY\}$ per bipartition. Hence, for the $n-1=4$ bipartitions, a total of twelve measurement settings would be required for this witnessing method. 
One can argue that this number is considerably higher than the  two settings needed in the SL witness. However, this method has a larger witness bound. For the SL method, we have a $\tilde{l}^{\rm{SL}}=5/6$, which is larger than the one required in the CL method, $l^{\rm{CL}}=1/2$. This implies that, with the SL method, we lose sensitivity in the verification of entanglement. There are states that lie within the convex hull of all biseparable states that would not be detected by using the SL witness~\cite{PRXQuantum.2.020304}. On the contrary, with the CL method, we build $n-1$ witnesses that can be tangent to the set of biseparable states, thus increasing the sensitivity on detecting  multipartite entanglement \cite{PRXQuantum.2.020304}. We would see later in the text that the CL method is more robust than the SL against the circuit noise models considered.

\section{\textbf{Error model for light-shift gates}}\label{Sec:non_ideal_conditions}

So far, we have presented the native trapped-ion gate set in Eqs.~\eqref{eq:zz}-\eqref{eq:XYrotation} from a high-level perspective. This suffices to find trapped-ion compilations of specific circuits and analyze the propagation of generic phenomenological errors, as discussed in the previous section. However, in order to give  a realistic account of the performance of the circuits and, in the present context, assess the detection of multipartite entanglement, one needs to delve into a lower-level description and derive a microscopic noise model that captures how the gates deviate in practice from the ideal unitaries~\eqref{eq:zz}-\eqref{eq:XYrotation}. In~\cite{PRXQuantum.2.020304}, we evaluated the robustness of the previous entanglement witnesses for a  phenomenological noise model that consisted of a depolarizing channel~\cite{nielsen00}, together with bit-flip errors in the measurements. Each of these noise sources was controlled by a specific error rate, and the robustness of the witness was studied as one increases the severity of both noise sources. We note that, however, it is possible to go beyond this simplified model and derive more realistic microscopic error channels with rates that are fixed by specific experimental parameters. As noted in the introduction, this
research direction has been developed to some extent for optical~\cite{PhysRevX.7.041061, PhysRevA.99.022330,PhysRevA.100.062307,ParradoRodriguez2021crosstalk} and hyperfine~\cite{Trout_2018, Debroy_2020,https://doi.org/10.48550/arxiv.2004.04706,9407237,Tinkey_2021} trapped-ion qubits operated with M\o lmer-Sorensen entangling gates~\cite{PhysRevA.62.022311}. The goal of this section is to derive a microscopic noise model for trapped-ion Zeeman qubits operated with single-qubit rotations and entangling light-shift gates. As mentioned before, the bottleneck is the performance of entangling gates, and we will thus focus on the error model for them.

\subsection{Ideal entangling light-shift gates}\label{sec:ideal_ZZgate}

\begin{figure}[t!]
  \includegraphics[width=1\columnwidth]{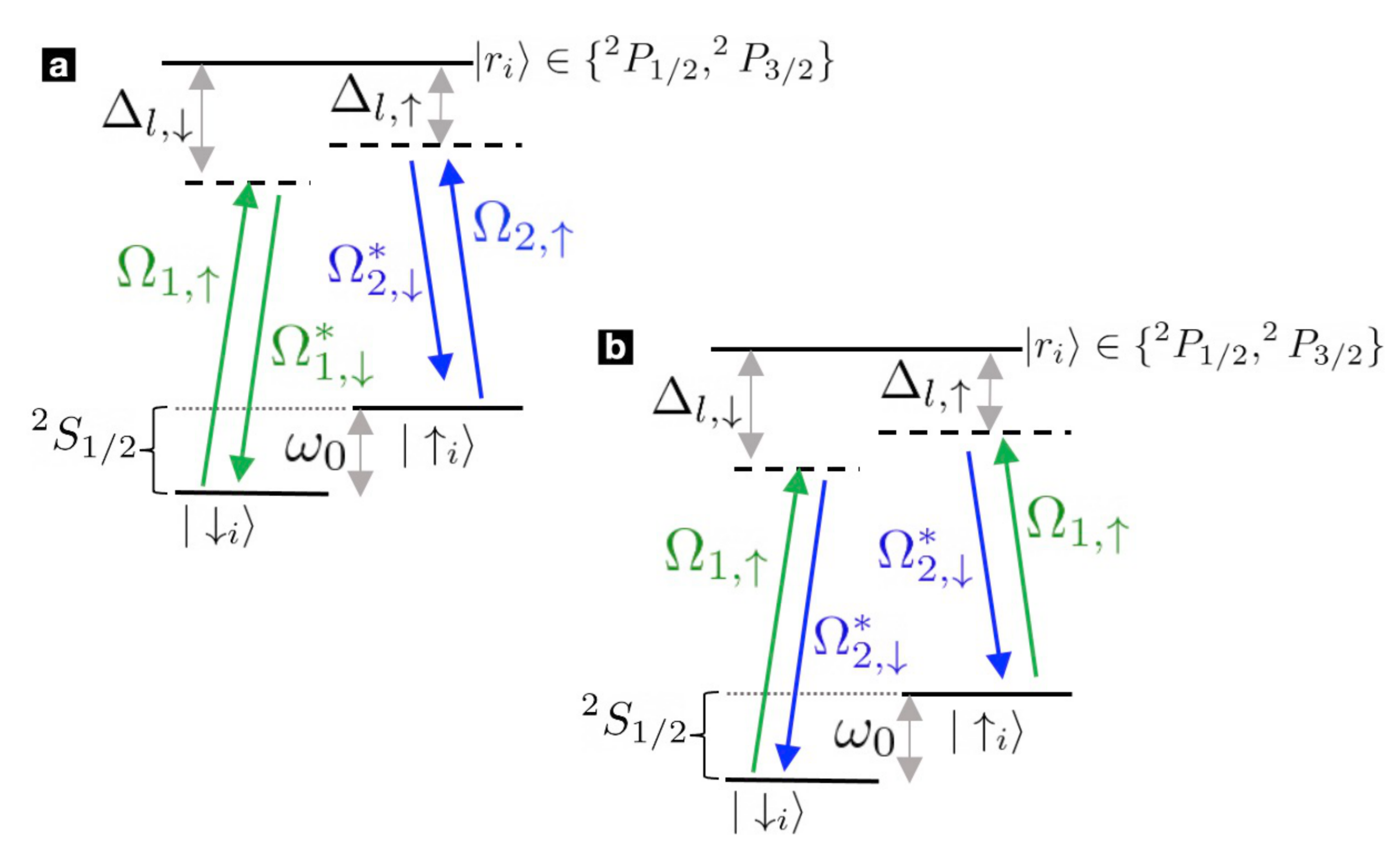}
  \caption{Atomic levels and laser beam configurations used to create light-shift gates in $^{40}\rm{Ca}^{+}$ ions with the qubit $s\in\{ \ket{\uparrow}, \ket{\downarrow} \}$ encoded in the Zeeman sublevels of the $S_{1/2}$ ground state. $\Omega_{l,s}$ are the Rabi frequencies of the dipole-allowed transitions, with $l\in\{ 1,2\}$ labeling a pair of non-copropagating laser beams. We consider that the detunings of these laser beams $\Delta_{l, s}$ are much bigger than the qubit frequency splitting  $\omega_{0} \ll \Delta_{l, \downarrow}\approx \Delta_{l, \uparrow}=\Delta$. The effect of this laser-beam configuration leads to second order (two-photon) process depicted  in {\bf (a)} where the ion virtually absorbs and emits a photon from and into the same laser, and {\bf (b)} where the ion virtually absorbs from one laser and emits into the other one.}
  \label{fig:2ndorder}
\end{figure}

In order to discuss sources of error, we first need to review the ideal scheme of a light-shift gate. We consider a crystal of $N$ trapped $^{40}\rm{Ca}^{+}$ ions, where the qubit is encoded in the two Zeeman sublevels of the $S_{1/2}$ ground-state~\cite{PhysRevX.12.011032}, with  $m_{J}=-1/2$ for $\ket{\downarrow}=\ket{0}$ and  $m_{J}=1/2$ for $\ket{\uparrow}=\ket{1}$, with $\omega_{0}/2\pi=10.5\,$MHz as the transition frequency splitting between those states obtained by applying an external magnetic field~\cite{PhysRevX.12.011032}. The ions' equilibrium positions $\vec{r}_i^0$ are confined along the null of the rf-fields of a segmented linear Paul trap ~\cite{PhysRevX.12.011032}.
The light-shift gates~\cite{Leibfried2003} can be obtained from a pair of non-copropagating laser beams of frequency $\omega_{l}$ for $l\in\{1,2\}$, which are far off-resonant from the dipole-allowed transitions $S_{1/2}\rightarrow P_{1/2}, P_{3/2}$ (see Fig.~\ref{fig:2ndorder}). The detuning of the  laser beams  from the dipole-allowed transition from  $s\in\{ \ket{\uparrow}, \ket{\downarrow} \}$, is assumed to be much larger than the qubit frequency splitting,  $\omega_{0} \ll \Delta_{l, \downarrow}\approx \Delta_{l, \uparrow}=\Delta$. It is also assumed in the last step that both detunings are approximately equal. Likewise, the detuning must be much larger than the linewidth of the dipole-allowed transition, $\Delta \gg \Gamma$, such that the residual spontaneous scattering is reduced. As we will see below, the effect of such laser beams is a collection of ac-Stark shifts that arise due to second-order (two-photon) processes.

In addition to the internal electronic states, the ions also vibrate around the crystal equilibrium  positions, giving rise to collective oscillations.

Following standard steps~\cite{RevModPhys.75.281,James1998}, let us define the beatnote frequency of the lasers $\Delta \omega_{L}=\omega_{L,1}-\omega_{L,2}$ and denote the three photon axes, the so-called radial $\alpha\in\{x,y\}$ and axial $\alpha=z$, with normal-mode frequencies $\omega_{\alpha,m}$ and creation and annihilation operators $a_{\alpha,m}^{\dagger}, a_{\alpha, m}^{\phantom{\dagger}}$. With $N=2$ trapped ions, we have $m=2$ frequency modes for each $\alpha$-axis. In particular, the center of mass (COM) frequency mode, and the zig-zag (ZZ) frequency mode.
Ion crystals, in general, have micromotions, i.e., periodic motions due to oscillations of the Paul trap's potential.
Excess micromotion are classically driven motions of the ions that off the rf null, and intrinsic micromotions which is a quantum-mechanical driven motion with the rf frequency~\cite{Bermudez_2017}.
Assuming the ions lie in the rf null and restricting to beat note frequencies far detuned from the rf frequencies, it is possible to neglect the excess and intrinsic micromotions and describe the ions' motional degrees of freedom as three decoupled branches.
The internal and motional Hamiltonian of ions is then  described by
\begin{equation}\label{eq:hamiltonian_0}
    H_{0}=\frac{1}{2}\sum_{i=1}^{N}\!\!\big(\omega_{0}+\delta\Tilde{\omega}_{0, i}(t)\big)Z_{i}+\!\!\sum_{m=1}^{N} \sum_{\alpha=x,y,z}\!\!\!\omega_{ \alpha,m}a_{\alpha, m}^{\dagger}a_{ \alpha, m},
\end{equation}
wherein
\begin{equation}\label{eq:ac_shifts}
\begin{split}
& \delta\Tilde{\omega}_{0,i}(t)=\delta \omega_{0,i}+ \delta \omega_{L,i}
\end{split}
\end{equation}
reflects the contribution of the differential ac-Stark shifts.
The first term in $\delta\Tilde{\omega}_{0,i}$ represents an energy shift of the qubit frequency due to an ion virtually absorbing and emitting a photon from and into the same laser beam (see Fig.~\ref{fig:2ndorder}{\bf (a)}).
The second term occurs when an ion virtually absorbs a photon from a laser beam and then emits it into the other (see Fig.~\ref{fig:2ndorder}{\bf (b)}), leading to crossed-beam oscillating ac-Stark shifts.

In more detail,
\begin{equation}
    \delta \omega_{0,i}=\sum_{l}(|\Omega_{l, \uparrow}^{i}|^{2}-|\Omega_{l,\downarrow}^{i}|^{2})/4 \Delta,
\end{equation}
where $\Omega_{l, s}^{i}$ is the Rabi frequency of the corresponding dipole-allowed transition for the $i$-th ion. 
Furthermore,
\begin{equation}
    \delta \omega_{L,i} = |\Tilde{\Omega}_{L,i}|\mathrm{cos}(\Delta \vec{k} \vec{r}_{i}^{0}-\Delta \omega_l t+\phi_{l}).
\end{equation}
In this expression, $\Delta \vec{k}_L=\vec{k}_{L,1}-\vec{k}_{L,2}$ is the differential wave-vector of the laser beams, and $\phi_{L}=\phi_{L,1}-\phi_{L,2}$ is the corresponding  phase mismatch. 
The $\Tilde{\Omega}_{L, i}=(\Omega_{1,\uparrow}\Omega_{2,\uparrow}^{*}-\Omega_{1,\downarrow}\Omega_{2,\downarrow}^{*}){\rm exp}\{-\frac{1}{2}\sum\limits_{m}(\eta_{\alpha,m}{M}_{i,m}^{\alpha})^{2}\}/({4\Delta})$ contains contains the correction of the differential ac-Stark shifts  by the Debye-Waller factors due to zero-point fluctuations of ion's vibrations~\cite{RevModPhys.75.281}, where $M_{i, m}^{\alpha}$ is the $i$-th ion displacement in the $m$-th mode along the $\alpha$-axis, and $\eta_{\alpha, m}=\Delta\vec{k}_L\cdot\textbf{e}_\alpha/\sqrt{2m_{\rm Ca}\omega_{\alpha,m}}\ll 1$ is the Lamb-Dicke parameter~\cite{RevModPhys.75.281} with $m_{\rm Ca}$ being the ion's mass.

The light-matter interaction Hamiltonian is given by
\begin{equation}
    H_{\rm int}\!=\!\!\sum_{\alpha, i, m}\!\!\ii\mathcal{F}_{i,m}^{\alpha}\ee^{\ii(\Delta \vec{k}_L\cdot\vec{r}_{i}^{0}-\Delta\omega_{L} t+\phi_{L})}Z_{i}(a_{\alpha,m}+a_{\alpha,m}^{\dagger})\! +{\rm H.c.},
\label{eq:hamiltonian_I}
\end{equation}
where $\mathcal{F}_{i,m}^{\alpha}$ is a state-dependent dipole force,
\begin{equation}\label{eq:F_strength}
\mathcal{F}_{i,m}^{\alpha}=\ii\frac{|\Tilde{\Omega}_{L,i}|}{2}\eta_{\alpha, m}M_{i,m}^{\alpha}.
\end{equation}
The total light-matter Hamiltonian of the trapped ions interacting with the laser beams is then $H=H_{0}+H_{\rm int}$ ($\hbar=1$).

For the ideal performance of the gate, we assume that $\Delta\vec{k}_L$ is perfectly aligned with one of the radial vibrational branches, e.g. $\Delta \vec{k}_L\parallel  \textbf{e}_{x}$. Note that we have focused on the radial modes as they are more robust to successive shuttling and reconfiguration operations which, as discussed in the introduction, form a crucial part of the native trapped-ion toolbox in scalable shuttling-based approaches for quantum information processing (QIP)~\cite{doi:10.1116/1.5126186}. Now, moving to the interaction picture with respect to $H_{0}$, the resulting Hamiltonian is  
\begin{equation}\label{eq:interaction_Hint}
\begin{split}
    H_{\rm int}(t)& \simeq \sum_{i,m}^{N, M}\ii\mathcal{F}_{i,m}^{x}Z_{i}a_{x,m}^{\dagger}e^{i(\Delta \bar{k}\vec{r}_{i}^{0} +\delta_{m}^{x}t+\phi_{L})}+ {\rm H.c.}
\end{split}
\end{equation}
Here, we have defined the detunings $\delta_{m}^{\alpha}=\omega_{x,m}-\Delta \omega_{L}\ll \omega_{x,m}$, and used the underlying  assumption  $|\Omega_{L, i}|\ll\Delta \omega_{L}\simeq \omega_{x, m}$ which is used to neglect other powers of the vibrational operators in the Lamb-Dicke expansion of the laser-ion interaction. One thus sees that, when the laser beat note is tuned close to the target vibrational branch, it  induces a state-dependent displacement that pushes the ions along different trajectories in phase space depending on the internal state. This results in a geometric phase gate capable of generating entanglement.\\

We can obtain the exact time evolution operator using the so-called Magnus expansion~\cite{https://doi.org/10.1002/cpa.3160070404,Blanes_2010} which, uUnder the assumption of perfectly aligned laser and the neglect of high-order vibrational operators, closes at second order. This results in
\begin{equation}\label{eq:unitary_Hint}
\begin{split}
    U(t)&=\ee^{-\ii H_{0}t}\ee^{\,\sum\limits_{i,m}(\Phi_{i,m}^{x}(t)a_{x, m}^{\dagger}-{\rm H.c.})Z_{i}}\\
    & \times \ee^{-\ii\sum\limits_{i,j}Z_{i}Z_{j}\left(\frac{J_{ij}t}{2}-2\sum\limits_{m}\textrm{Re}\{\Phi_{i,m}^{\alpha}(t)(\Phi_{i,m}^{\alpha}(0))^*\}\right)},
\end{split}    
\end{equation}
where we have defined
\begin{equation}\label{eq:displ_phase_space}
  \Phi_{i,m}^{x}(t)=\frac{\mathcal{F}^x_{i,m}}{\delta^x_{m}}\ee^{\ii(\Delta \vec{k}_L\cdot \vec{r}_{i}^{0}+\phi_{L})}(1-\ee^{\ii\delta^x_{m}t}).
\end{equation}
We thus see that the time-evolution operator contains state-dependent displacement operators, i.e. $\mathcal{D}_\pm(\Phi_{i,m}^{x}(t))={\rm exp}(\pm\Phi_{i,m}^{x}(t)a_{x, m}^{\dagger}\mp\Phi_{i,m}^{x}(t)^{*}a_{x, m})$ for $\ket{\uparrow},\ket{\downarrow}$, which can lead to residual qubit-phonon entanglement  if the trajectories are not closed, i.e. if $\Phi_{i,m}^{x}(t)\neq 0.$ The evolution operator also contains a qubit-qubit coupling term of strength 
\begin{equation}\label{eq:spin_spin_couplings}
    J_{ij}=-2\sum_{m}\frac{\mathcal{F}^x_{i,m}\mathcal{F}^x_{j,m}}{\delta^x_{m}}\mathrm{cos}(\phi_{ij}^{0}), 
\end{equation}
that generates entanglement through phonon-mediated interactions~\cite{PhysRevLett.92.207901,PhysRevLett.97.050505,Lin_2009,2010RPPh...73c6401B}. The phase $\phi_{ij}^{0}=\Delta \vec{k}_L\cdot(\vec{r}_{i}^{0}-\vec{r}_{j}^{0})$ causes the coupling strength $J_{ij}$ to oscillate with the distance between ions in a generic situation. However, for radial phonons, this oscillation disappears. We note that off-resonant terms additional to the force in  Eq.~\eqref{eq:interaction_Hint} can also be accounted for in the Magnus expansion, which lead to small perturbations to the spin-spin couplings~\cite{Lin_2009}.

An ideal implementation of the two-qubit light-shift gate exploits these  phonon-mediated interactions to generate entanglement and minimizes the residual qubit-phonon entanglement with the active phonons mediating the entangling operations. By active mode we refer to the mode $m$ along an $\alpha$-axis direction used to perform the entangling gate, the remaining $3N-1$ modes  are refereed to as the spectator modes, which in an ideal scenario are not involved in the entangling operation.. Using a single pulse per ion qubit, this can be achieved as follows. For radial modes, there is no large separation between the center-of-mass mode (COM, $m=1$)  and the remaining zig-zag (ZZ, $m=2$) mode. As a consequence, we can not divide the radial modes that the state-dependant force excites into active and spectator modes, as occurs for forces coupling to the axial modes~\cite{PhysRevA.62.022311}, and both modes mediate the interaction.  This means that we will excite two trajectories in the phase spaces of each mode with large excursions that both must be closed at the final gate time duration $t_g$, in order to minimize the residual qubit-phonon entanglement and achieve a high gate fidelity. This leads to  the following two conditions:
 \begin{equation}
 \begin{split}
      \Phi_{i,{\rm com}}^{x}(t_{g})&=0\implies\frac{\mathcal{F}_{i, {\rm com}}^{x}}{\delta_{\rm com}^{x}}(1-e^{\ii\delta_{\rm com}^{x}t_{g}})=0\\
      \Phi_{i,{\rm zz}}^{x}(t_{g})&=0\implies\frac{\mathcal{F}_{i, {\rm zz}}^{x}}{\delta_{\rm zz}^{x}}(1-e^{\ii\delta_{\rm zz}^{x}t_{g}})=0.
 \end{split}
 \end{equation}
The only possibility to satisfy both conditions with a single pulse is to have two detunings are multiples of each other, $\delta_{\rm zz}^{x}=p\delta_{\rm com}^{x}$, such that 
 \begin{equation}
 \begin{split}
     &\delta_{\rm com}^{x}t_{g}=2\pi r, \hspace{1ex}   \delta_{\rm zz}^{x}t_{g}= 2\pi rp, \hspace{1ex}r,p\in \mathbb{Z}
 \end{split}
 \end{equation} 
Here,  $r$ is the number of closed loops in the phase space of the COM mode, and $|pr|$ is the number of loops in the ZZ mode.
One possibility is to fix $r=1$, $p=-1$, such that, $\delta^{x}\equiv \delta_{\rm zz}^{x}=-\delta_{\rm com}^{x}$, and the state-dependent force lies midway from both resonances $\Delta \omega_{L} \simeq(\omega_{\mathit{x},\rm com}+   \omega_{\mathit{x},\rm zz})/2$. In this case, the trajectories correspond to two circles winding in opposite directions with $t_{g}=2\pi/\delta^{x}$. Using the normal-mode displacements $M_{1,\rm com}^{x}=M_{2,\rm com}^{x}=\frac{1}{\sqrt{2}}$ and $M_{1,\rm zz}^{x}=-M_{2,\rm zz}^{x}=\frac{1}{\sqrt{2}}$, the phonon-mediated interactions have a strength 
\begin{equation}
\begin{split}
 J_{12, x}& =-\frac{1}{2\delta^{x}}\Big(\frac{\Omega_{L}\eta_{x, \rm com}}{\sqrt{2}}\Big)^{2} \ee^{-\frac{1}{2}(\eta_{x, \rm com}^{2}+\eta_{x,\rm zz}^{2})} \Big(1+ \frac{\omega_{x, \rm com}}{\omega_{x, \rm zz}}\Big).
\end{split}
 \end{equation}
  In the interaction picture, the ideal unitary evolution operator  for $N=2$  ions becomes
\begin{equation}
\label{eq:ideal_gate}
\begin{split}
    U_{\rm id}(t_{g})&=  \ee^{-\ii(J_{x}t_{g})\sigma_{1}^{z}\sigma_{2}^{z}},
\end{split}    
\end{equation}
where  $J_{12,x}=J_{21,x}=J_{x}$. The unitary at multiples of this gate time describes decoupled  dynamics of the gate-mediated phonons and qubits and yields a maximally-entangling gate for $J_{x}t_{g}=\pi/4$, $U_{\rm id}=U_{\rm id}\left({\pi}/{4J_x}\right)$,  that generates an entangled pair from an initial separable state $\ket{\Psi_0}=\ket{++}$. In the interaction picture, this reads
\begin{equation}
\label{eq:target_state}
\begin{split}
   \ket{\psi_{\rm id}}&= U_{\rm id}\ket{\psi(0)}=\frac{1}{\sqrt{2}}\Big(\ket{--}-\ii\ket{++}\Big). 
\end{split}
\end{equation}
 In contrast to the use of axial modes, where $\delta$ is pretty much free~\cite{PhysRevA.62.022311},  here we have to place it between the COM and ZZ modes to close both trajectories simultaneously. 

In the following subsection, we study the effect of noise and experimental imperfections on this maximally entangling gate. As noted in the introduction, previous results~\cite{ballance_thesis,PhysRevA.105.022437} follow similar calculations of state infidelities in the M\o lmer-S\o rensen gate~\cite{PhysRevA.62.022311} and identify different contributions to deviations from the target state~\eqref{eq:target_state}. These deviations can be quantified through the state fidelity $\mathcal{F}_s=\bra{\Psi_{\rm id}(t_{g})}\mathcal{E}(\rho_0)\ket{\Psi_{\rm id}(t_{g})}$, where $\mathcal{E}(\rho_0)$ describes the noisy evolution of the initial state, which is typically considered to be in a product state of the internal and vibrational degrees of freedom $\rho_0=\ketbra{++}\otimes\rho_{\rm vib}$. In general, however, the gate embedded in a larger circuit will act on other states and, in order to  provide a more representative  of the gate performance, one should instead  compute the  gate error $\epsilon_g=1-\bar{\mathcal{F}}_{g}$, which is obtained by averaging the corresponding fidelities for all possible initial states $\rho_0=\ketbra{\Psi_0}\otimes\rho_{\rm vib}$. Below we give explicit expressions for these errors, investigating the influence of thermal fluctuations in all vibrational branches $\epsilon_{g}^{\rm th}$, off-resonant couplings $\epsilon_{g}^{\rm off}$, dephasing noise due to fluctuating unshielded magnetic fields of laser phase noise $\epsilon_{g}^{\rm d}$, and residual spontaneous photon scattering $\epsilon_{g}^{\rm sp}$. For small/weak errors, the effect of distinct sources of noise is additive ~\cite{nielsen00}. Therefore, 
\begin{equation}\label{eq:additive_errors}
    \epsilon_{g}^{{\rm tot}}={\epsilon}_{g}^{\rm th}+{\epsilon}_{g}^{\rm off}+{\epsilon}_{g}^{\rm deph}+{\epsilon}_{g}^{\rm scatt}.
\end{equation}

\subsection{Average gate fidelity }

 The light-shift gate fidelity can be calculated by averaging the state fidelities of the real time-evolved states for  all possible initial states of the two qubits \cite{nielsen00}, with respect to the ideal target state for each of them, namely $U_{\rm id}\ket{\Psi_0}$. Accordingly, the gate fidelity is
\beq\label{eq:avg_Fg}
\bar{\mathcal{F}}_{g}(U_{\rm id}, \mathcal{E})=\int \!\!{\rm d} \Psi_{\rm 0}\bra{\Psi_{\rm 0}} U_{\rm id}^{\dagger} \mathcal{E}({\rho}_{0})U_{\rm id}^{\phantom{\dagger}}\ket{\Psi_{\rm 0}}.
\eeq
where the integral is taken with respect to the Haar measure over the  two-qubit Hilbert space, and thus averages over all possible two-qubit states $\ket{\Psi_{\rm 0}}$.  Here, $\mathcal{E}({\rho}_{0})$ describes the real evolution of the  initial state ${\rho}_{0}$  under a noisy quantum channel. To account for sources of errors added to the thermal fluctuations of the vibrational modes, we consider  $\mathcal{E}(\rho_0)=\langle{\rm Tr}_{\rm aux}\{U_{g(t)}\ketbra{\Psi_0}\otimes\rho_{\rm aux}U_{g(t)}^{\dagger}\}\rangle_{\rm stoch}$, where  $\rho_{\rm aux}$ describes ``auxiliary'' degrees of freedom, including all the phonon branches, other electronic states such as the virtually-populated $P$ levels used to induce the state-dependent dipole forces, as well as the photons of the surrounding electromagnetic environment. Additionally,  the unitary evolution of the complete system $U_{g(t)}$ may depend on external fluctuations, such as un-shielded magnetic fields or fluctuating control parameters, which can be modeled by  stochastic processes $\{g(t)\}$. Accordingly, the quantum channel will also result from statistical averaging over such stochastic processes, which corresponds to $\langle\cdot\rangle_{\rm stoch}$ in the previous expression. As a result of this average and the partial trace, the evolution of the system will no longer be unitary as in the ideal case~\eqref{eq:ideal_gate}. The goal of this section is to find contributions to the gate infidelity in the limit of small errors, which can be calculated analytically, and then fed into effective noise models.

To avoid the integral over the  two-qubit Hilbert space, we can alternatively calculate the entanglement fidelity~\cite{PhysRevA.54.2614,PhysRevA.60.1888,NIELSEN2002249} given by
\beq
\label{eq:F_e_definition}
 \bar{\mathcal{F}}_{e}(U_{\rm id}, \mathcal{E})=\bra{\phi_{m}}{I}^{d}\otimes U_{\rm id}^{\dagger}\, \mathcal{E}(\rho_m)\,{I}^{d}\otimes U_{\rm id}\ket{\phi_{m}}.
\eeq
The entanglement fidelity is defined in an  enlarged  Hilbert space with partitions A and B, where A is auxiliary with the same number of qubits as B and $\ket{\phi_{m}}=\sum_{\xi=1}^{d}\ket{\xi}_A\otimes\ket{\xi}_B/\sqrt{d}$ is a maximally entangled state. Here,  $d=2^{N}$where $N$ is the number of physical qubits involved in the gate, and the set $\{\ket{\xi}\}$ of $d$ quantum states is chosen to form an orthonormal basis. For the current case of $N=2$, we have $d=4$ basis states $\{\ket{++},\ket{+-},\ket{-+},\ket{--}\}$. In the above equation~\eqref{eq:F_e_definition}, the auxiliary qubits are  not subjected to the ideal or noisy time evolution, such that 
$\mathcal{E}(\rho_m)=\langle{{\rm Tr}_{\rm aux}\{I^d\otimes U_{g(t)}\,\ketbra{\Phi_m}\otimes\rho_{\rm aux}\,I}^d\otimes U_{g(t)}^{\dagger}\}\rangle_{\rm stoch}$. By doubling the number of qubits, and exploiting their entanglement, one can prove~\cite{PhysRevA.54.2614, NIELSEN2002249} that a single state fidelity of the enlarged system can be used to infer the gate fidelity of the original physical system via
\begin{equation}\label{eq:avg_Fg_and_Fe}
    \begin{split}
       \bar{\mathcal{F}}_{\rm g}(U_{\rm id}, \mathcal{E})=\frac{d\bar{\mathcal{F}}_{e}(U_{\rm id}, \mathcal{E})+1}{d+1}.
    \end{split}
\end{equation}
Equipped with this formal tool, it is possible to derive analytical expressions for various sources of gate infidelities. As advanced previously, we are interested in a perturbative regime where different error sources are additive and thus can be thus discussed sequentially.

\subsection{Gate infidelity due to thermal errors} 
The light-shift gates, just like M\o lmer-S\o rensen gates~\cite{PhysRevLett.82.1835,PhysRevA.62.022311}, are robust against the thermal occupation of the active phonon modes in the Lamb-Dicke approximation. In this section, we analyze the effect of thermal occupation due to higher-order terms in the Lamb-Dicke expansion for both the active vibrational modes as well as the spectator  modes.

\subsubsection{Warm active phonons}  There are higher order terms in the Lamb-Dicke expansion of Eq.~\eqref{eq:hamiltonian_I}, there are additional corrections that stem from higher-order terms in the Lamb-Dicke parameters which are not far-off resonant and thus must be taken into account. Under these circumstances, ensuring that the beat note laser frequencies are still satisfying $|\Tilde{\Omega}_{L, i}|\ll\Delta \omega_{l}\simeq \omega_{x, m}$, the force parameter from Eq.(\ref{eq:F_strength}) transforms into 
\begin{equation}\label{eq:F_strength_thermal_active}
    \widetilde{\mathcal{F}_{i,m}^{x}}=\ii\frac{|\Tilde{\Omega}_{L,i}|}{2}\eta_{x, m}M_{i,m}^{x}\mathcal{G}_{i, m}^{x}(\{a^{\dagger}_{x,m}a_{x,m}^{\phantom{\dagger}} \}),
\end{equation}
where we have introduced the operator
\begin{equation}
\begin{split}
\mathcal{G}_{i, m}^{x}&=\left(\sum_{l=0}^{\infty}\frac{(-(\eta_{x, m}M_{i,m}^{x})^{2})^{l}}{(l!)^{2}(l+1)}(a_{x,m}^{\dagger}a_{x,m})^{l} \right)\\
&\prod_{m'\neq m}\left(\sum_{l=0}^{\infty}\frac{(-(\eta_{x, m'}M_{i,m'}^{x})^{2})^{l}}{(l!)^{2}}(a_{x,m'}^{\dagger}a_{x,m'})^{l} \right). 
\end{split}
\end{equation}
Depending on the motional state, this operator will result in  fluctuations in the amplitude of the force.

As highlighted before, one important difference between  axial and radial-mediated entangling gates is that in the former case, there is only one active vibrational mode. Typically, it is chosen to be the COM mode, whereas the other breathing mode is off-resonant, and merely acts as a spectator. In our case, since we are focusing on radial modes along the $\alpha=x$ axis, both COM and ZZ modes are active, and they contribute to the gate equally. We consider the initial vibrational state  described by the tensor product of two thermal Gibbs states $\rho_{{\rm vib},x}^{\rm th}=\rho_{\rm com}^{\rm th}\otimes\rho_{\rm zz}^{\rm th}$  of the form

\beq\label{eq:Gibbs}
  \rho_{m}^{\rm th}=\sum_{n_{m}=0}^{\infty}p_{m}(n_{m})\ket{n_{m}}\bra{n_{m}},
\eeq
 where the probability $p\s{m}(n\s{ m})$ is given by a thermal distribution
\begin{equation}
   p\s{m}(n\s{m})=\frac{1}{1+\bar{n}_{m}}\left(\frac{\bar{n}_{m}}{1+\bar{n}_{m}}\right)^{n_{m}}.
\end{equation}
Here, $\bar{n}_{m}=1/(\ee^{{k_{\rm B}T_{m}/\hbar \omega_{ m}}}-1)$ is the Bose-Einstein distribution,  $\omega_{m}$ is the frequency of the $m$ mode, and $T_{m}$ is the effective temperature for each mode  used to model situations in which resolved sideband cooling is used. 

By considering the new force operator~(\ref{eq:F_strength_thermal_active}), one can revisit the Magnus expansion that leads to the evolution in Eq.(\ref{eq:unitary_Hint}). We find that, the closure conditions $\Phi_{i,com}^{x}(t_{g})=\Phi_{i,zz}^{x}(t_{g})=0$, are not modified by thermal fluctuations. On the other hand,  the spin-spin interactions~\eqref{eq:spin_spin_couplings}  depend on the number of phonons. To the lowest order of the Lamb-Dicke parameters, they read
\begin{equation}\label{eq:J_warm_active}
\begin{split}
    J_{ij,x}&=-2\sum_{m}\frac{\mathcal{F}_{i,m}^{x}\mathcal{F}_{j,m}^{x}}{\delta_{m}}\bigg(1-\sum_{m'}\eta_{x, m'}^{2}a_{x,m'}^{\dagger}a_{x,m'}\bigg).
\end{split}
\end{equation}
Therefore, the $J_{x}t_{g}=\pi/4$ condition, which leads to the ideal maximally-entangling gate~\eqref{eq:ideal_gate} for phonons in the vacuum state, now leads to qubit-qubit interactions that depend on the thermal fluctuations of phonons, affecting the gate fidelity. Using  the entanglement fidelity~(\ref{eq:avg_Fg_and_Fe}) with the auxiliary qubits in the specific entangled state, it is possible to evaluate the gate fidelity for the initial thermal states analytically, arriving at an error due to warm active phonons, 

\begin{equation}\label{eq:gate_thermal_active}
\begin{split}
 \epsilon_{\rm th}^{\rm act} =\frac{\pi^{2}}{20}\sum_{m,m'}\eta_{x,m}^{2}\eta_{x,m'}^{2}\langle\hat{n}_{x,m}\hat{n}_{x,m'}\rangle,
\end{split}
\end{equation}
where $\langle\hat{n}_{x,m}\hat{n}_{x,m'}\rangle=(2\overline{n}^2_m+\overline{n}_m)\delta_{m,m'}+\overline{n}_m\overline{n}_{m'}(1-\delta_{m,m'})$ for the above thermal states. We note that the condition  $J_{x}t_{g}=\pi/4$ for the entangling gate could be modified to account for the mean number of phonons, and the above gate errors would only depend on the thermal fluctuations because of the variances on the phonon number. However, since we are interested in modeling shuttling-based QIP, the phonon state will change between different gates, and such optimal conditions will not hold for the whole circuit. Accordingly, we follow a conservative approach by considering that the  errors~\eqref{eq:gate_thermal_active} follow from setting the vacuum condition $J_{x}t_{g}=\pi/4$ throughout the circuit. In table~\ref{tab:table_gateinfidelity}, we give the specific error~\eqref{eq:gate_thermal_active} for $N=2$ ions.

\subsubsection{Warm spectator phonons} 
So far, we have assumed that the spectator modes (i.e. those along the remaining radial $\alpha=x$ direction, along the radial $\alpha=y$ axis, and those along the $\alpha=z$ axial direction) do not affect the gate performance. This is strictly so if the alignment of the beams is perfect $\Delta \vec{k}_L=\Delta k_L \textbf{e}_{x}$. However, if there is a small misalignment, such that the corresponding Lamb-Dicke parameters no longer vanish $\eta_{z,m},\eta_{y,m}\neq 0$,  the spectator modes could contribute to fluctuations of the effective Rabi frequency of the state-dependant dipole force, off-resonant drivings, and unclosed trajectories in phase space. Following a similar calculation as above, the force parameter, when considering the spectator phonons transforms into
\begin{equation}\label{eq:F_strength_thermal_spectators}
    \widetilde{\widetilde{\mathcal{F}_{i,m}^{x}}}=\ii\frac{|\widetilde{\tilde{\Omega}_{L,i}}|}{2}\eta_{x, m}M_{i,m}^{x}\mathcal{G}_{i, m}^{x}(\{a^{\dagger}a \})\mathcal{K}_{i}^{y,z}(\{a^{\dagger}a\}).
\end{equation}
where, in this case, the Debye-Waller factor must also account for the vacuum contributions of the additional vibrational modes,
\begin{equation}
\widetilde{\tilde{\Omega}_{L,i}}=\Omega_{L,i}e^{-\frac{1}{2}\big(\sum\limits_{ m}(\eta_{x,m}M_{i,m}^{x})^{2}+\sum\limits_{\alpha=y,z}\sum\limits_{m}(\eta_{\alpha,m}M_{i,m}^{\alpha})^{2}\big)}.
\end{equation}
In addition, the force can fluctuate depending on the moments of the number of spectator phonons,
\begin{equation}
    \mathcal{K}_{i}^{y,z}=\prod_{\alpha=y,z}\prod_{m}\left(\sum_{l=0}^\infty\frac{(-(\eta_{\alpha, m}M_{i,m}^{\alpha})^{2})^{l}}{(l!)^{2}}(a_{\alpha, m}^{\dagger}a_{\alpha, m})^{l} \right).
\end{equation}

In a shuttling-based approach to QIP, since the axial modes (along the trap axis) are more susceptible to heating by the crystal reconfigurations, the modes in $\alpha=z$ will have higher contributions to this source of error. To calculate it explicitly, we check how it affects the Magnus expansion and the spin-spin interactions. The closure conditions of the phase-space trajectories are the same in spite of  the additional spectator modes, that is, $\Phi_{i,\rm com}^{x}(t_{g})=\Phi_{i,\rm zz}^{x}(t_{g})=0$. However, as in the case of warm active phonons, the spin-spin strength will depend upon the phonon operators as 
\begin{equation}\label{eq:J_warm_spectators}
\begin{split}
    J_{ij,x}=-2\sum_{m}\frac{\mathcal{F}_{i,m}^{x}\mathcal{F}_{j,m}^{x}}{\delta_{m}}\Bigg(1-&\sum_{m'}\eta_{x, m'}^{2}a_{x,m'}^{\dagger}a_{x,m'}\\
    -\frac{1}{2}\sum_{\alpha=y,z}&\sum_{m'}\eta_{x, m'}^{2}a_{\alpha,m'}^{\dagger}a_{\alpha,m'}\Bigg).
\end{split}
\end{equation}
The calculation of the gate fidelity goes along the same lines as before, and 
 leads to the following error  due to spectator modes
\begin{equation}\label{eq:gate_thermal_spectator}
\begin{split}
 \epsilon_{\rm th}^{\rm spec} =\frac{\pi^{2}}{20}\sum_{m,m'}\sum_{\alpha,\alpha'}C_{\alpha,\alpha'}\eta_{\alpha,m}^{2}\eta_{\alpha',m'}^{2}\langle\hat{n}_{\alpha,m}\hat{n}_{\alpha',m'}\rangle,
\end{split}
\end{equation}
where we have introduced the symmetric matrix with coefficients $C_{x,y}=C_{x,z}=1/2$ and $C_{y,z}=C_{x,z}=1/4$, otherwise zero.
The total gate error due to thermal noise in Eq.~\eqref{eq:additive_errors} is then $\epsilon^{\rm th}_{g}=\epsilon_{\rm th}^{\rm ac}+\epsilon_{\rm th}^{\rm spec}$. In table~\ref{tab:table_gateinfidelity}, we give the specific error for $N=2$ ions. 

\subsection{Gate infidelity due to off-resonant forces}
When moving to the interaction picture in Eq.~(\ref{eq:interaction_Hint}), we neglected off-resonant contributions to the state-dependent forces of both active and spectator modes. In this section, we calculate the contribution of those errors to gate infidelity.

\subsubsection{Off-resonant forces on active phonons}
In Eq.~(\ref{eq:interaction_Hint}), we kept just the contributions of $e^{i(\Delta \vec{k}_L\cdot\vec{r}_{i}^{0} +\delta_{m}^{x}t+\phi_{L})}$ to the dipole force, where $\delta_{m}^{x}=\omega_{x,m}-\Delta \omega_{L}$. With the beat-note laser frequency set close to the vibrational modes $\Delta \omega_{l}\approx \omega_{x, m}$ so that the neglected terms rotate with $\omega_{\alpha,m}+\Delta \omega_{l}$, and are thus off-resonant contributing perturbatively to the gate. In order to account for the error,  we treat this perturbation 
\begin{equation}
    V_{I}=\sum_{i,m}\ii\mathcal{F}_{i,m}^{x}\ee^{\ii(\Delta \vec{k}_L\cdot\vec{r}_{i}^{0} +(\omega_{x, m}+\Delta\omega_{l})t+\phi_{L})}Z_{i}a_{x, m} +{\rm H.c.}
\end{equation}
in the interaction picture with respect to the ideal unitary evolution~(\ref{eq:unitary_Hint}), such that 
\begin{equation}\label{eq:unitary_off_warm_active}
    \tilde{U}(t_{g})=\ee^{-\ii H_{0}t_g}U(t_{g})\mathcal{T} \{\ee^{-\ii\int_{0}^{t_{g}}dt' V_{I}(t')}\}.
\end{equation}
  We evaluate the leading contribution to the error using the Dyson series expansion of the time-ordered operator. The leading error contribution is calculated by assuming  the symmetric closure conditions for the  ideal gate $t_{g}=2\pi/\delta^{x}$, with $\delta_{\rm com}^{x}=-\delta_{\rm zz}^{x}=\delta^{x}$. By substituting the Dyson expansion in the expression for entanglement fidelity~(\ref{eq:F_e_definition}),  we get the following contribution to the gate infidelity for $N=2$ ions:
\begin{equation}\label{eq:gate_off_active}
\begin{split}
\epsilon_{\rm off}^{{\rm act}}=\sum_{i,m}\frac{\Tilde{\Omega}_{L,i}^{2}}{5}
\eta^2_{x,m}\frac{2\pi}{(\omega_{x,m}^2-\Delta\omega_L^2)},
\end{split}
\end{equation}
as shown in table~\ref{tab:table_gateinfidelity}.
Let us note that, from the Dyson expansion, one finds two processes with opposite detunings contributing to the error, such that the dependence on the mean phonon numbers cancels. In that regard, the contribution to the off-resonant error~\eqref{eq:gate_off_active}    could indeed be accounted for by  a small renormalization of the phonon-mediated interaction strength, leading to an under-rotation. As such, one could  modify the required laser intensity needed to meet the constraint for a maximally-entangling gate $J_xt_g=\pi/4$, by using the qubit-qubit interactions that also take into account these off-resonant contributions~\cite {PhysRevLett.103.120502}. The situation is  different for  spectator modes.

\subsubsection{Off-resonant forces on spectator phonons}\label{eq:off_spectator}
The spectator modes will also be subject to  off-resonant forces when the laser wavevector is misaligned. We now include the spectator modes for $\alpha=y, z$ in  Eq.~(\ref{eq:interaction_Hint}), and consider the following perturbation due to off-resonant terms,
\begin{equation}
\begin{split}
    V_{I}&\simeq\sum_{\alpha=y,z}\sum_{i,m}\ii\mathcal{F}_{i,m}^{\alpha}\big(\ee^{\ii(\Delta \vec{k}_L\cdot\vec{r}_{i}^{0} +\delta_{m}^{\alpha}t+\phi_{L})}\\
    &+\ee^{-\ii(\Delta \vec{k}_L\cdot\vec{r}_{i}^{0} +(\omega_{\alpha, m}+\Delta\omega_{l})t+\phi_{L})}\big)Z_{i}a_{\alpha,m}^{\dagger} +{\rm H.c.}
\end{split}
\end{equation}
Proceeding similarly as in Eq.(\ref{eq:unitary_off_warm_active}), we estimate the leading error to the gate fidelity which, in contrast to the active modes, depends on the mean number of phonons. In this case, the contribution to the gate error is given by
\begin{equation}\label{eq:gate_off_spectator}
\begin{split}
\epsilon_{\rm off}^{{\rm spec}}&=\frac{4}{5}\sum_{\alpha=y,z}\sum_{m}\frac{(\Tilde{\Omega}_{L}\eta_{\alpha,m}^{2})^{2}}{\Delta\omega^{2}-\omega_{\alpha,m}^{2}}\\
&\times\left| \frac{\Delta\omega^{2}_{l}+\omega_{\alpha,m}^{2}}{\Delta\omega^{2}_{l}-\omega_{\alpha,m}^{2}}(2\bar{n}_{\alpha,m}+1)+\ii \frac{2\pi\omega_{\alpha,m}}{\omega_{\alpha,m}-\Delta\omega_{L}}\right|.
\end{split}
\end{equation}
Note that, the second contribution in the equation above is indeed caused by an under/over-rotation that could be accounted for by summing also over the spectator modes in the definition of the qubit-qubit couplings~\eqref{eq:spin_spin_couplings}, and by modifying the condition for a maximally-entangling gate. The thermal contribution, on the other hand, cannot be minimized and contributes to the $N=2$ error of Table~\ref{tab:table_gateinfidelity}. Altogether, the gate error from off-resonant forces is ${\epsilon}_{g}^{\rm off}={\epsilon}^{\rm spec}_{\rm act}+ {\epsilon}^{\rm spec}_{\rm off}$.

\subsection{Gate infidelity due to dephasing noise} 
As noted in the introduction to the light-shift gate \ref{sec:ideal_ZZgate},  there could also be possible errors due to magnetic field and laser-phase fluctuations on the Zeeman sub-levels forming our qubit. 
We can study these fluctuations in two regimes; when the magnetic field fluctuations act locally on the ions and they are fully uncorrelated, and when they act globally and affect both qubits in a correlated manner. In both cases, we assume that the stochastic process describing this noise is Markovian, and model its effect by a standard dephasing master equation in the Lindblad form~\cite{nielsen00}, namely
\begin{equation}\label{eq:master_dephasing}
    \dot{\rho}(t)=-\ii[H_{0}+H_{\rm int}, \rho(t)]+\mathcal{D}_{d}(\rho(t)).
\end{equation}
The form of the coherent evolution with $H_{0}$ and $H_{\rm int}$ is that of the Hamiltonians in Eq.~(\ref{eq:hamiltonian_0}) and Eq.~(\ref{eq:hamiltonian_I}), respectively. The super-operator $\mathcal{D}_{d}(\rho(t))$ stems from averaging the stochastic noise describing local and global magnetic field fluctuations in the weak-coupling limit.
\subsubsection{Individual dephasing noise}\label{sec:fully_uncorrelated}
In this scenario, the fluctuations are local, and there are no spatial correlations in the noise. We also assume that the noise is Markovian and the Lindblad operators for the dephasing are $L_{i}=\sqrt{\frac{\Gamma_{d}}{2}}Z_{i}$, where $\Gamma_{d}=1/T_{2}$ is the dephasing rate, and $T_{2}$ the dephasing time of the qubits.
Since all terms in the Hamiltonian and noise commute, it is straightforward to move to the interaction picture in Eq.(\ref{eq:master_dephasing}), to find that the ideal evolution of the form:
\begin{equation}\label{eq:deph_uncorrelated}
\begin{split}
\dot{\rho}_{I}(t)=\frac{\Gamma_{d}}{2}\sum_{i}(Z_{i}\rho_{I}(t)Z_{i}-\rho_{I}(t)),
\end{split}
\end{equation}
where $\rho_{I}(t)=U^{\dagger}(t)\rho(t)U(t)$ is the evolution of the system in an interaction picture with respect to the coherent terms $U(t)=\mathcal{T}\{{\rm exp}(-\ii\int_{0}^{t}{\rm d}t' (H_0+H_{\rm int}(t')))\}$. The solution to this equation is simply the composition of two dephasing channels which read
 \beq
 \begin{split}
 \rho_I(t_{g})=&(1-p_d(t_g))^2\rho(0) +p_d^2(t_g)Z_1Z_2\rho(0)Z_1Z_2\\
 & +(1-p_d(t_g))p_d(t_g) \sum_iZ_i\rho(0)Z_i,
  \end{split}
 \eeq
 where  $p_d(t_g)=\half(1-\ee^{-t_g/T_2})$ is the phase-flip error rate. We should bear in mind that, when moving back to the original picture, $\rho(t_{g})=U(t_g)\rho_I(t)U^{\dagger}(t_g)$, one gets the   additional evolution  under coherent terms~\eqref{eq:unitary_Hint}, which would lead to the  light-shift gate~\eqref{eq:ideal_gate} under idealised conditions.  Substituting this expression into the entanglement fidelity formula in Eq.(\ref{eq:F_e_definition}), and neglecting the combination of thermal and dephasing errors, as these will only appear in higher-order perturbations, we find that the leading error to the gate infidelity  is given by
\begin{equation}\label{eq:gate_deph_uncorrelated}
\begin{split}
\epsilon_{g}^{\rm deph, l}= \frac{4}{5} (2p_d(t_g)-p_d^2(t_g)).
\end{split}
\end{equation}
If one  assumes a weak dephasing during the extent of the gate $t_{g}\ll T_{2}$, with $t_{g}=2\pi/\delta^{x}$, the dephasing gate error can be approximated by $\epsilon_{g}^{\rm deph, l}\approx \frac{4 N}{5}\Big(t_{g}/2T_{2} \Big)$. In this case, the gate infidelity scales linearly with $N$. In this manuscript, we restrict to $N=2$ as summarised in Table~\ref{tab:table_gateinfidelity}.

\subsubsection{Correlated dephasing noise}

Let us now discuss the limit where the magnetic field fluctuations or the laser phase noise are global, and affect both qubits in a correlated manner. In this case, moving to the interaction picture in Eq.(\ref{eq:master_dephasing}), the master equation becomes 
\begin{equation}\label{eq:deph_correlated}
\begin{split}
\dot{\rho}_{I}(t)=2\Gamma_{d}\left(S_{z}\rho_{I}(t)S_{z}-\frac{1}{2}S_{z}^{2}\rho_{I}(t)-\frac{1}{2}\rho_{I}(t)S_{z}^{2}\right).
\end{split}
\end{equation}
with $S^{z}=(Z_{1}+Z_{2})/2$ being the total spin operator. The solution to this master equation can be written in terms of $\chi(t)\equiv\ee^{-Nt/T_2}$, which decays twice as fast as the case of individual dephasing. The quantum channel admits an analytical solution in Krauss form $\rho_{I}(t_{g})=\sum_n K_n^\dagger(t_g)\rho(0)K_n(t_g)$, 
\beq
\begin{split}
K_1(t_g)&=\half(\sqrt{\chi(t_g)}+1)I+\half(\sqrt{\chi(t_g)}-1)Z_1Z_2,\\
K_2(t_g)&=\half\sqrt{\chi(t_g)(1-\chi(t_g))}(Z_1+Z_2),\\
K_3(t_g)&=\fourth(1-\chi(t_g))(1+Z_1+Z_2+Z_1Z_2),\\
K_4(t_g)&=\fourth(1-\chi(t_g))(1-Z_1-Z_2+Z_1Z_2).\\
\end{split}
\eeq

It is worth mentioning that if one is interested in the state fidelity for the maximally-entangled pair obtained by  M\o lmer-S\o rensen gates~\cite{PhysRevX.7.041061}, it scales as $N^2$ due to the global nature of the noise. In the present case, however, we are  interested in light-shift gates and, moreover, not the fidelity of producing a single state but rather the average gate fidelity. Substituting the previous Krauss channel  n the expression of the entanglement fidelity~(\ref{eq:F_e_definition}) to estimate the contribution to the gate infidelity, we find that
\begin{equation}\label{eq:gate_deph_correlated}
\begin{split}
\epsilon_{g}^{\rm deph, c}\simeq \frac{4}{5} (2p_d(t_g)-3p_d^2(t_g)+4p_d^3(t_g)-2p_d^4(t_g)).
\end{split}
\end{equation}
Although this expression differs from the uncorrelated one in Eq.\eqref{eq:gate_deph_uncorrelated}, for weak dephasing during the gate time, one finds that
$\epsilon_{g}^{\rm deph, c}\approx \epsilon_{g}^{\rm deph, c}$.

\subsection{Gate infidelity due to residual photon scattering} 
So far, when using the expression of the state-dependent force~\eqref{eq:interaction_Hint}, we have neglected the possible irreversible dynamics due to the emission of photons from the excited $P$ levels to the background of electromagnetic modes (see Fig.~\ref{fig:2ndorder}). This is justified for very large detunings of the lasers, but, in general, it will have some contribution to the error. To determine this contribution, we consider the Markovian master equation 
\begin{equation}\label{eq:master_sp}
   \dot{\rho}(t)=-\ii[H_{0}+H_{\rm int}, \rho(t)]+\mathcal{D}_{sc}(\rho(t))
\end{equation}
which is obtained after tracing over the excited $P_{1/2},P_{3/2}$ levels of linewidth $1/\Gamma$ and the electromagnetic photons~\cite{nielsen00}. In addition to the coherent terms in  Eq.~(\ref{eq:master_sp}), we get a Lindblad-type term $\mathcal{D}_{sc}(\rho(t))$ describing the effect of the residual spontaneous emission  within the qubit computational subspace due to off-resonant  scattering of photons. Following~\cite{PhysRevA.85.032111, Bermudez_2012}, we get

\begin{equation}
    \mathcal{D}_{sc}(\rho)=\sum_{i}\sum_{n}\big(L_{i,n}\rho L_{i,n}^{\dagger}-\half\{L_{i,n}^{\dagger}L_{i,n},\rho\}\big).
\end{equation}
where one finds two possible decoherence effects, the so-called Raman and Rayleigh scattering of photons~\cite{PhysRevLett.95.030403,PhysRevLett.105.200401}. These are described by the following effective jump operators
\begin{equation}
\begin{split}
    L_{i,1}=\frac{\sqrt{\Gamma}}{\Delta}\sum_{l=1,2}\Omega_{l,\downarrow}^{i}\ee^{\ii(\vec{k}_{L,l}\cdot\vec{r}_{i}-\omega_{L,l}t)}(\ket{\downarrow_{i}}\bra{\downarrow_{i}}+\ket{\downarrow_{i}}\bra{\uparrow_{i}})\\
 L_{i,2}=\frac{\sqrt{\Gamma}}{\Delta}\sum_{l=1,2}\Omega_{l,\uparrow}^{i}\ee^{\ii(\vec{k}_{L,l}\cdot\vec{r}_{i}-\omega_{L,l}t)}(\ket{\uparrow_{i}}\bra{\uparrow_{i}}+\ket{\uparrow_{i}}\bra{\downarrow_{i}}).
 \end{split}
\label{eq:jump_operators} 
\end{equation}
These jump operators lead to 
an effective scattering rate of $\Gamma_{\rm eff}\sim\Gamma\big({\Omega_{l,s}^{i}}/{\Delta}\big)^2$ which, in the limit of very large detunings,  can be made much smaller than the dipole-force amplitudes in Eq.~\eqref{eq:F_strength} which scale with ${(\Omega_{l,s}^{i})^2}/{\Delta}$. 

In the expressions of the effective jump operators in Eq.~(\ref{eq:jump_operators}), the first addend terms on the right, which are proportional to $({I}\pm Z_{i})/2$ correspond to the elastic Rayleigh photon scattering that does not modify the internal qubit states. The second addend terms in Eq.~(\ref{eq:jump_operators}), which are proportional to the raising and lowering spin operators $\sigma_{i}^{+}=(X_{i}+\ii Y_i)/2$ and $\sigma_{i}^{-}=(X_{i}-\ii Y_{i})/2$,  correspond to the Raman scattering, where the qubit state is flipped after the emission of the photon. Let us now calculate the contribution of each scattering event to the light-shift gate fidelity.

\subsubsection{Rayleigh scattering errors}
Moving to the interaction picture with respect to the complete Hamiltonian, the contribution to Eq.(\ref{eq:master_sp}) from Rayleigh scattering is
\begin{equation}\label{eq:deph_Rayleigh}
\begin{split}
\mathcal{D}_{ sc}^{\rm Rai}(\rho_{I}(t))=\sum_{i}\Gamma^{{\rm Rai}}_{\rm sc, i}(Z_{i}\rho_{I}(t)Z_{i}-\rho_{I}(t))
\end{split}
\end{equation}
which induces pure dephasing as in Eq.~(\ref{eq:deph_uncorrelated}). $\Gamma^{{\rm Rai}}_{\rm sc, i}$ is the dephasing rate due to the elastic Rayleigh scattering given by
\begin{equation}\label{eq:Gamma_Rayleigh}
\Gamma_{{\rm sc
},i}^{\rm Rai}=\frac{\Gamma}{\Delta}\delta\Tilde{\omega}_{0,i}\approx \frac{\Gamma}{\Delta}\left(\sum_{l=1,2}\frac{|\Omega_{l, \downarrow}^{i}|^{2}+|\Omega_{l, \uparrow}^{i}|^{2}}{4\Delta}\right),
\end{equation}
where $\delta\Tilde{\omega}_{0,i}(t)$ as in Eq.~(\ref{eq:ac_shifts}). In the last step, since $({\Gamma}/{\Delta})\ll 1$, we can also assume that $\frac{\Gamma}{\Delta}(\frac{\Omega_{l,s}^{2}}{\Delta})\ll\Delta\omega_{l}\simeq\omega_{x}$, and thus, the oscillating part in $\delta\Tilde{\omega}_{0,i}(t)$ can be neglected in a rotating-wave approximation. Since the effect of Rayleigh scattering is analogous to uncorrelated dephasing, we can use the results of the previous subsection, and  express the gate infidelity as
\begin{equation}\label{eq:gate_deph_Rayleigh}
\begin{split}
\epsilon_{\rm Rai}^{\rm scatt}=\frac{4}{5} N\Big(t_{g}/2T_{2}^{\rm eff} \Big),
\end{split}
\end{equation}
with the particularity that now the coherence time is related to the elastic scattering rate as $T_{2}^{\rm eff}=1/ \Gamma_{{\rm Rai},i}^{\rm sc}$.

\begin{figure}[t!]
  \includegraphics[width=1\columnwidth]{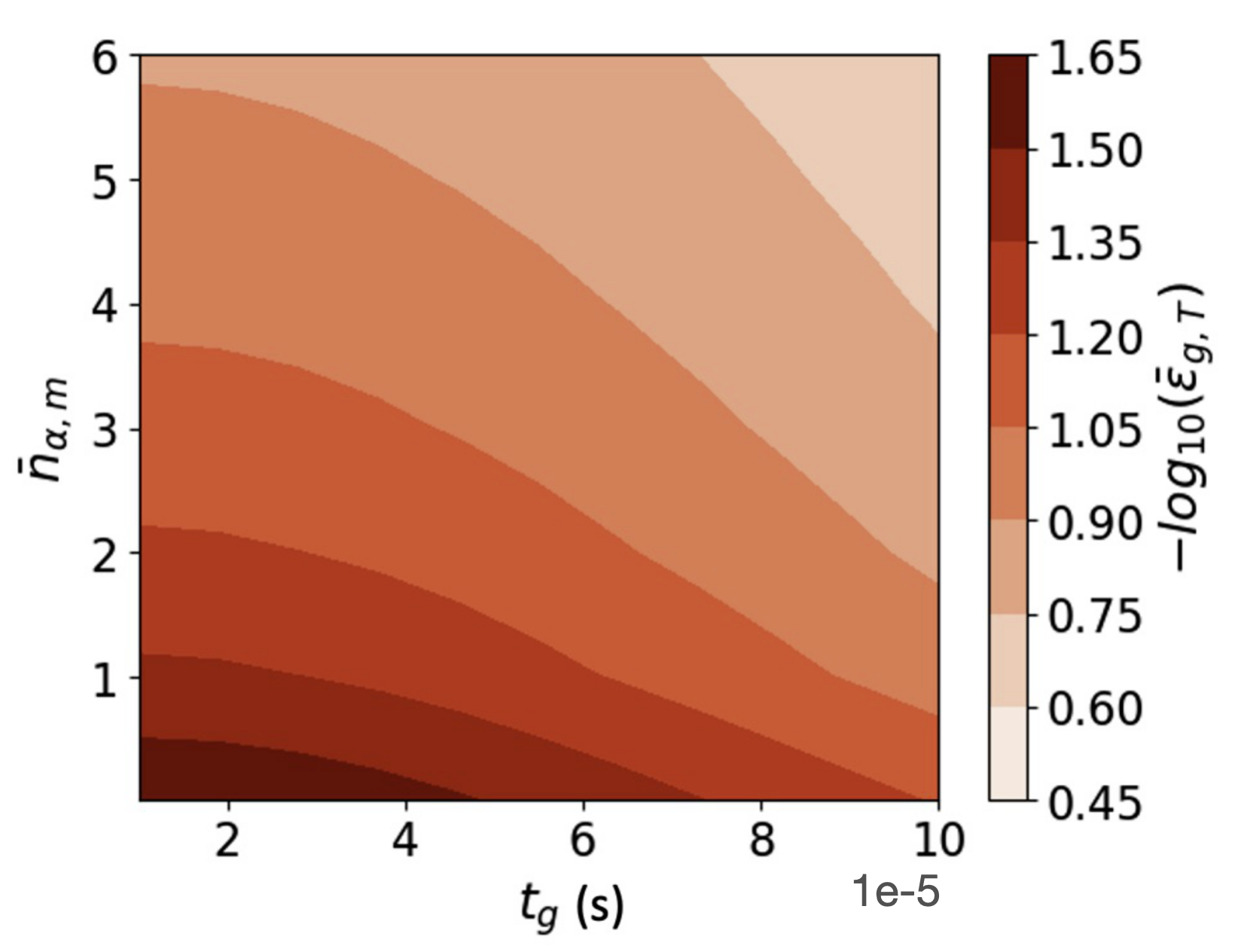}
  \caption{Total  gate error $\bar{\epsilon}_{g,T}$ as a function of the gate time $t_g$ and the initial average vibrational occupation number $\bar{n}_{\alpha, m}$. We assume here that for each axis $\alpha=x,y,z$, all modes $m$ are cooled down to the same $\bar{n}_{\alpha, m}$, before running the parity-check circuits.}
  \label{fig:avg_gate_infidelities}
\end{figure}

\subsubsection{Raman scattering errors}

Next, we calculate the leading error due to Raman scattering, which, in contrast to Rayleigh scattering, leads to the damping of the qubit populations.
The contribution to Eq.(\ref{eq:master_sp}) from Raman scattering in the interaction picture 
is
{\begin{equation}
\begin{split}
   \mathcal{D}_{ sc}^{\rm Ram}(\tilde{\rho})=\sum_{i=1}^{N}\Big(&\Gamma^{\rm Ram}_{{\rm em},i}\big(\tilde{\sigma}_{i}^{-}\tilde{\rho}\tilde{\sigma}_{i}^{+}-\half\{\tilde{\sigma}_{i}^{+}\tilde{\sigma}_{i}^{-}, \tilde{\rho} \}\big)\\
    +&\Gamma^{\rm Ram}_{{\rm ab},i}\big(\tilde{\sigma}_{i}^{+}\tilde{\rho}\tilde{\sigma}_{i}^{-}-\half\{\tilde{\sigma}_{i}^{-}\tilde{\sigma}_{i}^{+}, \tilde{\rho}\}\big)\Big)
\end{split}
\end{equation}
with $\tilde{\sigma}_{i}^{\pm}=U_{id}(t)\sigma_{i}^{\pm}U^{\dagger}_{id}(t)$ being the rotating-frame spin ladder operators,  $\tilde{\rho}=U_{id}(t)\rho U^{\dagger}_{id}(t)$ the evolution operator in the interaction picture, and $U_{id}(t)$ describing the perfect unitary evolution (\ref{eq:unitary_Hint}).  $\Gamma^{\rm Ram}_{{\rm abs},i}$ and $\Gamma^{\rm Ram}_{{\rm em},i} $ are the absorption and emission rates, respectively, given by
\begin{equation}
    \Gamma^{\rm Ram}_{{\rm em}, i}=\frac{\Gamma}{\Delta}\left(\sum_{l=1,2}\frac{\Omega_{l, \downarrow}^{i}\Omega^{i *}_{l, \uparrow}}{\Delta}\right)=(\Gamma^{\rm Ram}_{{\rm abs}, i})^{*},
\end{equation}
which are approximately equal. 

The effect of the Raman absorption and emission on the gate error can be calculated following a strategy as with other weak sources of noise. In this case, the solution to the interaction-picture master equation $ \dot{\rho}_{I}(t_{g})=\mathcal{D}_{sc}^{\rm Ram}(\rho_{I}(t_{g}))$ for the specific matrix element that appears in the calculation of entanglement fidelity~(\ref{eq:F_e_definition}), allows us to find the following contribution to the gate error
\begin{equation}\label{eq:gate_deph_Raman}
\begin{split}
\epsilon_{\rm Ram}^{\rm scatt}=\frac{6}{5} N\Big(t_{g}/T_{1}\Big),
\end{split}
\end{equation}
where the amplitude damping time is defined as $T_{1}=1/2|\Gamma_{\rm em}^{\rm Ram}|$.
The total error due to the residual photon scattering involving the contributions from Rayleigh and Raman scattering is $\epsilon_{g}^{\rm scatt}=\epsilon_{\rm Rai}^{\rm scatt}+\epsilon_{\rm Ram}^{\rm scatt}$.
In Table I, we provide further details specifically for  $N = 2$ ions

In Fig.~\ref{fig:avg_gate_infidelities}, we present the total average gate infidelity  $\bar{\epsilon}_{g}^{{\rm tot}}$ as a function of the gate time $t_{g}$, and of the initial average vibrational occupation number $\bar{n}_{\alpha, m}$.  For the  displayed results, we have assumed that all the modes $m$ for the different axes $\alpha=x,y,z$ are initially Doppler- and sideband-cooled up to the same average phonon number $\bar{n}_{\alpha, m}$. The gate infidelity values shown in Fig.~\ref{fig:avg_gate_infidelities} range  between $2.4\times10^{-2}$ and  $0.30$, which are higher than the gate infidelity values expected for performing FT-QEC. However, to show $\text{tr}(W_{n}\rho)= 0$ we consider higher possible errors. Recall that we are focusing on two-qubit light-shift gates mediated by the radial phonons along the $x$-axis. The values for the trapping frequencies are $\omega_{x, m}/2\pi=\{4.64,4.37   \}$ MHz,  $\omega_{y, m}/2\pi=\{3.88, 3.57\}$ MHz and $\omega_{z, m}/2\pi=\{1.49, 2.57\}$ MHz for the two normal modes $m=\{\rm com, zz\}$ ~\cite{PhysRevX.12.011032}. The laser beams are slightly misaligned with respect to the $\alpha=x$ axis,  $\Delta \vec{k}_L \parallel \boldsymbol{e}_{x}=-0.5$ rad.  The Rabi frequency of the dipole-allowed transitions is set to $\Omega_{l}=0.1\omega_{x, \rm{com}}$, and the detuning of both lasers from the dipole-allowed transitions are set to $\Delta_{l}/2\pi = -100$ GHz, with the decay from the $P$-manifold set by the decay rate $\Gamma/2\pi\approx 22$ MHz. 

For the light-shift gate, the beat-note frequency of the laser $\Delta\omega_{L}$ is set to $ f_{1}\omega_{\mathit{x},{\rm com}}+ f_{2}\omega_{\mathit{x},{\rm zz}}$ where $f_{1}$  and $f_{2}$ are constants. The detunings to close the loops in phase space  $\delta^{x}_{m}=\omega_{x, { m}}-\Delta \omega_{L}$ satisfy $\delta^{x}_{\rm com}=2\pi r$ and $\delta^{x}_{\rm zz}=2\pi p$ with $r, p \in \mathbb{Z}$. In Fig.~\ref{fig:fracs_and_loops}, we show the way we adjusted the detunings when sweeping across gate times from $10 \rm{\mu} s$ to $100\rm{\mu} s$. If the gate time increases, the values of the frequency modes have to be reduced to achieve smaller detunings $\delta^{x}_{m}$. However, to keep values consistent with the experiment, we can not reduce the trap frequencies indefinitely. What we do instead is to change where we set the beat-note frequency in, $1/2, 2/3, 3/4,...$, and so on. This helps to decrease the detuning values without the need to reduce the trap frequencies, at the expense of increasing the number of loops in one of the modes. We consider a coherence time $T_{2}=2.1$ s \cite{Ruster2016} and the $T_{1}$-time which is set at $T_{1}=1/2\Gamma^{\rm Ram}_{{\rm em}, i}\approx[ 2(\frac{\Gamma}{\Delta_{l}})\frac{\Omega_{l}^{2}}{\Delta_{l}}]^{-1}$ according to the Raman scattering rate. We also ensure that $T_{2}^{\rm eff}$ $({\Gamma}/{\Delta_{l}})\ll 1$ is fulfilled.

\begin{table*}[th] 
\centering
 \begin{tabular}{|c | c  c |} 
 \hline
 Thermal errors & warm active phonons (\ref{eq:gate_thermal_active}) &  $\epsilon_{\rm th}^{\rm act} =\frac{\pi^{2}}{20}\sum_{m,m'}\eta_{x,m}^{2}\eta_{x,m'}^{2}\langle\hat{n}_{x,m}\hat{n}_{x,m'}\rangle$   \\ 
 & warm spectator phonons (\ref{eq:gate_thermal_spectator}) & $\epsilon_{\rm th}^{\rm spec} =\frac{\pi^{2}}{20}\sum_{m,m'}\sum_{\alpha,\alpha'}C_{\alpha,\alpha'}\eta_{\alpha,m}^{2}\eta_{\alpha',m'}^{2}\langle\hat{n}_{\alpha,m}\hat{n}_{\alpha',m'}\rangle$  \\ \hline
 Off-resonant coupling & active phonons (\ref{eq:gate_off_active}) & $ \epsilon_{\rm off}^{{\rm act}}=\sum_{i,m}\frac{\Tilde{\Omega}_{L,i}^{2}}{5}
\eta^2_{x,m}\frac{2\pi}{(\omega_{x,m}^2-\Delta\omega_L^2)}$  \\ 
  & spectator phonons (\ref{eq:gate_off_spectator}) &  $\epsilon_{\rm off}^{{\rm spec}}=\frac{4}{5}\sum_{\alpha=y,z}\sum_{m}\frac{(\Tilde{\Omega}_{L}\eta_{\alpha,m}^{2})^{2}}{\Delta\omega^{2}-\omega_{\alpha,m}^{2}}\times\left| \frac{\Delta\omega^{2}_{l}+\omega_{\alpha,m}^{2}}{\Delta\omega^{2}_{l}-\omega_{\alpha,m}^{2}}(2\bar{n}_{\alpha,m}+1)+\ii \frac{2\pi\omega_{\alpha,m}}{\omega_{\alpha,m}-\Delta\omega_{L}}\right|$ \\ \hline
 Dephasing noise & Local B-field (\ref{eq:gate_deph_uncorrelated}) & $\epsilon_{g}^{\rm deph, l}\approx \frac{8}{5}\Big(t_{g}/2T_{2} \Big)$ \\ 
  & Correlated B-field (\ref{eq:gate_deph_correlated}) & $\epsilon_{g}^{\rm deph, c}\approx \frac{8}{5}\Big(t_{g}/2T_{2} \Big)$   \\ 
  \hline
 Residual photon scattering & Rayleigh (\ref{eq:gate_deph_Rayleigh})  & $\epsilon_{\rm Rai}^{\rm scatt}=\frac{8}{5}\Big(t_{g}/2T_{2}^{\rm eff} \Big)$ \\ 
   & Raman (\ref{eq:gate_deph_Raman})  & $\epsilon_{\rm Ram}^{\rm scatt}=\frac{12}{5}\Big(t_{g}/T_{1}\Big)$ \\ \hline
 \end{tabular}
 \caption{{ Summary of the different error source contributions to the gate infidelity. For low values of $\bar{n}_{\alpha, m}$, the leading error contribution is due to off-resonant forces. However, for increasing $\bar{n}_{\alpha, m}$, the thermal error contribution becomes more predominant. For $\bar{n}_{\alpha, m}\approx 10$ becomes the leading error source. The contribution to the infidelity due to dephasing and residual photon scattering is much  lower than the other two sources, with the one due to dephasing one order of magnitude higher.}}
 \label{tab:table_gateinfidelity}
\end{table*}

\begin{figure}[t!]
  \includegraphics[width=1\columnwidth]{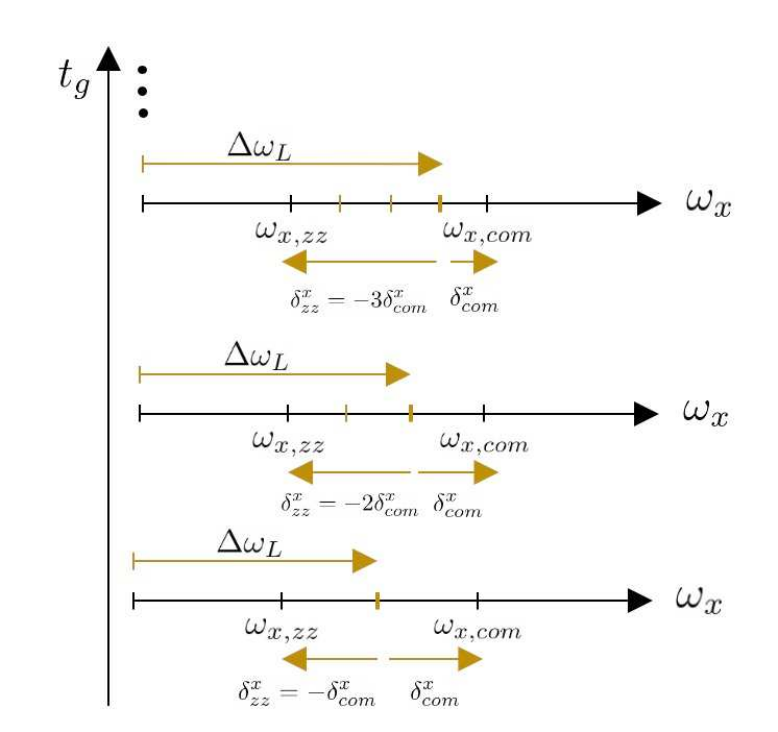}
  \caption{Adjusting phase-loop detunings for the radial $\alpha=x$ trap frequency modes. For a target gate time $t_{g}$, we need to set the detuning $\delta^{x}_{m}$ for the $m=\{\rm com, zz \}$ modes such that $\delta^{x}_{\rm com}=2\pi r$ and  $\delta^{x}_{\rm zz}=\delta^{x}_{\rm com} p$ with $r, p \in \mathbb{Z}$ to ensure that both trajectories are closed in phase space. The detuning  $\delta^{x}_{m}=\omega_{x,m}-\Delta\omega_{L}$, can be adjusted by setting $\Delta\omega_{L}$ the beat-note frequency to different fractions of the frequency difference between $\omega_{x,{\rm com}}$ and $\omega_{x,{\rm zz}}$, that is $\Delta \omega_{L}=f_{1}\omega_{x, {\rm zz}}+f_{2}\omega_{x, {\rm zz}} $ with $f_{1}$ and $f_{2}$ the fractional quantities.}
  \label{fig:fracs_and_loops}
\end{figure}

\subsection{Effective error channels}

 During QEC cycles, the individual parity-check readout circuits consist of sequences of single-qubit gates and two-qubit light-shift gates that map the information from data qubits onto ancillary qubits. Typically, the circuit noise is modelled by introducing a quantum noise channel after each  gate  which, with some probability $p$, leads to a single or two-qubit error on the physical qubits participating in the gate, whereas with a probability $1-p$, the qubits are left unaffected.\\

In this section, we discuss different effective error models for two-qubit light-shift gates, which are the bottleneck in the current trapped-ion implementations. We use all the microscopic error contributions to the $U_{id}$ gate calculated in Sec.~\ref{Sec:non_ideal_conditions} to extract the error rate $p$ that is fed into different two-qubit error channels, such as the depolarizing channel and the two-qubit dephasing channel, instead of assuming  arbitrary values of $p$.

We can achieve this by establishing a connection between the analytical expression of the microscopic gate infidelity $(i)$ $\bar{\epsilon}_{g, T}(\epsilon_{g}^{\rm th}+{\epsilon}_{g}^{\rm off}+{\epsilon}_{g}^{\rm deph}+{\epsilon}_{g}^{\rm scatt})$, and $(ii)$ the gate infidelity we get when introducing an effective two-qubit noise channel (e.g. depolarizing $\bar{\epsilon}_{g, T}(p_{dp})$ or dephasing $\bar{\epsilon}_{g, T}(p_{deph})$) following the ideal unitary, with $p_{dp}$ and $p_{deph}$ being the error rates for each channel, respectively. By matching expressions $(i)$ and $(ii)$, the error rates $p_{dp}$ and $p_{deph}$ can be written in terms of the microscopic error parameters, i.e.,  $p_{dp}, p_{deph} \rightarrow {\epsilon}_{g}^{\rm th}+{\epsilon}_{g}^{\rm off}+{\epsilon}_{g}^{\rm deph}+\epsilon_{g}^{\rm scatt}$. \\
\subsubsection{Two-qubit depolarizing noise}\label{sec:depo_channel}
For a two-qubit depolarizing noise channel, each pair of ions involved in a light-shift gate can undergo 15 possible one and two-qubit Pauli errors, so that the total error channel is described as
\begin{equation}\label{eq:noisedepo2}
\begin{split}
    \epsilon_{dp}(\rho)=(1-p_{dp})\rho+ \frac{p_{dp}}{15}\sum_{(\sigma_i,\kappa_j)\in \mathcal{Y}}\sigma_{i}\kappa_{j}\rho\kappa_{j}\sigma_{i}.
\end{split}
\end{equation} 
Here, $i$ and $j$ with $i\neq j$ denote the active ions involved in each two-qubit entangling gate, $p_{dp}$ is the  error probability, and the sum on the right runs over the 6 nontrivial single-qubit and the 9 nontrivial two-qubit Pauli operators in $ \mathcal{Y}=\{I,X,Y,Z\}^{2}/(I, I)$. Here, we have considered a symmetric two-qubit depolarizing channel, where the probability of single and two-qubit errors is the same. Next, by making $\epsilon_{dp}(\rho)= \epsilon(\ket{\phi_{m}}\bra{\phi_{m}})$ we calculate the entanglement fidelity (\ref{eq:F_e_definition}) in terms of $p_{dp}$. One finds  
\beq
\bar{\mathcal{F}}_{e}= |1-p_{dp}|^{2}=1-2p_{dp}+O(p_{dp}^{2})
\label{eq:entanglement_fidelity_depo}
\eeq
which means that the effect of all the possible errors perturbing the target state, i.e. bit and phase flip errors, is on average null. In this case, we find that the two-qubit errors $\{XX, YZ, ZY\}$ contribute positively (+1) or negatively (-1) to the probability $p_{dp}$ depending on the initial input state  
$\ket{\alpha} \in \{\ket{++}, \ket{+-}, \ket{-+}, \ket{--}\}$ so that, on average, their effects cancel out. Finally, the average gate fidelity from Eq.~(\ref{eq:avg_Fg_and_Fe}) allows us to fix the channel error rate to
\beq \label{eq:av_gate_fidelity_depo}
\bar{\mathcal{F}}_{g}= 1-\frac{8}{5}p_{dp} \rightarrow p_{dp}=\frac{5}{8}\bar{\epsilon}_{g, T}
\eeq
{where we have used $\bar{\epsilon}_{g, T}=1-\bar{\mathcal{F}}_{g}$.}

\subsubsection{Two-qubit dephasing noise}\label{sec:deph_channel}

As per discussion in the previous section, there is a privileged basis for error channels (except the Raman scattering), namely, the $Z$ basis. With the exception of the Raman scattering, all the other sources of error act on such a phase-flip basis. Accordingly, it is expected that dephasing-type errors be more dominant than bit-flip errors, or a combination of both. Thus, it is more reasonable to consider  a two-qubit dephasing channel 
\begin{equation}\label{eq:two_deph_channel}
\begin{split}
    \epsilon_{deph}(\rho)=(1-p_{deph})\rho+ \frac{p_{deph}}{3}\sum_{(\sigma_i,\kappa_j)\in\mathcal{I}}\sigma_{i}\kappa_{j}\rho\kappa_{j}\sigma_{i},
\end{split}
\end{equation}
where $i$ and $j$ represent the active ions involved in the two-qubit gate, and $\mathcal{I}=\{I,Z\}^2/(I,I)$. The quantum channel in Eq.~(\ref{eq:two_deph_channel}) is also symmetric, where all the errors $\{IZ, ZI, ZZ\}$ occur with the same probability $p_{deph}$. As before, we can compute the entanglement fidelity for this channel by introducing $\epsilon_{deph}=\epsilon(\ket{\phi_{m}}\bra{\phi_{m}})$ into the entanglement fidelity Eq.(\ref{eq:F_e_definition}). We arrive at a similar result as for the depolarizing channel, $\bar{\mathcal{F}}_{e}\approx 1-p_{deph}$. In this case, all single and two-qubit dephasing errors affect the expected output state. Thus, the relation between the dephasing error rate $p_{deph}$ and the average gate infidelity is
\beq \label{eq:av_gate_fidelity_deph}
\bar{\mathcal{F}}_{g}= 1-\frac{8}{5}p_{deph} \rightarrow p_{deph}=\frac{5}{8}\bar{\epsilon}_{g, T},
\eeq
leading to a similar relation as in the depolarizing noise model. Note, however, that the effect of these channels will be different for the specific circuits and target $n$-qubit entangled states.

We have thus shown how to relate the error probabilities for different effective noise channels with the average gate infidelities of the light-shift gates that arise from detailed microscopic  calculations. Here, even though the weight of the error probability with the average gate fidelity is the same for both cases, i.e $5/8$, the structure of the noise is different and would propagate differently through the circuit ~\cite{PRXQuantum.2.020304}. In the following section, we address this question: can the noise structure of both channels lead to better or worse performance of the corresponding circuit? And can one modify the circuits to leverage our detailed knowledge of the microscopic noise? 

\section{\bf  Characterization of noisy trapped-ion QEC circuits via entanglement witnesses}\label{Sec:results}

In this section, taking previous works~\cite{PRXQuantum.2.020304,PhysRevX.12.011032} a step further we evaluate the performance of the  parity check measurement circuits in terms of their ability to generate GME states in the presence of realistic microscopic noise.
For the numerical simulations, the chosen error model in this work includes perfect light-shift gate operations followed  by a  two-qubit depolarizing~\eqref{eq:noisedepo2}  or a two-qubit dephasing~\eqref{eq:two_deph_channel} channel on the active qubits. We include measurement errors for the Pauli operator readouts and neglect errors on single-qubit gates since it has been shown that for this particular experimental setup~\cite{doi:10.1116/1.5126186}, they are negligible in comparison to the two-qubit  and measurement errors.\\

We focus on experiments with  trapped-ion crystals of a single atomic species, which are shuttled around in order to perform the sequence of single- and two-qubit gates by bringing the corresponding ions to operation zones of a segmented trap where they can be coupled to lasers. In this work, we assume that all the vibrational modes, active and spectator, have the same average number of phonons  after the Doppler and sideband cooling processes. Since we do not have mixed species at our disposal, no sympathetic cooling is considered, and the entangling gates will get worse due to the heating caused by the intermediate shuttling operations. This effect is directly accounted for by our expressions of the gate errors i.e. thermal errors contributions in Table.\ref{tab:table_gateinfidelity} and their relations to the error-channel rates. Finally, we note that
in Ref.~\cite{Ruster2016}, it was shown that stable magnetic fields lead to a negligible dephasing of qubits during the idle-time intervals in which they are not acted upon. Thus, we assume that during gates and shuttling operations in between, idle qubits do not suffer from dephasing noise.  

\begin{figure*}[t!]
  \includegraphics[width=1\textwidth]{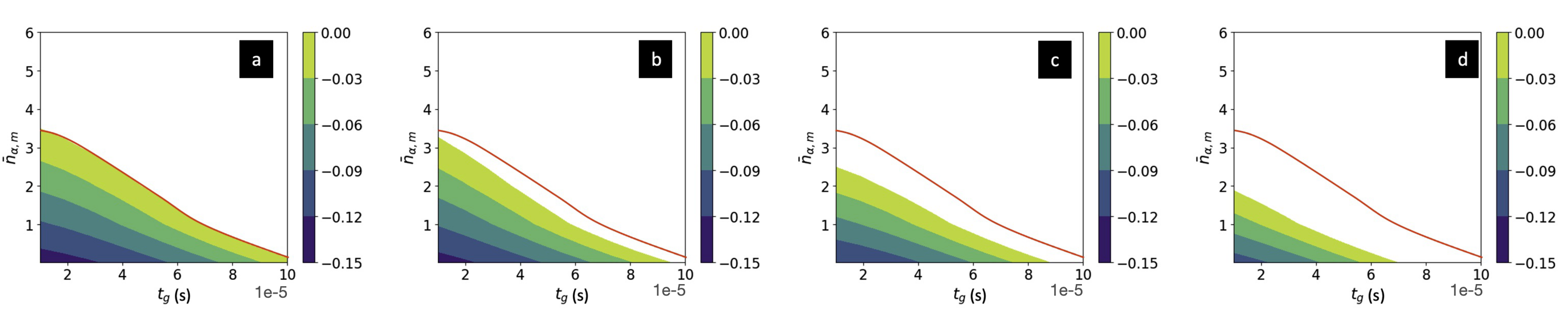}
  \caption{SL witnessing for a  $X$-type parity-check readout circuit under effective microscopic noise channels. The color bar indicates the value of the witness. For all the colors represented, the witness takes a negative value signalling the presence of entanglement. The white area represents all positive values of the witness values where the verification is inconclusive. The red solid line fits the perimeter area of the (a) non-FT $X$-type circuit under a two-qubit depolarizing noise channel. Then, the solid line is plotted on top of the following graphs to benchmark the size of the enclosed colored area on the (b) flag-based FT $X$- type circuit under depolarizing noise, and for (c) the non-FT and (d) FT X-type circuits under dephasing noise, respectively. This witness is reconstructed using the following 5-qubit stabilizer generators $g_{1}=Z_{1}Z_{2}, g_{2}=Z_{2}Z_{3}, g_{3}=Z_{3}Z_{4}, g_{4}=Z_{4}Z_{5},  g_{5}=X_{1}X_{2}X_{3}X_{4}X_{5}$ of the state $(\ket{00000}+\ket{11111})/\sqrt{2}$.}
  \label{fig:XXXX}
\end{figure*}

For the numerical simulations, we model the errors during measurement by including an effective noise channel for the projection operators. The single-qubit Pauli operator $\sigma\in\{X,Y,Z\}$ can be measured using the  corresponding error-free positive operator-valued measure (POVM)~\cite{nielsen00} 
$\{P^\sigma_\pm={I \pm \sigma \over 2}\}$.
However, for each measurement outcome, a bit-flip error can take place with probability $p\s{me}$, which gives rise to the erroneous effect operators 
\begin{equation}
\begin{split}
    P^{\sigma}_+ &  \mapsto  (1 - p\s{me}) P^\sigma_+ + p\s{me} P^\sigma_-,
\end{split}
\end{equation}
and
\begin{equation}
\begin{split}
    P^{\sigma}_- &   \mapsto(1 - p\s{me}) P^\sigma_- + p\s{me} P^\sigma_+. 
\end{split}
\end{equation}
For all the simulation results shown in this manuscript, the measurement error probability is set to $p_{me}=10^{-3}$ \cite{doi:10.1063/1.5088164}. We now analyze  the numerical results for the SL and the CL witnessing methods.

\subsubsection{Numerics for the standard linear witnessing}\label{sec:results_SL_witness}

In Fig.~\ref{fig:XXXX}, we show the entanglement witnessing results for the $X$-type parity-check readout circuits of Figs.~\ref{fig:parity_check_measurement} and~\ref{fig:parity_check_measurement_FT} following the standard linear method described in Sec.~\ref{sec:SL_witness}. The results for the $Z$-type parity checks measurement circuits are practically the same since the error propagation of one and two-qubit Pauli operators from either depolarizing or depahasing channel through the native gates leads to a similar result in terms of  number of expectation values of the stabilizer generators flipped, that is same profile and the number of $\expec{g_{i}}=-1$ than when the propagation of errors is evaluated in an $X$-type parity-check circuit. Note that when the stabilizers~\eqref {eq:CSS_stabilizers} flip, the corresponding values of the test operators described above  will  decrease, which can make the result of the witness measurement inconclusive.

For the simulation, we have used the same parameters used to generate  Fig.~\ref{fig:avg_gate_infidelities}. We plot the results for different gate times $t_{g}$ and for different initial average phonon numbers $\bar{n}_{\alpha, m}$, which will change the gate error and the corresponding channel parameters. Recall that all modes are cooled down to the same initial phonon occupation number. However, after each entangling gate, the phonon number increases, which can be estimated to contribute with $3.9$ phonons for the $z$-axial direction (shuttling axis), and  $0.255$ for the radial $x$- and $y$-axis directions~\cite{Bermudez_2017}.\\
Fig.~\ref{fig:XXXX} {\bf (a)} and {\bf (b)} show the results for the non-FT and flag-based $X$-type circuits under the effective depolarizing channel described in Sec.~\ref{sec:depo_channel}. Fig.~\ref{fig:XXXX} {\bf (c)} and {\bf (d)} show the results for the same circuits in the presence of the effective dephasing channel of Sec.~\ref{sec:deph_channel}. The red solid line represents the boundary between the region in parameter space where errors are low, and we can conclusively detect GME via the witness, and an inconclusive outer region. One can notice that the entanglement witnessed area reduces from left to right. With the SL witness built from stabilizers $\langle g_1=Z_{1}Z_{2}, g_2=Z_{2}Z_{3}, g_3=Z_{3}Z_{4}, g_4=Z_{4}Z_{5}, g_5=X_{1}X_{2}X_{3}X_{4}X_{5}\rangle$ we can detect more entangled stated for the same physical parameters when the circuits  are affected  by  the depolarizing noise channel~\eqref{eq:noisedepo2} than when they are subjected to the dephasing channel~\eqref{eq:two_deph_channel}. We will refer to as the nearest-neighbor (NN) generators.\\

To better understand the entanglement witnessing areas and the noise channels, we count the number of stabilizers flipped by the error operators (cf. Table \ref{table:5q_depo_Xtype}), that is, how many turns into $\expec{g_i}=-1$, when a single or a two-qubit error from the depolarizing channel propagates through the circuit of Fig.~\ref{fig:parity_check_measurement}. Let us note that these tables are expected to provide a rough estimate of the performance of the GME witness for finite error rates, which should become more and more reliable as the error rate decreases, as errors are only linear in $p$. In Table.~\ref{table:5q_deph_Xtype}, we show the same results but for the single and two-qubit phase errors. We calculate the ratio between the number of stabilizer generators that change sign when an error occurs, versus the total number of generators measured for all different types of errors. These error propagations and stabilizer flips show that using a dephasing channel after each entangling gate leads to more stabilizer flips than a depolarizing channel. In fact, Table. \ref{table:5q_depo_Xtype}, the percentage of "$-1$'s" goes as $\%16, \%32, \%32, \%32$, for the depolarizing errors applied after the first, second, third, and fourth entangling gate, respectively. However, for the dephasing-type errors in Table. \ref{table:5q_deph_Xtype}, that percentage increases up to $\%40$. A larger number of $\expec{g_i}=-1$, consequently reduces the value of the witness test operator against the entanglement bound (see Eq.~\eqref{eq:normalizedwitness}), and thus, the negativity of the witness. 

\begin{table}
\begin{center}
\begin{tabular}{ |c|c| c|c | c|c|} 
\hline
Error & $g_1$ $g_2$ $g_3$ $g_4$ $g_5$  & \# -1's &   Error & $g_1$ $g_2$ $g_3$ $g_4$ $g_5$ & \# -1's  \\
\hline
$X_{1}I_{5}$ & -1 1 1 1 -1 & 2 & $X_{2}I_{5}$ &  -1 -1 1 1 -1 & 3 \\  
$I_{1}X_{5}$ & -1 1 1 1 -1 & 2 & $I_{2}X_{5}$ & 1 -1 1 1 1 & 1 \\ 
$X_{1}X_{5}$ & 1 1 1 1 1 &  0 &  $X_{2}X_{5}$ &  -1 1 1 1 -1 & 2\\
$X_{1}Y_{5}$ & 1 1 1 1 -1  &1 & $X_{2}Y_{5}$ & -1 1 1 1 1 & 1\\
$Y_{1}X_{5}$ & -1 1 1 1 1& 1 & $Y_{2}X_{5}$ &  1 -1 1 1 -1 & 2\\ 
$X_{1}Z_{5}$ &  -1 1 1 1 1 & 1 & $X_{2}Z_{5}$ &  -1 -1 1 1 1 & 2\\
$Z_{1}X_{5}$ & 1 1 1 1 -1 & 1& $Z_{2}X_{5}*$ &  -1 1 1 1 1 & 1\\
$Y_{1}I_{5}$ & 1 1 1 1 -1 &1 & $Y_{2}I_{5}$ & 1 1 1 1 -1 & 1\\
$I_{1}Y_{5}$ & -1 1 1 1 1 & 1& $I_{2}Y_{5}$ & 1 -1 1 1 -1 & 2\\
$Y_{1}Y_{5}$ & -1 1 1 1 -1 & 2 & $Y_{2}Y_{5}$ &  1 -1 1 1 1 & 1\\
$Y_{1}Z_{5}$ & 1 1 1 1 1 & 0  & $Y_{2}Z_{5}$ & 1 1 1 1 1 & 0\\
$Z_{1}Y_{5}$ & 1 1 1 1 1 & 0 & $Z_{2}Y_{5}$ & -1 1 1 1 -1 & 2\\
$Z_{1}I_{5}$ & -1 1 1 1 1 & 1 & $Z_{2}I_{5}$ &  -1 -1 1 1 1 & 2\\
$I_{1}Z_{5}$ & 1 1 1 1 -1 &1 & $I_{2}Z_{5}$ &  1 1 1 1 -1 & 1\\
$Z_{1}Z_{5}$ & -1 1 1 1 -1 &2 & $Z_{2}Z_{5}$ & -1 -1 1 1 -1 & 3\\
\hline
\multicolumn{3}{| c |}{Total: 16/75; \%21} & \multicolumn{3}{| c |}{Total: 22/75; \%32 }\\
\hline
$X_{3}I_{5}$ & 1 -1 -1 1 -1 & 3  & $X_{4}I_{5}$ & 1 1 -1 -1 -1 & 3\\ 
$I_{3}X_{5}$ & 1 1 -1 1 -1 & 2 & $I_{4}X_{5}$ & 1 1 1 -1 1  & 1\\ 
$X_{3}X_{5}$ & 1 -1 1 1 1 & 1   & $X_{4}X_{5}$ & 1 1 -1 1 -1  & 2\\
$X_{3}Y_{5}$ & 1 -1 1 1 -1 & 2 & $X_{4}Y_{5}$ & 1 1 -1 1 1 & 1\\ 
$Y_{3}X_{5}$ & 1 1 -1 1 1 & 1 & $Y_{4}X_{5}$ & 1 1 1 -1 -1 & 2\\ 
$X_{3}Z_{5}$ & 1 -1 -1 1 1 & 2  &  $X_{4}Z_{5}$ & 1 1 -1 -1 1 & 2\\
$Z_{3}X_{5}$ & 1 -1 1 1 -1 & 2&  $Z_{4}X_{5}$ & 1 1 -1 1 1  & 1 \\ 
$Y_{3}I_{5}$ & 1 1 1 1 -1 & 1 &  $Y_{4}I_{5}$ & 1 1 1 1 -1 & 1\\ 
$I_{3}Y_{5}$ & 1 1 -1 1 1 & 1   &  $I_{4}Y_{5}$ & 1 1 1 -1 -1 & 2\\
$Y_{3}Y_{5}$ & 1 1 -1 1 -1 & 2 &  $Y_{4}Y_{5}$ & 1 1 1 -1 1 & 1 \\ 
$Y_{3}Z_{5}$ & 1 1 1 1 1 & 0  &  $Y_{4}Z_{5}$ & 1 1 1 1 1& 0\\ 
$Z_{3}Y_{5}$ & 1 -1 1 1 1 & 1   &  $Z_{4}Y_{5}$ & 1 1 -1 1 -1 & 2\\
$Z_{3}I_{5}$ & 1 -1 -1 1 1& 2  & $Z_{4}I_{5}$ &  1 1 -1 -1 1 & 2\\
$I_{3}Z_{5}$ & 1 1 1 1 -1  & 1 &  $I_{4}Z_{5}$ & 1 1 1 1 -1 & 1\\
$Z_{3}Z_{5}$ & 1 -1 -1 1 -1 & 3 &  $Z_{4}Z_{5}$ & 1 1 -1 -1 -1 & 3\\
\hline
\multicolumn{3}{| c |}{Total: 24/75; \%32} & \multicolumn{3}{| c |}{Total: 24/75;  \%32}\\
\hline
\end{tabular}
\end{center}
\caption{ One and two-qubit depolarizing errors propagating through a non-FT $X$-type parity-check circuit. The subscripts on the Pauli operators refer to the qubits affected by the errors after the application of the entangling light-shift gate on those two qubits. $\pm 1$ numbers refers to the expectation of the stabilizer generators $g_{1}=Z_{1}Z_{2}$,  $g_{2}=Z_{2}Z_{3}$, $g_{3}=Z_{3}Z_{4}$, 
$g_{4}=Z_{4}Z_{5}$, $g_{5}=X_{1}X_{2}X_{3}X_{4}X_{5}$. }
\label{table:5q_depo_Xtype}
\end{table}

\begin{table}
\begin{center}
\begin{tabular}{ |c|c| c|c | c|c|} 
\hline
Error &  $g_1$ $g_2$ $g_3$ $g_4$ $g_5$  & \# -1's &   Error &  $g_1$ $g_2$ $g_3$ $g_4$ $g_5$  & \# -1's  \\
\hline

$Z_{1}I_{5}$ & -1 1 1 1 1 & 1 & $Z_{2}I_{5}$ &  -1 -1 1 1 1 & 2\\
$I_{1}Z_{5}$ & 1 1 1 1 -1&1 & $I_{2}Z_{5}$ &  1 1 1 1 -1 & 1\\
$Z_{1}Z_{5}$ & -1 1 1 1 -1 &2 & $Z_{2}Z_{5}$ & -1 -1 1 1 -1 & 3\\
\hline
\multicolumn{3}{| c |}{Total: 4/15; \%27} & \multicolumn{3}{| c |}{Total:6/15; \%40}\\
\hline
$Z_{3}I_{5}$ & 1 -1 -1 1 1& 2  & $Z_{4}I_{5}$ &  1 1 -1 -1 1 & 2\\
$I_{3}Z_{5}$ & 1 1 1 1 -1  & 1 &  $I_{4}Z_{5}$ & 1 1 1 1 -1 & 1\\
$Z_{3}Z_{5}$ & 1 -1 -1 1 -1 & 3 &  $Z_{4}Z_{5}$ & 1 1 -1 -1 -1 & 3\\
\hline
\multicolumn{3}{| c |}{Total: 6/15; \%40} & \multicolumn{3}{| c |}{Total: 6/15; \%40}\\
\hline
\end{tabular}
\end{center}
\caption{One and two-qubit dephasing errors propagating through a non-FT $X$-type parity-check circuit. The subscripts on the Pauli operators refer to the qubits affected by the errors after the application of the entangling light-shift gate on those two qubits. The $\pm 1$ numbers refers to the expectation of the stabilizer generators $g_{1}=Z_{1}Z_{2}$,  $g_{2}=Z_{2}Z_{3}$, $g_{3}=Z_{3}Z_{4}$, 
$g_{4}=Z_{4}Z_{5}$, $g_{5}=X_{1}X_{2}X_{3}X_{4}X_{5}$. }
\label{table:5q_deph_Xtype}
\end{table}

From the perspective of the different behavior of the errors with respect to the flipped stabilizers,  it is possible to try to construct witnesses that are more robust for a particular type of noise by just looking at its error propagation through the circuit. In Fig.~\ref{fig:XXXX_Z5type}, we show the numerical results and the error propagation tables when  using different stabilizer generators  $\langle g'_1=Z_{1}Z_{5}, g'_2=Z_{2}Z_{5}, g'_3=Z_{3}Z_{5}, g'_4=Z_{4}Z_{5}, g_5=X_{1}X_{2}X_{3}X_{4}X_{5}\rangle$, which will be referred the in the text as $Z_{5}$-biased generators. With these new generators, we are able to construct a witness that exhibits a larger witnessed entanglement area for the dephasing channel (see Fig.\ref{fig:XXXX_Z5type} {\bf{(c)}})  than it does for the depolarizing channel, as shown in Fig.\ref{fig:XXXX_Z5type} {\bf{(a)}}. This reverts the behavior observed in Fig.\ref{fig:XXXX}.
The fact that, under the dephasing channel, we obtain better results for the $Z_5$-biased generators than for the NN ones can be justified as follows: $(i)$ the $Z_{d}I_5$ errors with $d=1,2,3,4,$ transforms into $X_{d}I_{5}$ type of errors at the end of the circuit, causing just one "-1" sign flip for the $Z_5$-biased stabilizers. On the other hand, for the NN-type generators, it can contribute with up to two, e.g. $Z_{2}I_5 \rightarrow X_{2}I_5$ flips the $Z_1Z_2$ and $Z_2Z_3$ generators. $(ii)$ $I_{d}Z_5$ errors change the sign of $g_5=X_{1}X_{2}X_{3}X_{4}X_{5}$ for both type of generators. Finally, $(iii)$ $Z_{d}Z_{5}$ errors, which are the product of the other two types, contribute to up to three sign flips of the NN-type generators. On the other hand,  at most two sign flips can occur for the $Z_5$-biased generators. In other words, the $Z_5$-biased stabilizers commute with more dephasing errors than the NN-type generators do, and consequently, there are more "+1" values helping to increase the difference between the entanglement test operator and its entanglement bound, and consequently, the negativity of the witness. We can also corroborate these results by looking at the individual error propagation for the depolarizing and dephasing channels of the $Z_5$-biased generators in Table.\ref{Tab: Z5_type_depoalrizing} and \ref{Tab:Z5_type_dephasing}, respectively. We note that the percentage of "-1" sign flips for the dephasing channel is around $\%27$, whilst it is, in the majority of cases, higher for the depolarizing channel.
We note, then, that the numerical and tabulated results agree for the case of the non-FT circuits. By looking at Fig.\ref{fig:XXXX_Z5type} {\bf{(b)}} and {\bf{(d)}}, we confirm again that the FT circuit design with the flag-qubit is not advisable for gates with high failure rates.
Thus, if one wants to qualitatively characterize the performance of QEC circuits in terms of SL witnesses, it is important to construct an optimal witness, that is, one whose generators  commute with the largest number of  errors propagated through the circuit.

\begin{figure*}[t!]
  \includegraphics[width=1\textwidth]{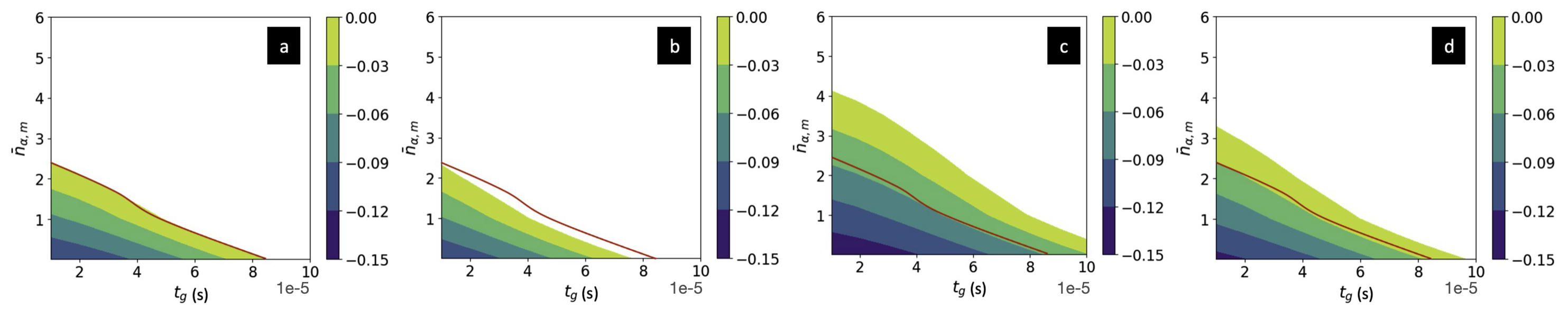}
  \caption{SL witnessing for a  $X$-type parity-check readout circuit under effective microscopic noise channels. The color bar indicates the value of the witness. For all the colors represented, the witness takes a negative value signalling the presence of entanglement. The white area represents all positive values of the witness values where the verification is inconclusive. The red solid line fits the perimeter area of the (a) non-FT $X$-type circuit under a two-qubit depolarizing noise channel. Then, the solid line is plotted on top of the following graphs to benchmark the size of the enclosed colored area on the (b) flag-based FT $X$- type circuit under depolarizing noise, and for (c) the non-FT and (d) FT $X$-type circuits under dephasing noise, respectively. This witness is reconstructed using the following 5-qubit stabilizer generators $g_{1}=Z_{1}Z_{5}, g_{2}=Z_{2}Z_{5}, g_{3}=Z_{3}Z_{5}, g_{4}=Z_{4}Z_{5},  g_{5}=X_{1}X_{2}X_{3}X_{4}X_{5}$ of the state $(\ket{00000}+\ket{11111})/\sqrt{2}$.}
  \label{fig:XXXX_Z5type}
\end{figure*}

\begin{table}
\begin{center}
\begin{tabular}{ |c|c| c|c | c|c|} 
\hline
Error & Syndrome & \# -1's &   Error & Syndrome & \# -1's  \\
\hline
$X_{1}I_{5}$ & -1 1 1 1 -1  & 2 & $X_{2}I_{5}$ &  1 -1 1 1 -1 & 2 \\  
$I_{1}X_{5}$ & -1 1 1 1 -1 & 2 & $I_{2}X_{5}$ & -1 -1 1 1 1 & 2 \\ 
$X_{1}X_{5}$ & 1 1 1 1 1 &  0 &  $X_{2}X_{5}$ &  -1 1 1 1 -1 & 2\\
$X_{1}Y_{5}$ & 1 1 1 1 -1  &1 & $X_{2}Y_{5}$ & -1 1 1 1 1 & 1\\
$Y_{1}X_{5}$ & -1 1 1 1 1& 1 & $Y_{2}X_{5}$ &  -1 -1 1 1 -1 & 3\\ 
$X_{1}Z_{5}$ &  -1 1 1 1 1 & 1 & $X_{2}Z_{5}$ &  1 -1 1 1 1 & 1\\
$Z_{1}X_{5}$ & 1 1 1 1 -1 & 1& $Z_{2}X_{5}$ &  -1 1 1 1 1 & 1\\
$Y_{1}I_{5}$ & 1 1 1 1 -1 &1 & $Y_{2}I_{5}$ & 1 1 1 1 -1 & 1\\
$I_{1}Y_{5}$ & -1 1 1 1 1 & 1& $I_{2}Y_{5}$ & -1 -1 1 1 -1 & 3\\
$Y_{1}Y_{5}$ & -1 1 1 1 -1 & 2 & $Y_{2}Y_{5}$ &  -1 -1 1 1 1 & 2\\
$Y_{1}Z_{5}$ & 1 1 1 1 1 & 0 & $Y_{2}Z_{5}$ & 1 1 1 1 1 & 0\\
$Z_{1}Y_{5}$ & 1 1 1 1 1 & 0 & $Z_{2}Y_{5}$ & -1 1 1 1 -1 & 2\\
$Z_{1}I_{5}$ & -1 1 1 1 1 & 1 & $Z_{2}I_{5}$ &  1 -1 1 1 1 & 1\\
$I_{1}Z_{5}$ & 1 1 1 1 -1 &1 & $I_{2}Z_{5}$ &  1 1 1 1 -1 & 1\\
$Z_{1}Z_{5}$ & -1 1 1 1 -1 &2 & $Z_{2}Z_{5}$ & 1 -1 1 1 -1 & 2\\
\hline
\multicolumn{3}{| c |}{Total: 16/75; \%21} & \multicolumn{3}{| c |}{Total: 24/75; \%32 }\\
\hline
$X_{3}I_{5}$ & 1 1 -1 1 -1 & 2  & $X_{4}I_{5}$ & 1 1 1 -1 -1 & 2\\ 
$I_{3}X_{5}$ & -1 -1 -1 1 -1 & 4 & $I_{4}X_{5}$ & -1 -1 -1 -1 1  & 4\\ 
$X_{3}X_{5}$ & -1 -1 1 1 1 & 1   & $X_{4}X_{5}$ & -1 -1 -1 1 -1  & 4\\
$X_{3}Y_{5}$ & -1 -1 1 1 -1 & 3 & $X_{4}Y_{5}$ & -1 -1 -1 1 1 & 3\\ 
$Y_{3}X_{5}$ & -1 -1 -1 1 1 & 3 & $Y_{4}X_{5}$ & -1 -1 -1 -1 -1 & 5\\ 
$X_{3}Z_{5}$ & 1 1 -1 1 1 & 1  &  $X_{4}Z_{5}$ & 1 1 1 -1 1 & 1\\
$Z_{3}X_{5}$ & -1 -1 1 1 -1 & 3&  $Z_{4}X_{5}$ & -1 -1 -1 1 1  & 3 \\ 
$Y_{3}I_{5}$ & 1 1 1 1 -1 & 1 &  $Y_{4}I_{5}$ & 1 1 1 1 -1 & 1\\ 
$I_{3}Y_{5}$ & -1 -1 -1 1 1 & 3   &  $I_{4}Y_{5}$ & -1 -1 -1 -1 -1 & 5\\
$Y_{3}Y_{5}$ & -1 -1 -1 1 -1 & 4 &  $Y_{4}Y_{5}$ & -1 -1 -1 -1 1 & 4 \\ 
$Y_{3}Z_{5}$ & 1 1 1 1 1 & 0  &  $Y_{4}Z_{5}$ & 1 1 1 1 1& 0\\ 
$Z_{3}Y_{5}$ & -1 -1 1 1 1 & 2   &  $Z_{4}Y_{5}$ & -1 -1 -1 1 -1 & 4\\
$Z_{3}I_{5}$ & 1 1 -1 1 1& 1  & $Z_{4}I_{5}$ &  1 1 1 -1 1 & 1\\
$I_{3}Z_{5}$ & 1 1 1 1 -1  & 1 &  $I_{4}Z_{5}$ & 1 1 1 1 -1 & 1\\
$Z_{3}Z_{5}$ & 1 1 -1 1 -1 & 2 &  $Z_{4}Z_{5}$ & 1 1 1 -1 -1 & 2\\
\hline
\multicolumn{3}{| c |}{Total: 31/75; \%41} & \multicolumn{3}{| c |}{Total: 40/75; \%53}\\
\hline
\end{tabular}
\end{center}
\caption{One and two-qubit depolarizing errors propagating through a non-FT $X$-type parity-check circuit. The subscripts on the Pauli operators refer to the qubits affected by the errors after the application of the entangling light-shift gate on those two qubits. $\pm 1$ numbers refer to the expectation of the stabilizer generators $g'_{1}=Z_{1}Z_{5}$,  $g'_{2}=Z_{2}Z_{5}$, $g'_{3}=Z_{3}Z_{5}$, $g'_{4}=Z_{4}Z_{5}$, $g_{5}=X_{1}X_{2}X_{3}X_{4}X_{5}$. }
\label{Tab: Z5_type_depoalrizing}
\end{table}

\begin{table}
\begin{center}
\begin{tabular}{ |c|c| c|c | c|c|} 
\hline
Error & Syndrome & \# -1's &   Error & Syndrome & \# -1's  \\
\hline
$Z_{1}I_{5}$ & -1 1 1 1 1 & 1 & $Z_{2}I_{5}$ &  1 -1 1 1 1 & 1\\
$I_{1}Z_{5}$ & 1 1 1 1 -1 &1 & $I_{2}Z_{5}$ &  1 1 1 1 -1 & 1\\
$Z_{1}Z_{5}$ & -1 1 1 1 -1 &2 & $Z_{2}Z_{5}$ & 1 -1 1 1 -1 & 2\\
\hline
\multicolumn{3}{| c |}{Total: 4/15; \%27} & \multicolumn{3}{| c |}{Total: 4/15; \%27}\\
\hline
$Z_{3}I_{5}$ & 1 1 -1 1 1& 1  & $Z_{4}I_{5}$ &  1 1 1 -1 1 & 1\\
$I_{3}Z_{5}$ & 1 1 1 1 -1  & 1 &  $I_{4}Z_{5}$ & 1 1 1 1 -1 & 1\\
$Z_{3}Z_{5}$ & 1 1 -1 1 -1 & 2 &  $Z_{4}Z_{5}$ & 1 1 1 -1 -1 & 2\\
\hline
\multicolumn{3}{| c |}{Total: 4/15; \%27} & \multicolumn{3}{| c |}{Total: 4/15; \%27}\\
\hline
\end{tabular}
\end{center}
\caption{One and two-qubit dephasing errors propagating through a non-FT $X$-type parity-check circuit. The subscripts on the Pauli operators refer to the qubits affected by the errors after the application of the entangling light-shift gate on those two qubits. $\pm 1$ numbers refer to the expectation of the stabilizer generators $g'_{1}=Z_{1}Z_{5}$,  $g'_{2}=Z_{2}Z_{5}$, $g'_{3}=Z_{3}Z_{5}$, 
$g'_{4}=Z_{4}Z_{5}$, $g_{5}=X_{1}X_{2}X_{3}X_{4}X_{5}$.}
\label{Tab:Z5_type_dephasing}
\end{table}

Regarding the flag-based FT circuits, by looking at Fig.~\ref{fig:XXXX} {\bf (b)} and {\bf (d)}, it looks like the flag qubit and the associated post-selection does not help to increase the robustness of the GME witness. In fact, the additional entangling gates between the syndrome and flag qubits, which are included to catch dangerous weight-2 errors to allow for an FT readout, actually add more noise in terms of spin flips for the stabilizer generators of the GHZ-type state. To show this, we have proceeded similarly as before, performing the  error propagation for the flag-based circuits of Fig.~\ref{fig:parity_check_measurement_FT}. In Tables \ref{table:6q_depo_Xtype} and \ref{table:6q_deph_Xtype}, we show the error propagation of the depolarizing and dephasing channels,  counting only the flipped $\expec{g_i}=-1$ instances in which the flag-qubit has been measured, and post-selected in the $\ket{+}$ state, that is, when $M_{f}(X)=+1$ and we get right of dangerous weight-2 errors.  We can clearly verify the increased number of stabilizer flips  for the dephasing noise, by comparing these tables with  those of the non-FT circuits, i.e. Tables~\ref{table:5q_deph_Xtype} and~\ref{table:6q_deph_Xtype}. Even if the post-selection gets rid of some weight-2 errors that would deteriorate the GHZ-type state, the price to pay is to introduce more faulty light-shift gates that result in more stabilizer flips and thus do not reduce the percentage of "$-1$" values, but instead add an extra \%6 per entangling gate between syndrome and flag. This trend matches qualitatively  the simulation results shown in Fig.\ref{fig:XXXX} {\bf (c)} and {\bf (d)}. 

All the errors in the Z-basis are non-detectable by the flag, and, consequently, adding more gates, will just introduce more errors in the circuit. On the other hand, for the depolarizing channel, the simulations and the tabulated results seem contradictory. The flag qubit in Table~\ref{table:6q_depo_Xtype} helps to reduce the percentage of errors for the intermediate gates almost in half, at the expense, though, of putting in an extra \%11 per entangling gate between syndrome and flag qubit. Despite this extra noise, the net result of having the flag is favorable, a lower number of "$-1$" are produced in the presence of the flag qubit. This apparently contradicts the results displayed in Fig.\ref{fig:XXXX} {\bf (a)} and {\bf (b)}, where the enclosed GME-witnessing  area for the flag-based circuit is smaller than for the non-FT case.

This discrepancy can be primarily due to the fact that we are considering high error probabilities, whereas the tables only consider erroneous events linear in such a probability.  For the depolarising channel, we are considering $p_{dp}>10^{-2}$, which increases due to shuttling as the circuit proceeds, and similarly for the  dephasing channel with $p_{deph}>10^{-2}$. For the dephasing channel, as discussed above,  the flag qubit under a dephasing channel would always add more noise than it could help to remove, even for low-error probabilities. On the other hand, for the depolarizing channel, this is not always the case. For lower failure probabilities, it has been shown that the flag-qubit helps to increase pseudo-thresholds in FT-QEC protocols \cite{Chamberland2018flagfaulttolerant}. For the GME-witnessing studied  here, we need to use higher values of $p_{dp}$  to show the boundary between the entanglement witnessed and the non-witnessed regions. For such values, terms of order $\mathcal{O}(p_{dp}^{2})$ and higher become relevant, and the flag can not cope with them. If we go to low failure probabilities with one error propagating at most in the circuit, that is, of the order of  $\mathcal{O}(p_{dp})$, the flag-qubit has indeed a positive effect, as it is shown in Table~\ref{table:6q_depo_Xtype}. On the other hand, having a more negative witness cannot be used to ascertain that the circuit is performing better, as we can only detect when the state has GME. On the other hand, if we compute the state fidelity, its quantitative value can give us further confirmation that the above reasoning is correct.

\begin{figure}[ht!]
\centering
  \includegraphics[width=0.9\columnwidth]{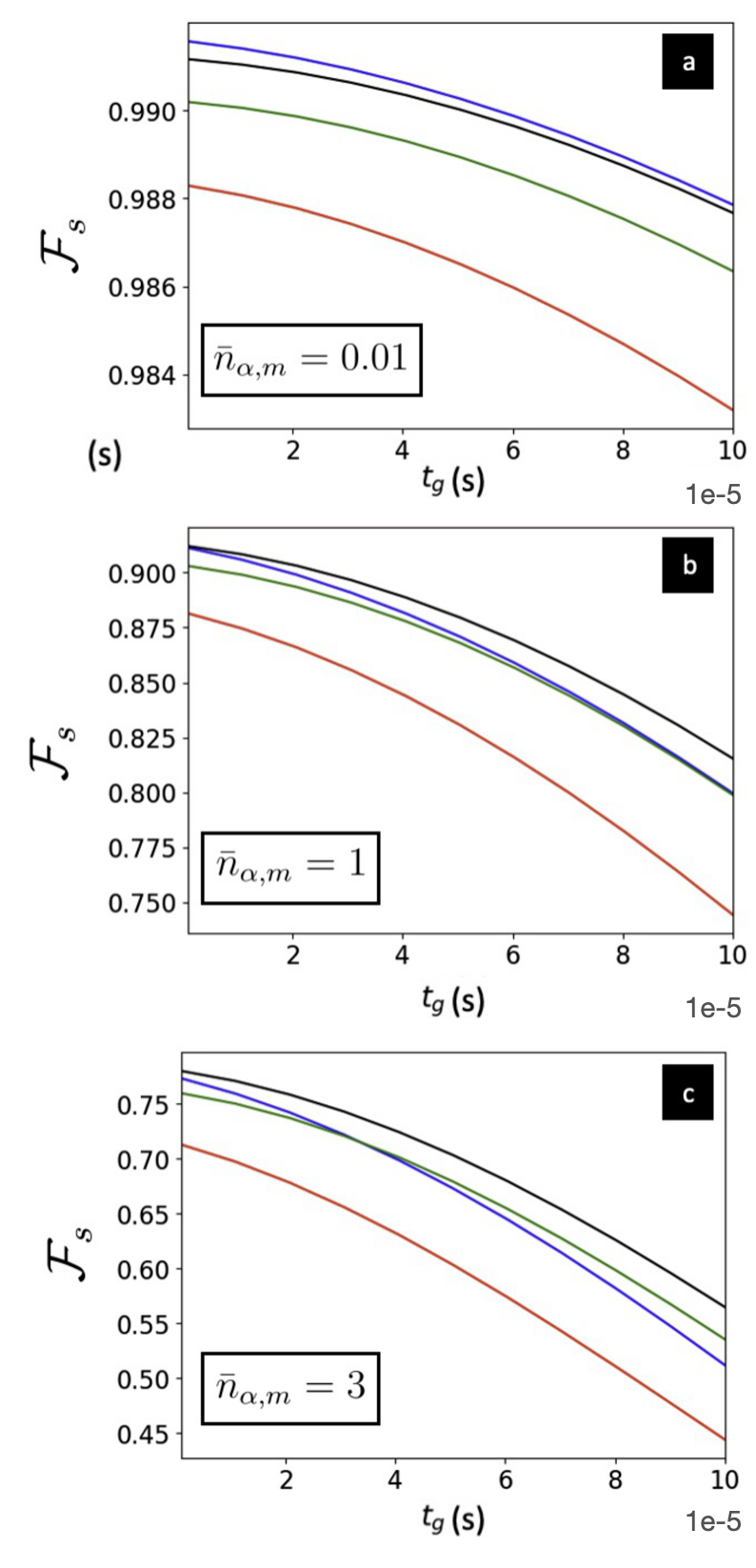}
  \caption{{Fidelity $\mathcal{F}_{s}$ between the output state generated by an ideal performance of the $X$- type circuit and the output state generated in the presence of two-qubit depolarizing and two-qubit dephasing error channels for different gate times and initial average phonon number. All modes are initially cooled down to the same $\bar{n}$ (inset). The three figures show the results for gate times $t_{g}$ ranging from $1 {\rm{\mu}}$s up to $100 {\rm{\mu}}$s for different initial $\bar{n}$. Figure {\bf (a)} shows the result for an initial phonon occupation number of $\bar{n}_{\alpha, m}=0.01$,  figure {\bf (b)} for $\bar{n}_{\alpha, m}=1$ and {\bf (c)} does it for $\bar{n}_{\alpha, m}=3$. The black solid line represents the state fidelity in a non-FT $X-$ type circuit run in the presence of the depolarizing channel. The blue line does the same but, for a flag-based FT circuit. The green line represents the state fidelity for the non-FT $X$-type parity check circuit in the presence of the dephasing channel, and  the red line shows the results for the flag-based version under dephasing errors. }}
  \label{fig:state_fidelity}
\end{figure}

In Fig.~\ref{fig:state_fidelity}, we study a low-to-high error probability regime for $p_{dp}$ and $p_{deph}$ in terms of the state fidelity  $\mathcal{F}_{s}$ of the output state generated by the $X$- type circuit. The three figures show the results for gate times $t_{g}$ ranging from $1 {\rm{\mu}}$s up to $100 {\rm{\mu}}$s. From top to bottom, we consider an initial phonon occupation number of $\bar{n}_{\alpha, m}=0.01$ {\bf (a)}, $\bar{n}_{\alpha, m}=1$ {\bf (b)} and $\bar{n}_{\alpha, m}=3$ {\bf (c)}, respectively, leading to increasing error probabilities. The black solid line represents the state fidelity of the ideal output GME state against the output state generated by the non-FT $X$-type circuit in the presence of a depolarizing channel. The blue line does the same but, for a flag-based FT circuit. The green line represents the state fidelity values of the output state generated with a non-FT $X$-type parity check circuit in the ideal case and in the presence of a dephasing channel. The red line shows the results for the flag-based version under dephasing errors. We note that, just in Fig.~\ref{fig:state_fidelity}{\bf (a)}, which represent the lowest error probabilities, the flag qubit mitigates the effect of the depolarizing channel, yielding higher state fidelity results than for the non-FT version. However, for increasing error probabilities, just as the $\bar{n}_{\alpha, m}$ increases, the flag-qubit has a detrimental effect, as it can be seen in Fig.~\ref{fig:state_fidelity} {\bf (b)} and Fig.~\ref{fig:state_fidelity} {\bf (c)}. That is the reason why there was an apparent discrepancy between the tabulated results in Table.\ref{table:6q_depo_Xtype}, which represent the low-error rate regime, and the ones in Figs.~\ref{fig:XXXX} {\bf (a)} and {\bf (b)}, which represent the higher-error  regime.

Let us close this sub-section by drawing some conclusions. We have characterized trapped-ion QEC circuits by their ability to generate multipartite entanglement. We have shown that it is important to incorporate realistic microscopic modeling of the noise to draw general conclusions about the benefits of an FT circuit design. From the perspective of GME witnessing, the added complexity of FT circuit designs is detrimental, as the region in parameter space where  GME can be inferred from the measurements decreases when moving from non-FT to FT designs. Despite the fact that the post-selection on the flag avoids certain weight-2 errors that are dangerous regarding  the stabilizers of the QEC code, the extra gates lead to further errors that induce more sign flips of the stabilizers of the GHZ-type GME state, which is detrimental for the performance of the entanglement witness. The severity of this trend actually depends on the noise channel that is used to model errors. For the more realistic dephasing channel of trapped-ion light-shift gates, this trend is generic: the flag qubit does not help at mitigating the effects of dephasing noise on the entanglement witness   for both low- and high-error probabilities. On the other hand, for a depolarising channel, the capability of the flag-qubit FT circuit to generate GME states can actually improve for low-error rates, even if the region of witnessed GME that is characterized by larger errors does also decrease for the FT construction.  From these results, it follows that one would be overestimating the power of the FT design if a depolarising channel is used in the simulation of a trapped-ion device. On the other hand, as also discussed above, one can modify the circuits to measure other stabilizers that are more robust to a dephasing noise model. In this respect, information about the microscopic noise model can actually be used to improve the design of the QEC circuits, both with respect to the low-error regime for QEC or with respect to the GME witnessing region at higher errors.

\begin{table*}
\begin{center}
\begin{tabular}{ |c|c| c|c |c| c|c|c|c| c|c|c|} 
\hline
Error &  $g_1$ $g_2$ $g_3$ $g_4$ $g_5$  & \# -1's & $M_{f}(X)$ & Error &  $g_1$ $g_2$ $g_3$ $g_4$ $g_5$  & \# -1's & $M_{f}(X)$ &   Error &  $g_1$ $g_2$ $g_3$ $g_4$ $g_5$  & \# -1's & $M_{f}(X)$ \\
\hline
$X_{1}I_{5}$ & -1 1 1 1 -1  & 2 & +1& $X_{5}I_{6}$ & -1 1 1 1 1 &1     & -1 &  $X_{2}I_{5}$ &  -1 -1 1 1 -1 & 3 & +1\\ 
$I_{1}X_{5}$ & -1 1 1 1 -1 & 2 & +1 & $I_{5}X_{6}$ & 1 1 1 1 -1 & 1 & -1 & $I_{2}X_{5}$ & 1 -1 1 1 -1 & 2 & -1\\ 
$X_{1}X_{5}$ & 1 1 1 1 1 &  0 & +1 & $X_{5}X_{6}$ & -1 1 1 1 -1 & 2   & +1 & $X_{2}X_{5}$ &  -1 1 1 1 1 & 1 & -1\\
$X_{1}Y_{5}$ & 1 1 1 1 -1  &1 & +1 & $X_{5}Y_{6}$ & -1 1 1 1 -1 & 2   & -1 &$X_{2}Y_{5}$ & -1 1 1 1 -1 & 2 & -1\\
$Y_{1}X_{5}$ & -1 1 1 1 1& 1 & +1 & $Y_{5}X_{6}$ &  -1 1 1 1 1  & 1 & +1 & $Y_{2}X_{5}$ &  1 -1 1 1 1 & 1& -1\\ 
$X_{1}Z_{5}$ &  -1 1 1 1 1 & 1 & +1& $X_{5}Z_{6}$ &   -1 1 1 1 1 & 1&+1& $X_{2}Z_{5}$ &  -1 -1 1 1 1 & 2& +1\\
$Z_{1}X_{5}$ & 1 1 1 1 -1 & 1& +1 & $Z_{5}X_{6}$ &   1 1 1 1 1  & 0& -1& $Z_{2}X_{5}$ &  -1 1 1 1 -1 & 2 &-1 \\
$Y_{1}I_{5}$ & 1 1 1 1 -1 &1 & +1& $Y_{5}I_{6}$ &   -1 1 1 1 -1& 2 & -1&$Y_{2}I_{5}$ & 1 1 1 1 -1 & 1 &+1\\
$I_{1}Y_{5}$ & -1 1 1 1 1 & 1& +1 &$I_{5}Y_{6}$ &  1 1 1 1 -1 & 1&+1& $I_{2}Y_{5}$ & 1 -1 1 1 1 & 1 &-1\\
$Y_{1}Y_{5}$ & -1 1 1 1 -1 & 2 & +1&$Y_{5}Y_{6}$ &   -1 1 1 1 1& 1&-1& $Y_{2}Y_{5}$ &  1 -1 1 1 -1 & 2 &-1\\
$Y_{1}Z_{5}$ & 1 1 1 1 1 & 0 & +1& $Y_{5}Z_{6}$ &   -1 1 1 1 -1&2 & +1& $Y_{2}Z_{5}$ & 1 1 1 1 1 & 0 &+1\\
$Z_{1}Y_{5}$ & 1 1 1 1 1 & 0 & +1& $Z_{5}Y_{6}$ &  1 1 1 1 1&  0 &+1 & $Z_{2}Y_{5}$ & -1 1 1 1 1 & 1 & -1\\
$Z_{1}I_{5}$ & -1 1 1 1 1 & 1 &+1 & $Z_{5}I_{6}$ &  1 1 1 1 -1& 1&+1 & $Z_{2}I_{5}$ &  -1 -1 1 1 1 & 2 &+1\\
$I_{1}Z_{5}$ & 1 1 1 1 -1 &1 & +1& $I_{5}Z_{6}$ &   1 1 1 1 1&0 &-1& $I_{2}Z_{5}$ &  1 1 1 1 -1 & 1 &+1\\
$Z_{1}Z_{5}$ & -1 1 1 1 -1 &2 &+1 & $Z_{5}Z_{6}$ &  1 1 1 1 -1&1 &-1& $Z_{2}Z_{5}$ & -1 -1 1 1 -1 & 3 &+1\\
\hline
 \multicolumn{2}{|c|}{Total} & &16/75; \%21  &\multicolumn{2}{|c|}{Total}  &  &8/75; \%11 &\multicolumn{2}{|c|}{Total} &  &12/75; \%16\\
\hline
$X_{3}I_{5}$ & 1 -1 -1 1 -1 & 3 & +1& $X_{5}I_{6}$ & 1 1 -1 1 -1   & 2 &+1 & $X_{4}I_{5}$ & 1 1 -1 -1 -1 & 3 & +1 \\ 
$I_{3}X_{5}$ & 1 1 -1 1 1 & 1 & -1& $I_{5}X_{6}$ & 1 1 1 1 1   & 0&+1 & $I_{4}X_{5}$ & 1 1 1 -1 1  & 1 & +1\\ 
$X_{3}X_{5}$ & 1 -1 1 1 -1 & 2   &-1 & $X_{5}X_{6}$ & 1 1 -1 1 -1  &2 & +1& $X_{4}X_{5}$ & 1 1 -1 1 -1  & 2 & +1\\
$X_{3}Y_{5}$ & 1 -1 1 1 1 & 1 &-1& $X_{5}Y_{6}$ & 1 1 -1 1 -1  &2 & -1& $X_{4}Y_{5}$ & 1 1 -1 1 1 & 1 & +1\\ 
$Y_{3}X_{5}$ & 1 1 -1 1 -1 & 2 & -1& $Y_{5}X_{6}$ & 1 1 -1 1 1& 1& +1&  $Y_{4}X_{5}$ & 1 1 1 -1 -1 & 2 & +1\\ 
$X_{3}Z_{5}$ & 1 -1 -1 1 1 & 2  & +1& $X_{5}Z_{6}$ & 1 1 -1 1 -1  & 2& -1&$X_{4}Z_{5}$ & 1 1 -1 -1 1 & 2 & +1\\
$Z_{3}X_{5}$ & 1 -1 1 1 1 & 1& -1& $Z_{5}X_{6}$ & 1 1 1 1 -1 & 1 & +1& $Z_{4}X_{5}$ & 1 1 -1 1 1  & 1 & +1 \\ 
$Y_{3}I_{5}$ & 1 1 1 1 -1 & 1 & +1& $Y_{5}I_{6}$ & 1 1 -1 1 1 & 1 &+1  &$Y_{4}I_{5}$ & 1 1 1 1 -1 & 1 & +1\\ 
$I_{3}Y_{5}$ & 1 1 -1 1 -1 & 2 & -1 & $I_{5}Y_{6}$ &1 1 1 1 1 & 0& -1& $I_{4}Y_{5}$ & 1 1 1 -1 -1 & 2 & +1\\
$Y_{3}Y_{5}$ & 1 1 -1 1 1 & 1 & -1& $Y_{5}Y_{6}$ &1 1 -1 1 1 & 1  & -1 & $Y_{4}Y_{5}$ & 1 1 1 -1 1 & 1 & +1\\ 
$Y_{3}Z_{5}$ & 1 1 1 1 1 & 0  & +1 &  $Y_{5}Z_{6}$ & 1 1 -1 1 1 &1 &-1 &$Y_{4}Z_{5}$ & 1 1 1 1 1& 0 & +1\\ 
$Z_{3}Y_{5}$ & 1 -1 1 1 -1 & 2& -1& $Z_{5}Y_{6}$ & 1 1 1 1 -1 & 1 &-1 &   $Z_{4}Y_{5}$ & 1 1 -1 1 -1 & 2 & +1\\
$Z_{3}I_{5}$ & 1 -1 -1 1 1& 2 & +1 & $Z_{5}I_{6}$ & 1 1 1 1  -1&1 & +1 &  $Z_{4}I_{5}$ &  1 1 -1 -1 1 & 2 & +1\\
$I_{3}Z_{5}$ & 1 1 1 1 -1  & 1 & +1 &$I_{5}Z_{6}$ & 1 1 1 1 1 &0& -1& $I_{4}Z_{5}$ & 1 1 1 1 -1 & 1 & +1\\
$Z_{3}Z_{5}$ & 1 -1 -1 1 -1 & 3 & +1 & $Z_{5}Z_{6}$ & 1 1 1 1 -1&1 & -1&  $Z_{4}Z_{5}$ & 1 1 -1 -1 -1 & 3 & +1\\
\hline
 \multicolumn{2}{|c|}{Total} & &12/75; \%16  &\multicolumn{2}{|c|}{Total}  &  &8/75; \%11 &\multicolumn{2}{|c|}{Total}  &  & 24/75; \%32\\
\hline
\end{tabular}
\end{center}
\caption{One and two-qubit depolarizing errors propagating through a flag-based FT $X$-type parity-check circuit. The subscripts on the Pauli operators refer to the qubits affected by the errors after the application of the entangling light-shift gate on those two qubits. $\pm 1$ numbers refers to the expectation of the stabilizer generators $g_{1}=Z_{1}Z_{2}$,  $g_{2}=Z_{2}Z_{3}$, $g_{3}=Z_{3}Z_{4}$, 
$g_{4}=Z_{4}Z_{5}$, $g_{5}=X_{1}X_{2}X_{3}X_{4}X_{5}$.}
\label{table:6q_depo_Xtype}
\end{table*}

\begin{table*}
\begin{center}
\begin{tabular}{ |c|c| c|c |c| c|c|c|c| c|c|c|} 
\hline
Error &  $g_1$ $g_2$ $g_3$ $g_4$ $g_5$  & \# -1's & $M_{f}(X)$ & Error &  $g_1$ $g_2$ $g_3$ $g_4$ $g_5$  & \# -1's & $M_{f}(X)$ &   Error &  $g_1$ $g_2$ $g_3$ $g_4$ $g_5$  & \# -1's & $M_{f}(X)$ \\
\hline
$Z_{1}I_{5}$ & -1 1 1 1 1 & 1 &+1 & $Z_{5}I_{6}$ &  1 1 1 1 -1& 1&+1 & $Z_{2}I_{5}$ &  -1 -1 1 1 1 & 2 &+1\\
$I_{1}Z_{5}$ & 1 1 1 1 -1 &1 & +1& $I_{5}Z_{6}$ &   1 1 1 1 1&0 &-1& $I_{2}Z_{5}$ &  1 1 1 1 -1 & 1 &+1\\
$Z_{1}Z_{5}$ & -1 1 1 1 -1 &2 &+1 & $Z_{5}Z_{6}$ &  1 1 1 1 -1&1 &-1& $Z_{2}Z_{5}$ & -1 -1 1 1 -1 & 3 &+1\\
\hline
 \multicolumn{2}{|c|}{Total} & &4/15; \%27  &\multicolumn{2}{|c|}{Total}  &  &1/15; \%6 &\multicolumn{2}{|c|}{Total} &  &6/15; \%40\\
\hline
$Z_{3}I_{5}$ & 1 -1 -1 1 1& 2 & +1 & $Z_{5}I_{6}$ & 1 1 1 1  -1&1 & +1 &  $Z_{4}I_{5}$ &  1 1 -1 -1 1 & 2 & +1\\
$I_{3}Z_{5}$ & 1 1 1 1 -1  & 1 & +1 &$I_{5}Z_{6}$ & 1 1 1 1 1 &0& -1& $I_{4}Z_{5}$ & 1 1 1 1 -1 & 1 & +1\\
$Z_{3}Z_{5}$ & 1 -1 -1 1 -1 & 3 & +1 & $Z_{5}Z_{6}$ & 1 1 1 1 -1&1 & -1 &  $Z_{4}Z_{5}$ & 1 1 -1 -1 -1 & 3 & +1\\
\hline
 \multicolumn{2}{|c|}{Total} & &6/15; \%40  &\multicolumn{2}{|c|}{Total}  &  &1/15; \%6 &\multicolumn{2}{|c|}{Total}  &  & 6/15; \%40\\
\hline
\end{tabular}
\end{center}
\caption{One and two-qubit dephasing errors propagating through a flag-based FT $X$-type parity-check circuit. The subscripts on the Pauli operators refer to the qubits affected by the errors after the application of the entangling light-shift gate on those two qubits. $\pm 1$ numbers refers to the expectation of the stabilizer generators $g_{1}=Z_{1}Z_{2}$,  $g_{2}=Z_{2}Z_{3}$, $g_{3}=Z_{3}Z_{4}$, 
$g_{4}=Z_{4}Z_{5}$, $g_{5}=X_{1}X_{2}X_{3}X_{4}X_{5}$. }
\label{table:6q_deph_Xtype}
\end{table*}

\subsubsection{Numerics for the conditional linear witnessing}\label{sec:results_CL_witness}
In this section, we  characterize the ability to generate GME of the $X$-type parity check circuits of Figs.~\ref{fig:parity_check_measurement} and~\ref{fig:parity_check_measurement_FT} using the conditional linear witnessing method.
Recall from Sec.~\ref{subsec:GMEwitness_flag} that this  method is also  efficient  in the verification of ME. In contrast to the standard linear witnessing method, which uses $n$ stabilizer generators of the $n$-partite systems to construct the  test operators, the conditional method follows a slightly different approach. It was shown in~\cite{PRXQuantum.2.020304} that, for the certification of  ME of an $n$-partite system, it is sufficient  to verify entanglement in just $n-1$ bi-partitions. In this case, we choose to certify entanglement between the subsystems $s_5$ and $s_x$ for all $s_x\in\{s_5\}\us{c}=\{s_1,\cdots,s_4\}$ with $s_5$ being the syndrome qubit. This is achieved by post-selecting on the state $\ketbra{+}$ for the remaining $n-2=3$ qubits  that are not involved in the specific bi-partition. Altogether, the $X$-basis witness test operator in Eq.~\eqref{eq:condWitness_5_numeric} can be expressed as a linear combination of two-point correlations~\eqref{eq:CL_test_operator}, requiring three measurement basis per bi-partition.

Fig.~\ref{fig:CL_witnessing} shows the numerical simulations  for the non-FT circuit in the presence of the effective two-qubit depolarizing  {\bf{(a)}} and dephasing  {\bf{(b)}} channels. We only display the values of the  CL  witness  for the bi-partition $[s_5|s_1]$, which actually  has the lowest performance under the noise channels in comparison to the other bipartitions. The red solid line is the boundary between the entanglement witnesses region and the  inconclusive outer region for the depolarizing channel. This line is then plotted on top of the results for the dephasing channel for comparison, as shown in Fig.~\ref{fig:CL_witnessing} {\bf{(b)}}. We note that the conditioned  ME witnessed region is greater when the effective noise channel is a dephasing channel than when it is a depolarizing channel. However, this may not be the case for other bi-partitions.

In fact, for the rest of the bipartitions, which we do not show here to avoid redundancy, the region of witnessed entanglement in the presence of dephasing noise barely changes its size, whilst the depolarizing area increases. For the bi-partition $[s_5|s_4]$, the witnessed entanglement area, in the presence of depolarizing noise, surpasses the one obtained under the dephasing channel. The majority of the depolarizing errors with a Pauli-operator of the type $X_5$ or $Y_5$, occurring on the syndrome, propagate to multiple errors through the light-shift gates, and consequently, it is more likely to find errors that have propagated  on $s_3$, $s_4$ and $s_5$. This implies that the Bell pair generators with support in these subsystem qubits commute more times with the propagated errors, yielding better results for the witnessed entanglement regions.

\begin{figure}[h]
    \includegraphics[width=0.9\columnwidth]{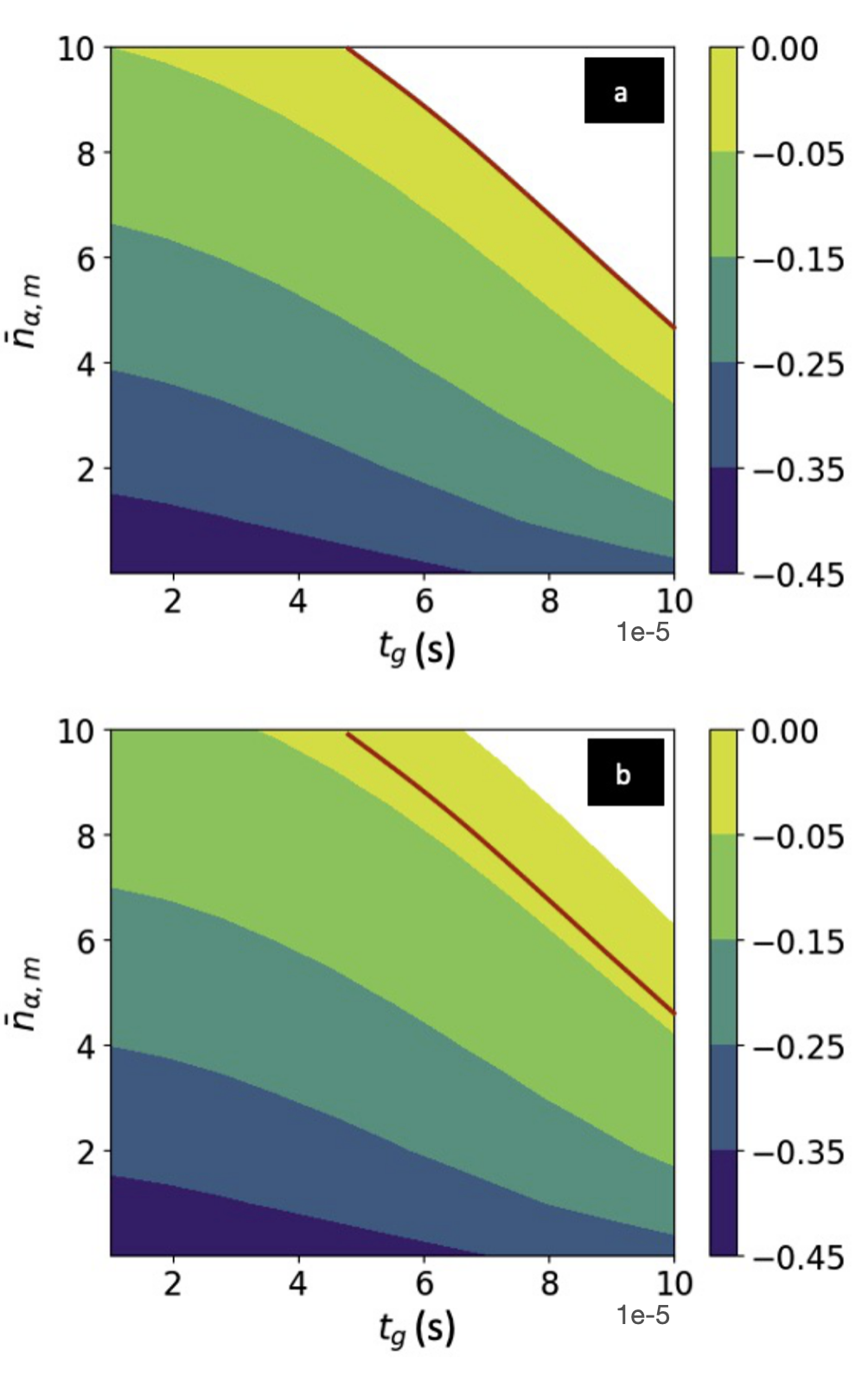}
  \caption{CL witnessing for a  $X$-type parity-check readout circuit. The color bar indicates the value of the witness under effective microscopic noise channels. For all the colors represented, the witness takes a negative value signaling the presence of entanglement. The white area represents all positive values of the witness values where the verification is inconclusive. The red solid line fits the perimeter area of the (a) non-FT $X$-type circuit under a two-qubit depolarizing noise channel. Then, the solid line is plotted on top of the following graphs to benchmark the size of the enclosed colored area on (b), the non-FT $X$-type circuits under dephasing noise, respectively. This witness is reconstructed using the products of the Bell-pair stabilizer generators for the $[s_5|s_1]$ bi-partition output state $(\ket{00}+\ket{11})/\sqrt{2}$ which are $g_{1}=X_{1}X_{5}, g_{2}=Z_{1}Z_{5}$. The resulting Bell pair is obtained after post-selecting on $\ketbra{+}\oprod{3}$ for $s_2, s_3$ and $s_4$ subsystem qubits.}
  \label{fig:CL_witnessing}
\end{figure}

The main conclusion that can be drawn from the results from Fig.~\ref{fig:XXXX} and Fig.\ref{fig:CL_witnessing} is that, for both types of noise channels, the conditional linear witness encloses a larger area of  witnessed entanglement than the one obtained via the standard linear witness. For brevity, in Fig.~\ref{fig:CL_witnessing} we do not show the results for the CL method using the flag-based FT circuit, as the behavior is the same as we extensively discussed in Sec.~\ref{sec:results_SL_witness}. For the high probability error rates needed to show the performance of the witnesses, the flag-based circuits would always add more noise and perform worse than the bare non-FT circuits.

\section{\bf Conclusions and Outlook}
In this manuscript, we have presented a detailed microscopic error model for the trapped-ion light-shift gates in terms of the average gate error. We have discussed a microscopic derivation  of this error in terms of entanglement fidelity,  which can be condensed  into an analytical formula that contains various  possible error sources  and experimental parameters.  We then fed this gate error  into effective noise models for the two-qubit light shift gates. In this way, the dynamical quantum map that describes  the evolution of the imperfect entangling gate has error rates  that connect to  realistic
microscopic calculations, instead of giving them arbitrary values from 0 to 1, as we had previously done in~\cite{PRXQuantum.2.020304}.

We have   used this noise model to assess the performance of the parity-check QEC circuits in terms of their ability to generate $n$-qubit entangled output states. 
Multipartite entanglement, then, is detected by  two entanglement witnessing operators, which require a number of bipartitions in which  entanglement must be checked that grows linearly with the qubit register size. These methods correspond to  
the standard linear (SL) method used experimentally in this context in~\cite{PhysRevX.12.011032}, and the conditional linear (CL) method that we introduced in~\cite{PRXQuantum.2.020304}.
We verify that the CL method is again more robust than the standard linear method for the current, more realistic, noise model. In spite of this improvement, it should be noted that the CL witness can lead to an increase in statistical errors due to the conditional evaluation or post-selection of the remaining subsystems. Besides, the CL needs twelve measurement settings, whilst the SL only needs two, so that, in order to choose one method with respect to the other, one must first evaluate the trade-off between robustness and experimental feasibility. 

Regarding the flag-based FT results, we have shown that for high failure rates on the two-qubit entangling gates, the flag-qubit does not have a positive effect on the GME witnessed certification. We have argued that  the two additional entangling gates between syndrome and flag qubit designed to catch weight-2 dangerous errors, introduce additional errors which, at high failure rates, are not eliminated after post-selecting on the flag, and lead to more sign flips of the stabilizers that reduce the expectation value of the witness operator. 

An interesting direction for further research could be to work on the design of optimal SL witnesses by a thorough analysis of the particular output state generated by the QEC primitives, and the propagation of the effective noise channel through it. We have shown in the text that, by changing the stabilizer generators for the SL method, the region of witnessed GME can be greater for the biased dephasing noise. Therefore, if one knows  the noise structure, and how the propagation through the circuit affects the resulting witness test operators, it is possible to optimize the design of entanglement witnesses to choose the most robust option. We believe that our work can also help experimentalists working with these gates to realize which experimental parameters can be modified to minimize the two-qubit gate infidelity values in their setups.

\acknowledgments 
We acknowledge support from   PID2021-127726NB-I00  (MCIU/AEI/FEDER, UE), from the Grant IFT Centro de Excelencia Severo Ochoa CEX2020-001007-S, funded by MCIN/AEI/10.13039/501100011033, from the grant QUITEMAD+ S2013/ICE-2801, and from the CSIC Research Platform on Quantum Technologies PTI-001.
A.R.B acknowledges support from the Universidad Complutense de Madrid-Banco Santander Predoctoral Fellowship. 
A.R.B also thanks J.G.F.U for providing access to computing capabilities to carry out numerical simulations.
F.S. acknowledges the Research Fellowship by the Royal Commission for the Exhibition of 1851. A. B. acknowledges support from the EU Quantum Technology Flagship grant AQTION under grant agreement number 820495, and by the US Army Research Office through grant number W911NF-21-1-0007. The project leading to this application/publication has received funding from the European Union’s Horizon Europe research and innovation programme under grant agreement No 101114305 (“MILLENION-SGA1” EU Project).

\appendix

\bibliography{EntInQEC_2}

\begin{thebibliography}{116}%
\makeatletter
\providecommand \@ifxundefined [1]{%
 \@ifx{#1\undefined}
}%
\providecommand \@ifnum [1]{%
 \ifnum #1\expandafter \@firstoftwo
 \else \expandafter \@secondoftwo
 \fi
}%
\providecommand \@ifx [1]{%
 \ifx #1\expandafter \@firstoftwo
 \else \expandafter \@secondoftwo
 \fi
}%
\providecommand \natexlab [1]{#1}%
\providecommand \enquote  [1]{``#1''}%
\providecommand \bibnamefont  [1]{#1}%
\providecommand \bibfnamefont [1]{#1}%
\providecommand \citenamefont [1]{#1}%
\providecommand \href@noop [0]{\@secondoftwo}%
\providecommand \href [0]{\begingroup \@sanitize@url \@href}%
\providecommand \@href[1]{\@@startlink{#1}\@@href}%
\providecommand \@@href[1]{\endgroup#1\@@endlink}%
\providecommand \@sanitize@url [0]{\catcode `\\12\catcode `\$12\catcode `\&12\catcode `\#12\catcode `\^12\catcode `\_12\catcode `\%12\relax}%
\providecommand \@@startlink[1]{}%
\providecommand \@@endlink[0]{}%
\providecommand \url  [0]{\begingroup\@sanitize@url \@url }%
\providecommand \@url [1]{\endgroup\@href {#1}{\urlprefix }}%
\providecommand \urlprefix  [0]{URL }%
\providecommand \Eprint [0]{\href }%
\providecommand \doibase [0]{http://dx.doi.org/}%
\providecommand \selectlanguage [0]{\@gobble}%
\providecommand \bibinfo  [0]{\@secondoftwo}%
\providecommand \bibfield  [0]{\@secondoftwo}%
\providecommand \translation [1]{[#1]}%
\providecommand \BibitemOpen [0]{}%
\providecommand \bibitemStop [0]{}%
\providecommand \bibitemNoStop [0]{.\EOS\space}%
\providecommand \EOS [0]{\spacefactor3000\relax}%
\providecommand \BibitemShut  [1]{\csname bibitem#1\endcsname}%
\let\auto@bib@innerbib\@empty
\bibitem [{\citenamefont {Rodriguez-Blanco}\ \emph {et~al.}(2021)\citenamefont {Rodriguez-Blanco}, \citenamefont {Bermudez}, \citenamefont {M\"uller},\ and\ \citenamefont {Shahandeh}}]{PRXQuantum.2.020304}%
  \BibitemOpen
  \bibfield  {author} {\bibinfo {author} {\bibfnamefont {A.}~\bibnamefont {Rodriguez-Blanco}}, \bibinfo {author} {\bibfnamefont {A.}~\bibnamefont {Bermudez}}, \bibinfo {author} {\bibfnamefont {M.}~\bibnamefont {M\"uller}}, \ and\ \bibinfo {author} {\bibfnamefont {F.}~\bibnamefont {Shahandeh}},\ }\bibfield  {title} {\emph {\bibinfo {title} {Efficient and robust certification of genuine multipartite entanglement in noisy quantum error correction circuits},\ }}\href {\doibase 10.1103/PRXQuantum.2.020304} {\bibfield  {journal} {\bibinfo  {journal} {PRX Quantum}\ }\textbf {\bibinfo {volume} {2}},\ \bibinfo {pages} {020304} (\bibinfo {year} {2021})}\BibitemShut {NoStop}%
\bibitem [{\citenamefont {Hilder}\ \emph {et~al.}(2022)\citenamefont {Hilder}, \citenamefont {Pijn}, \citenamefont {Onishchenko}, \citenamefont {Stahl}, \citenamefont {Orth}, \citenamefont {Lekitsch}, \citenamefont {Rodriguez-Blanco}, \citenamefont {M\"uller}, \citenamefont {Schmidt-Kaler},\ and\ \citenamefont {Poschinger}}]{PhysRevX.12.011032}%
  \BibitemOpen
  \bibfield  {author} {\bibinfo {author} {\bibfnamefont {J.}~\bibnamefont {Hilder}}, \bibinfo {author} {\bibfnamefont {D.}~\bibnamefont {Pijn}}, \bibinfo {author} {\bibfnamefont {O.}~\bibnamefont {Onishchenko}}, \bibinfo {author} {\bibfnamefont {A.}~\bibnamefont {Stahl}}, \bibinfo {author} {\bibfnamefont {M.}~\bibnamefont {Orth}}, \bibinfo {author} {\bibfnamefont {B.}~\bibnamefont {Lekitsch}}, \bibinfo {author} {\bibfnamefont {A.}~\bibnamefont {Rodriguez-Blanco}}, \bibinfo {author} {\bibfnamefont {M.}~\bibnamefont {M\"uller}}, \bibinfo {author} {\bibfnamefont {F.}~\bibnamefont {Schmidt-Kaler}}, \ and\ \bibinfo {author} {\bibfnamefont {U.~G.}\ \bibnamefont {Poschinger}},\ }\bibfield  {title} {\emph {\bibinfo {title} {Fault-tolerant parity readout on a shuttling-based trapped-ion quantum computer},\ }}\href {\doibase 10.1103/PhysRevX.12.011032} {\bibfield  {journal} {\bibinfo  {journal} {Phys. Rev. X}\ }\textbf {\bibinfo {volume} {12}},\ \bibinfo {pages} {011032} (\bibinfo {year} {2022})}\BibitemShut {NoStop}%
\bibitem [{\citenamefont {Nielsen}\ and\ \citenamefont {Chuang}(2000)}]{nielsen00}%
  \BibitemOpen
  \bibfield  {author} {\bibinfo {author} {\bibfnamefont {M.~A.}\ \bibnamefont {Nielsen}}\ and\ \bibinfo {author} {\bibfnamefont {I.~L.}\ \bibnamefont {Chuang}},\ }\href@noop {} {\emph {\bibinfo {title} {Quantum Computation and Quantum Information}}}\ (\bibinfo  {publisher} {Cambridge University Press},\ \bibinfo {year} {2000})\BibitemShut {NoStop}%
\bibitem [{\citenamefont {Montanaro}(2016)}]{Montanaro2016}%
  \BibitemOpen
  \bibfield  {author} {\bibinfo {author} {\bibfnamefont {A.}~\bibnamefont {Montanaro}},\ }\bibfield  {title} {\emph {\bibinfo {title} {Quantum algorithms: an overview},\ }}\href {\doibase 10.1038/npjqi.2015.23} {\bibfield  {journal} {\bibinfo  {journal} {npj Quantum Information}\ }\textbf {\bibinfo {volume} {2}},\ \bibinfo {pages} {15023} (\bibinfo {year} {2016})}\BibitemShut {NoStop}%
\bibitem [{\citenamefont {Ladd}\ \emph {et~al.}(2010)\citenamefont {Ladd}, \citenamefont {Jelezko}, \citenamefont {Laflamme}, \citenamefont {Nakamura}, \citenamefont {Monroe},\ and\ \citenamefont {O'Brien}}]{Ladd2010}%
  \BibitemOpen
  \bibfield  {author} {\bibinfo {author} {\bibfnamefont {T.~D.}\ \bibnamefont {Ladd}}, \bibinfo {author} {\bibfnamefont {F.}~\bibnamefont {Jelezko}}, \bibinfo {author} {\bibfnamefont {R.}~\bibnamefont {Laflamme}}, \bibinfo {author} {\bibfnamefont {Y.}~\bibnamefont {Nakamura}}, \bibinfo {author} {\bibfnamefont {C.}~\bibnamefont {Monroe}}, \ and\ \bibinfo {author} {\bibfnamefont {J.~L.}\ \bibnamefont {O'Brien}},\ }\bibfield  {title} {\emph {\bibinfo {title} {Quantum computers},\ }}\href {\doibase 10.1038/nature08812} {\bibfield  {journal} {\bibinfo  {journal} {Nature}\ }\textbf {\bibinfo {volume} {464}},\ \bibinfo {pages} {45} (\bibinfo {year} {2010})}\BibitemShut {NoStop}%
\bibitem [{\citenamefont {Preskill}(2018)}]{Preskill2018quantumcomputingin}%
  \BibitemOpen
  \bibfield  {author} {\bibinfo {author} {\bibfnamefont {J.}~\bibnamefont {Preskill}},\ }\bibfield  {title} {\emph {\bibinfo {title} {Quantum {C}omputing in the {NISQ} era and beyond},\ }}\href {\doibase 10.22331/q-2018-08-06-79} {\bibfield  {journal} {\bibinfo  {journal} {{Quantum}}\ }\textbf {\bibinfo {volume} {2}},\ \bibinfo {pages} {79} (\bibinfo {year} {2018})}\BibitemShut {NoStop}%
\bibitem [{\citenamefont {Arute}\ \emph {et~al.}(2019)\citenamefont {Arute} \emph {et~al.}}]{Arute2019}%
  \BibitemOpen
  \bibfield  {author} {\bibinfo {author} {\bibfnamefont {F.}~\bibnamefont {Arute}} \emph {et~al.},\ }\bibfield  {title} {\emph {\bibinfo {title} {Quantum supremacy using a programmable superconducting processor},\ }}\href {\doibase 10.1038/s41586-019-1666-5} {\bibfield  {journal} {\bibinfo  {journal} {Nature}\ }\textbf {\bibinfo {volume} {574}},\ \bibinfo {pages} {505} (\bibinfo {year} {2019})}\BibitemShut {NoStop}%
\bibitem [{\citenamefont {Zhong}\ \emph {et~al.}(2020)\citenamefont {Zhong}, \citenamefont {Wang}, \citenamefont {Deng}, \citenamefont {Chen}, \citenamefont {Peng}, \citenamefont {Luo}, \citenamefont {Qin}, \citenamefont {Wu}, \citenamefont {Ding}, \citenamefont {Hu}, \citenamefont {Hu}, \citenamefont {Yang}, \citenamefont {Zhang}, \citenamefont {Li}, \citenamefont {Li}, \citenamefont {Jiang}, \citenamefont {Gan}, \citenamefont {Yang}, \citenamefont {You}, \citenamefont {Wang}, \citenamefont {Li}, \citenamefont {Liu}, \citenamefont {Lu},\ and\ \citenamefont {Pan}}]{doi:10.1126/science.abe8770}%
  \BibitemOpen
  \bibfield  {author} {\bibinfo {author} {\bibfnamefont {H.-S.}\ \bibnamefont {Zhong}}, \bibinfo {author} {\bibfnamefont {H.}~\bibnamefont {Wang}}, \bibinfo {author} {\bibfnamefont {Y.-H.}\ \bibnamefont {Deng}}, \bibinfo {author} {\bibfnamefont {M.-C.}\ \bibnamefont {Chen}}, \bibinfo {author} {\bibfnamefont {L.-C.}\ \bibnamefont {Peng}}, \bibinfo {author} {\bibfnamefont {Y.-H.}\ \bibnamefont {Luo}}, \bibinfo {author} {\bibfnamefont {J.}~\bibnamefont {Qin}}, \bibinfo {author} {\bibfnamefont {D.}~\bibnamefont {Wu}}, \bibinfo {author} {\bibfnamefont {X.}~\bibnamefont {Ding}}, \bibinfo {author} {\bibfnamefont {Y.}~\bibnamefont {Hu}}, \bibinfo {author} {\bibfnamefont {P.}~\bibnamefont {Hu}}, \bibinfo {author} {\bibfnamefont {X.-Y.}\ \bibnamefont {Yang}}, \bibinfo {author} {\bibfnamefont {W.-J.}\ \bibnamefont {Zhang}}, \bibinfo {author} {\bibfnamefont {H.}~\bibnamefont {Li}}, \bibinfo {author} {\bibfnamefont {Y.}~\bibnamefont {Li}}, \bibinfo {author} {\bibfnamefont {X.}~\bibnamefont {Jiang}}, \bibinfo {author}
  {\bibfnamefont {L.}~\bibnamefont {Gan}}, \bibinfo {author} {\bibfnamefont {G.}~\bibnamefont {Yang}}, \bibinfo {author} {\bibfnamefont {L.}~\bibnamefont {You}}, \bibinfo {author} {\bibfnamefont {Z.}~\bibnamefont {Wang}}, \bibinfo {author} {\bibfnamefont {L.}~\bibnamefont {Li}}, \bibinfo {author} {\bibfnamefont {N.-L.}\ \bibnamefont {Liu}}, \bibinfo {author} {\bibfnamefont {C.-Y.}\ \bibnamefont {Lu}}, \ and\ \bibinfo {author} {\bibfnamefont {J.-W.}\ \bibnamefont {Pan}},\ }\bibfield  {title} {\emph {\bibinfo {title} {Quantum computational advantage using photons},\ }}\href {\doibase 10.1126/science.abe8770} {\bibfield  {journal} {\bibinfo  {journal} {Science}\ }\textbf {\bibinfo {volume} {370}},\ \bibinfo {pages} {1460} (\bibinfo {year} {2020})},\ \Eprint {http://arxiv.org/abs/https://www.science.org/doi/pdf/10.1126/science.abe8770} {https://www.science.org/doi/pdf/10.1126/science.abe8770} \BibitemShut {NoStop}%
\bibitem [{\citenamefont {Calderbank}\ and\ \citenamefont {Shor}(1996)}]{PhysRevA.54.1098}%
  \BibitemOpen
  \bibfield  {author} {\bibinfo {author} {\bibfnamefont {A.~R.}\ \bibnamefont {Calderbank}}\ and\ \bibinfo {author} {\bibfnamefont {P.~W.}\ \bibnamefont {Shor}},\ }\bibfield  {title} {\emph {\bibinfo {title} {Good quantum error-correcting codes exist},\ }}\href {\doibase 10.1103/PhysRevA.54.1098} {\bibfield  {journal} {\bibinfo  {journal} {Phys. Rev. A}\ }\textbf {\bibinfo {volume} {54}},\ \bibinfo {pages} {1098} (\bibinfo {year} {1996})}\BibitemShut {NoStop}%
\bibitem [{\citenamefont {Steane}(1996{\natexlab{a}})}]{PhysRevLett.77.793}%
  \BibitemOpen
  \bibfield  {author} {\bibinfo {author} {\bibfnamefont {A.~M.}\ \bibnamefont {Steane}},\ }\bibfield  {title} {\emph {\bibinfo {title} {Error correcting codes in quantum theory},\ }}\href {\doibase 10.1103/PhysRevLett.77.793} {\bibfield  {journal} {\bibinfo  {journal} {Phys. Rev. Lett.}\ }\textbf {\bibinfo {volume} {77}},\ \bibinfo {pages} {793} (\bibinfo {year} {1996}{\natexlab{a}})}\BibitemShut {NoStop}%
\bibitem [{\citenamefont {Terhal}(2015)}]{RevModPhys.87.307}%
  \BibitemOpen
  \bibfield  {author} {\bibinfo {author} {\bibfnamefont {B.~M.}\ \bibnamefont {Terhal}},\ }\bibfield  {title} {\emph {\bibinfo {title} {Quantum error correction for quantum memories},\ }}\href {\doibase 10.1103/RevModPhys.87.307} {\bibfield  {journal} {\bibinfo  {journal} {Rev. Mod. Phys.}\ }\textbf {\bibinfo {volume} {87}},\ \bibinfo {pages} {307} (\bibinfo {year} {2015})}\BibitemShut {NoStop}%
\bibitem [{\citenamefont {Aharonov}\ and\ \citenamefont {Ben-Or}(1998)}]{FTQEC}%
  \BibitemOpen
  \bibfield  {author} {\bibinfo {author} {\bibfnamefont {D.}~\bibnamefont {Aharonov}}\ and\ \bibinfo {author} {\bibfnamefont {M.}~\bibnamefont {Ben-Or}},\ }\bibfield  {title} {\emph {\bibinfo {title} {Fault-tolerant quantum computation with constant error},\ }}\href {\doibase 10.1137/S0097539799359385} {\bibfield  {journal} {\bibinfo  {journal} {SIAM J. Comput.}\ }\textbf {\bibinfo {volume} {38}},\ \bibinfo {pages} {1207} (\bibinfo {year} {1998})}\BibitemShut {NoStop}%
\bibitem [{\citenamefont {Svore}\ \emph {et~al.}(2005)\citenamefont {Svore}, \citenamefont {Cross}, \citenamefont {Chuang},\ and\ \citenamefont {Aho}}]{https://doi.org/10.48550/arxiv.quant-ph/0508176}%
  \BibitemOpen
  \bibfield  {author} {\bibinfo {author} {\bibfnamefont {K.~M.}\ \bibnamefont {Svore}}, \bibinfo {author} {\bibfnamefont {A.~W.}\ \bibnamefont {Cross}}, \bibinfo {author} {\bibfnamefont {I.~L.}\ \bibnamefont {Chuang}}, \ and\ \bibinfo {author} {\bibfnamefont {A.~V.}\ \bibnamefont {Aho}},\ }\bibfield  {title} {\emph {\bibinfo {title} {A flow-map model for analyzing pseudothresholds in fault-tolerant quantum computing},\ }}\href {\doibase 10.48550/ARXIV.QUANT-PH/0508176} {\  (\bibinfo {year} {2005}),\ 10.48550/ARXIV.QUANT-PH/0508176}\BibitemShut {NoStop}%
\bibitem [{\citenamefont {Cross}\ \emph {et~al.}(2007)\citenamefont {Cross}, \citenamefont {DiVincenzo},\ and\ \citenamefont {Terhal}}]{https://doi.org/10.48550/arxiv.0711.1556}%
  \BibitemOpen
  \bibfield  {author} {\bibinfo {author} {\bibfnamefont {A.~W.}\ \bibnamefont {Cross}}, \bibinfo {author} {\bibfnamefont {D.~P.}\ \bibnamefont {DiVincenzo}}, \ and\ \bibinfo {author} {\bibfnamefont {B.~M.}\ \bibnamefont {Terhal}},\ }\bibfield  {title} {\emph {\bibinfo {title} {A comparative code study for quantum fault-tolerance},\ }}\href {\doibase 10.48550/ARXIV.0711.1556} {\  (\bibinfo {year} {2007}),\ 10.48550/ARXIV.0711.1556}\BibitemShut {NoStop}%
\bibitem [{\citenamefont {Bruzewicz}\ \emph {et~al.}(2019)\citenamefont {Bruzewicz}, \citenamefont {Chiaverini}, \citenamefont {McConnell},\ and\ \citenamefont {Sage}}]{doi:10.1063/1.5088164}%
  \BibitemOpen
  \bibfield  {author} {\bibinfo {author} {\bibfnamefont {C.~D.}\ \bibnamefont {Bruzewicz}}, \bibinfo {author} {\bibfnamefont {J.}~\bibnamefont {Chiaverini}}, \bibinfo {author} {\bibfnamefont {R.}~\bibnamefont {McConnell}}, \ and\ \bibinfo {author} {\bibfnamefont {J.~M.}\ \bibnamefont {Sage}},\ }\bibfield  {title} {\emph {\bibinfo {title} {Trapped-ion quantum computing: Progress and challenges},\ }}\href {\doibase 10.1063/1.5088164} {\bibfield  {journal} {\bibinfo  {journal} {Applied Physics Reviews}\ }\textbf {\bibinfo {volume} {6}},\ \bibinfo {pages} {021314} (\bibinfo {year} {2019})}\BibitemShut {NoStop}%
\bibitem [{\citenamefont {Harty}\ \emph {et~al.}(2014)\citenamefont {Harty}, \citenamefont {Allcock}, \citenamefont {Ballance}, \citenamefont {Guidoni}, \citenamefont {Janacek}, \citenamefont {Linke}, \citenamefont {Stacey},\ and\ \citenamefont {Lucas}}]{PhysRevLett.113.220501}%
  \BibitemOpen
  \bibfield  {author} {\bibinfo {author} {\bibfnamefont {T.~P.}\ \bibnamefont {Harty}}, \bibinfo {author} {\bibfnamefont {D.~T.~C.}\ \bibnamefont {Allcock}}, \bibinfo {author} {\bibfnamefont {C.~J.}\ \bibnamefont {Ballance}}, \bibinfo {author} {\bibfnamefont {L.}~\bibnamefont {Guidoni}}, \bibinfo {author} {\bibfnamefont {H.~A.}\ \bibnamefont {Janacek}}, \bibinfo {author} {\bibfnamefont {N.~M.}\ \bibnamefont {Linke}}, \bibinfo {author} {\bibfnamefont {D.~N.}\ \bibnamefont {Stacey}}, \ and\ \bibinfo {author} {\bibfnamefont {D.~M.}\ \bibnamefont {Lucas}},\ }\bibfield  {title} {\emph {\bibinfo {title} {High-fidelity preparation, gates, memory, and readout of a trapped-ion quantum bit},\ }}\href {\doibase 10.1103/PhysRevLett.113.220501} {\bibfield  {journal} {\bibinfo  {journal} {Phys. Rev. Lett.}\ }\textbf {\bibinfo {volume} {113}},\ \bibinfo {pages} {220501} (\bibinfo {year} {2014})}\BibitemShut {NoStop}%
\bibitem [{\citenamefont {Ruster}\ \emph {et~al.}(2016)\citenamefont {Ruster}, \citenamefont {Schmiegelow}, \citenamefont {Kaufmann}, \citenamefont {Warschburger}, \citenamefont {Schmidt-Kaler},\ and\ \citenamefont {Poschinger}}]{Ruster2016}%
  \BibitemOpen
  \bibfield  {author} {\bibinfo {author} {\bibfnamefont {T.}~\bibnamefont {Ruster}}, \bibinfo {author} {\bibfnamefont {C.~T.}\ \bibnamefont {Schmiegelow}}, \bibinfo {author} {\bibfnamefont {H.}~\bibnamefont {Kaufmann}}, \bibinfo {author} {\bibfnamefont {C.}~\bibnamefont {Warschburger}}, \bibinfo {author} {\bibfnamefont {F.}~\bibnamefont {Schmidt-Kaler}}, \ and\ \bibinfo {author} {\bibfnamefont {U.~G.}\ \bibnamefont {Poschinger}},\ }\bibfield  {title} {\emph {\bibinfo {title} {A long-lived zeeman trapped-ion qubit},\ }}\href {\doibase 10.1007/s00340-016-6527-4} {\bibfield  {journal} {\bibinfo  {journal} {Applied Physics B}\ }\textbf {\bibinfo {volume} {122}},\ \bibinfo {pages} {254} (\bibinfo {year} {2016})}\BibitemShut {NoStop}%
\bibitem [{\citenamefont {Keselman}\ \emph {et~al.}(2011)\citenamefont {Keselman}, \citenamefont {Glickman}, \citenamefont {Akerman}, \citenamefont {Kotler},\ and\ \citenamefont {Ozeri}}]{Keselman_2011}%
  \BibitemOpen
  \bibfield  {author} {\bibinfo {author} {\bibfnamefont {A.}~\bibnamefont {Keselman}}, \bibinfo {author} {\bibfnamefont {Y.}~\bibnamefont {Glickman}}, \bibinfo {author} {\bibfnamefont {N.}~\bibnamefont {Akerman}}, \bibinfo {author} {\bibfnamefont {S.}~\bibnamefont {Kotler}}, \ and\ \bibinfo {author} {\bibfnamefont {R.}~\bibnamefont {Ozeri}},\ }\bibfield  {title} {\emph {\bibinfo {title} {High-fidelity state detection and tomography of a single-ion zeeman qubit},\ }}\href {\doibase 10.1088/1367-2630/13/7/073027} {\bibfield  {journal} {\bibinfo  {journal} {New Journal of Physics}\ }\textbf {\bibinfo {volume} {13}},\ \bibinfo {pages} {073027} (\bibinfo {year} {2011})}\BibitemShut {NoStop}%
\bibitem [{\citenamefont {Emerson}\ \emph {et~al.}(2005)\citenamefont {Emerson}, \citenamefont {Alicki},\ and\ \citenamefont {{\.{Z}}yczkowski}}]{Emerson_2005}%
  \BibitemOpen
  \bibfield  {author} {\bibinfo {author} {\bibfnamefont {J.}~\bibnamefont {Emerson}}, \bibinfo {author} {\bibfnamefont {R.}~\bibnamefont {Alicki}}, \ and\ \bibinfo {author} {\bibfnamefont {K.}~\bibnamefont {{\.{Z}}yczkowski}},\ }\bibfield  {title} {\emph {\bibinfo {title} {Scalable noise estimation with random unitary operators},\ }}\href {\doibase 10.1088/1464-4266/7/10/021} {\bibfield  {journal} {\bibinfo  {journal} {Journal of Optics B: Quantum and Semiclassical Optics}\ }\textbf {\bibinfo {volume} {7}},\ \bibinfo {pages} {S347} (\bibinfo {year} {2005})}\BibitemShut {NoStop}%
\bibitem [{\citenamefont {Monz}\ \emph {et~al.}(2016)\citenamefont {Monz}, \citenamefont {Nigg}, \citenamefont {Martinez}, \citenamefont {Brandl}, \citenamefont {Schindler}, \citenamefont {Rines}, \citenamefont {Wang}, \citenamefont {Chuang},\ and\ \citenamefont {Blatt}}]{doi:10.1126/science.aad9480}%
  \BibitemOpen
  \bibfield  {author} {\bibinfo {author} {\bibfnamefont {T.}~\bibnamefont {Monz}}, \bibinfo {author} {\bibfnamefont {D.}~\bibnamefont {Nigg}}, \bibinfo {author} {\bibfnamefont {E.~A.}\ \bibnamefont {Martinez}}, \bibinfo {author} {\bibfnamefont {M.~F.}\ \bibnamefont {Brandl}}, \bibinfo {author} {\bibfnamefont {P.}~\bibnamefont {Schindler}}, \bibinfo {author} {\bibfnamefont {R.}~\bibnamefont {Rines}}, \bibinfo {author} {\bibfnamefont {S.~X.}\ \bibnamefont {Wang}}, \bibinfo {author} {\bibfnamefont {I.~L.}\ \bibnamefont {Chuang}}, \ and\ \bibinfo {author} {\bibfnamefont {R.}~\bibnamefont {Blatt}},\ }\bibfield  {title} {\emph {\bibinfo {title} {Realization of a scalable shor algorithm},\ }}\href {\doibase 10.1126/science.aad9480} {\bibfield  {journal} {\bibinfo  {journal} {Science}\ }\textbf {\bibinfo {volume} {351}},\ \bibinfo {pages} {1068} (\bibinfo {year} {2016})},\ \Eprint {http://arxiv.org/abs/https://www.science.org/doi/pdf/10.1126/science.aad9480} {https://www.science.org/doi/pdf/10.1126/science.aad9480}
  \BibitemShut {NoStop}%
\bibitem [{\citenamefont {Hempel}\ \emph {et~al.}(2018)\citenamefont {Hempel}, \citenamefont {Maier}, \citenamefont {Romero}, \citenamefont {McClean}, \citenamefont {Monz}, \citenamefont {Shen}, \citenamefont {Jurcevic}, \citenamefont {Lanyon}, \citenamefont {Love}, \citenamefont {Babbush}, \citenamefont {Aspuru-Guzik}, \citenamefont {Blatt},\ and\ \citenamefont {Roos}}]{PhysRevX.8.031022}%
  \BibitemOpen
  \bibfield  {author} {\bibinfo {author} {\bibfnamefont {C.}~\bibnamefont {Hempel}}, \bibinfo {author} {\bibfnamefont {C.}~\bibnamefont {Maier}}, \bibinfo {author} {\bibfnamefont {J.}~\bibnamefont {Romero}}, \bibinfo {author} {\bibfnamefont {J.}~\bibnamefont {McClean}}, \bibinfo {author} {\bibfnamefont {T.}~\bibnamefont {Monz}}, \bibinfo {author} {\bibfnamefont {H.}~\bibnamefont {Shen}}, \bibinfo {author} {\bibfnamefont {P.}~\bibnamefont {Jurcevic}}, \bibinfo {author} {\bibfnamefont {B.~P.}\ \bibnamefont {Lanyon}}, \bibinfo {author} {\bibfnamefont {P.}~\bibnamefont {Love}}, \bibinfo {author} {\bibfnamefont {R.}~\bibnamefont {Babbush}}, \bibinfo {author} {\bibfnamefont {A.}~\bibnamefont {Aspuru-Guzik}}, \bibinfo {author} {\bibfnamefont {R.}~\bibnamefont {Blatt}}, \ and\ \bibinfo {author} {\bibfnamefont {C.~F.}\ \bibnamefont {Roos}},\ }\bibfield  {title} {\emph {\bibinfo {title} {Quantum chemistry calculations on a trapped-ion quantum simulator},\ }}\href {\doibase 10.1103/PhysRevX.8.031022} {\bibfield
  {journal} {\bibinfo  {journal} {Phys. Rev. X}\ }\textbf {\bibinfo {volume} {8}},\ \bibinfo {pages} {031022} (\bibinfo {year} {2018})}\BibitemShut {NoStop}%
\bibitem [{\citenamefont {Sanz-Fernandez}\ \emph {et~al.}(2021)\citenamefont {Sanz-Fernandez}, \citenamefont {Hernandez}, \citenamefont {Marciniak}, \citenamefont {Pogorelov}, \citenamefont {Monz}, \citenamefont {Benfenati}, \citenamefont {Mugel},\ and\ \citenamefont {Orus}}]{https://doi.org/10.48550/arxiv.2111.14970}%
  \BibitemOpen
  \bibfield  {author} {\bibinfo {author} {\bibfnamefont {C.}~\bibnamefont {Sanz-Fernandez}}, \bibinfo {author} {\bibfnamefont {R.}~\bibnamefont {Hernandez}}, \bibinfo {author} {\bibfnamefont {C.~D.}\ \bibnamefont {Marciniak}}, \bibinfo {author} {\bibfnamefont {I.}~\bibnamefont {Pogorelov}}, \bibinfo {author} {\bibfnamefont {T.}~\bibnamefont {Monz}}, \bibinfo {author} {\bibfnamefont {F.}~\bibnamefont {Benfenati}}, \bibinfo {author} {\bibfnamefont {S.}~\bibnamefont {Mugel}}, \ and\ \bibinfo {author} {\bibfnamefont {R.}~\bibnamefont {Orus}},\ }\href {\doibase 10.48550/ARXIV.2111.14970} {\bibinfo {title} {Quantum portfolio value forecasting},\ } (\bibinfo {year} {2021})\BibitemShut {NoStop}%
\bibitem [{\citenamefont {M\o{}lmer}\ and\ \citenamefont {S\o{}rensen}(1999)}]{PhysRevLett.82.1835}%
  \BibitemOpen
  \bibfield  {author} {\bibinfo {author} {\bibfnamefont {K.}~\bibnamefont {M\o{}lmer}}\ and\ \bibinfo {author} {\bibfnamefont {A.}~\bibnamefont {S\o{}rensen}},\ }\bibfield  {title} {\emph {\bibinfo {title} {Multiparticle entanglement of hot trapped ions},\ }}\href {\doibase 10.1103/PhysRevLett.82.1835} {\bibfield  {journal} {\bibinfo  {journal} {Phys. Rev. Lett.}\ }\textbf {\bibinfo {volume} {82}},\ \bibinfo {pages} {1835} (\bibinfo {year} {1999})}\BibitemShut {NoStop}%
\bibitem [{\citenamefont {S\o{}rensen}\ and\ \citenamefont {M\o{}lmer}(2000)}]{PhysRevA.62.022311}%
  \BibitemOpen
  \bibfield  {author} {\bibinfo {author} {\bibfnamefont {A.}~\bibnamefont {S\o{}rensen}}\ and\ \bibinfo {author} {\bibfnamefont {K.}~\bibnamefont {M\o{}lmer}},\ }\bibfield  {title} {\emph {\bibinfo {title} {Entanglement and quantum computation with ions in thermal motion},\ }}\href {\doibase 10.1103/PhysRevA.62.022311} {\bibfield  {journal} {\bibinfo  {journal} {Phys. Rev. A}\ }\textbf {\bibinfo {volume} {62}},\ \bibinfo {pages} {022311} (\bibinfo {year} {2000})}\BibitemShut {NoStop}%
\bibitem [{\citenamefont {Gaebler}\ \emph {et~al.}(2016)\citenamefont {Gaebler}, \citenamefont {Tan}, \citenamefont {Lin}, \citenamefont {Wan}, \citenamefont {Bowler}, \citenamefont {Keith}, \citenamefont {Glancy}, \citenamefont {Coakley}, \citenamefont {Knill}, \citenamefont {Leibfried},\ and\ \citenamefont {Wineland}}]{PhysRevLett.117.060505}%
  \BibitemOpen
  \bibfield  {author} {\bibinfo {author} {\bibfnamefont {J.~P.}\ \bibnamefont {Gaebler}}, \bibinfo {author} {\bibfnamefont {T.~R.}\ \bibnamefont {Tan}}, \bibinfo {author} {\bibfnamefont {Y.}~\bibnamefont {Lin}}, \bibinfo {author} {\bibfnamefont {Y.}~\bibnamefont {Wan}}, \bibinfo {author} {\bibfnamefont {R.}~\bibnamefont {Bowler}}, \bibinfo {author} {\bibfnamefont {A.~C.}\ \bibnamefont {Keith}}, \bibinfo {author} {\bibfnamefont {S.}~\bibnamefont {Glancy}}, \bibinfo {author} {\bibfnamefont {K.}~\bibnamefont {Coakley}}, \bibinfo {author} {\bibfnamefont {E.}~\bibnamefont {Knill}}, \bibinfo {author} {\bibfnamefont {D.}~\bibnamefont {Leibfried}}, \ and\ \bibinfo {author} {\bibfnamefont {D.~J.}\ \bibnamefont {Wineland}},\ }\bibfield  {title} {\emph {\bibinfo {title} {High-fidelity universal gate set for ${^{9}\mathrm{Be}}^{+}$ ion qubits},\ }}\href {\doibase 10.1103/PhysRevLett.117.060505} {\bibfield  {journal} {\bibinfo  {journal} {Phys. Rev. Lett.}\ }\textbf {\bibinfo {volume} {117}},\ \bibinfo {pages} {060505}
  (\bibinfo {year} {2016})}\BibitemShut {NoStop}%
\bibitem [{\citenamefont {Erhard}\ \emph {et~al.}(2019)\citenamefont {Erhard}, \citenamefont {Wallman}, \citenamefont {Postler}, \citenamefont {Meth}, \citenamefont {Stricker}, \citenamefont {Martinez}, \citenamefont {Schindler}, \citenamefont {Monz}, \citenamefont {Emerson},\ and\ \citenamefont {Blatt}}]{Erhard2019}%
  \BibitemOpen
  \bibfield  {author} {\bibinfo {author} {\bibfnamefont {A.}~\bibnamefont {Erhard}}, \bibinfo {author} {\bibfnamefont {J.~J.}\ \bibnamefont {Wallman}}, \bibinfo {author} {\bibfnamefont {L.}~\bibnamefont {Postler}}, \bibinfo {author} {\bibfnamefont {M.}~\bibnamefont {Meth}}, \bibinfo {author} {\bibfnamefont {R.}~\bibnamefont {Stricker}}, \bibinfo {author} {\bibfnamefont {E.~A.}\ \bibnamefont {Martinez}}, \bibinfo {author} {\bibfnamefont {P.}~\bibnamefont {Schindler}}, \bibinfo {author} {\bibfnamefont {T.}~\bibnamefont {Monz}}, \bibinfo {author} {\bibfnamefont {J.}~\bibnamefont {Emerson}}, \ and\ \bibinfo {author} {\bibfnamefont {R.}~\bibnamefont {Blatt}},\ }\bibfield  {title} {\emph {\bibinfo {title} {Characterizing large-scale quantum computers via cycle benchmarking},\ }}\href {\doibase 10.1038/s41467-019-13068-7} {\bibfield  {journal} {\bibinfo  {journal} {Nature Communications}\ }\textbf {\bibinfo {volume} {10}},\ \bibinfo {pages} {5347} (\bibinfo {year} {2019})}\BibitemShut {NoStop}%
\bibitem [{\citenamefont {Bermudez}\ \emph {et~al.}(2012{\natexlab{a}})\citenamefont {Bermudez}, \citenamefont {Schmidt}, \citenamefont {Plenio},\ and\ \citenamefont {Retzker}}]{PhysRevA.85.040302}%
  \BibitemOpen
  \bibfield  {author} {\bibinfo {author} {\bibfnamefont {A.}~\bibnamefont {Bermudez}}, \bibinfo {author} {\bibfnamefont {P.~O.}\ \bibnamefont {Schmidt}}, \bibinfo {author} {\bibfnamefont {M.~B.}\ \bibnamefont {Plenio}}, \ and\ \bibinfo {author} {\bibfnamefont {A.}~\bibnamefont {Retzker}},\ }\bibfield  {title} {\emph {\bibinfo {title} {Robust trapped-ion quantum logic gates by continuous dynamical decoupling},\ }}\href {\doibase 10.1103/PhysRevA.85.040302} {\bibfield  {journal} {\bibinfo  {journal} {Phys. Rev. A}\ }\textbf {\bibinfo {volume} {85}},\ \bibinfo {pages} {040302} (\bibinfo {year} {2012}{\natexlab{a}})}\BibitemShut {NoStop}%
\bibitem [{\citenamefont {Lemmer}\ \emph {et~al.}(2013)\citenamefont {Lemmer}, \citenamefont {Bermudez},\ and\ \citenamefont {Plenio}}]{Lemmer_2013}%
  \BibitemOpen
  \bibfield  {author} {\bibinfo {author} {\bibfnamefont {A.}~\bibnamefont {Lemmer}}, \bibinfo {author} {\bibfnamefont {A.}~\bibnamefont {Bermudez}}, \ and\ \bibinfo {author} {\bibfnamefont {M.~B.}\ \bibnamefont {Plenio}},\ }\bibfield  {title} {\emph {\bibinfo {title} {Driven geometric phase gates with trapped ions},\ }}\href {\doibase 10.1088/1367-2630/15/8/083001} {\bibfield  {journal} {\bibinfo  {journal} {New Journal of Physics}\ }\textbf {\bibinfo {volume} {15}},\ \bibinfo {pages} {083001} (\bibinfo {year} {2013})}\BibitemShut {NoStop}%
\bibitem [{\citenamefont {Tan}\ \emph {et~al.}(2013)\citenamefont {Tan}, \citenamefont {Gaebler}, \citenamefont {Bowler}, \citenamefont {Lin}, \citenamefont {Jost}, \citenamefont {Leibfried},\ and\ \citenamefont {Wineland}}]{PhysRevLett.110.263002}%
  \BibitemOpen
  \bibfield  {author} {\bibinfo {author} {\bibfnamefont {T.~R.}\ \bibnamefont {Tan}}, \bibinfo {author} {\bibfnamefont {J.~P.}\ \bibnamefont {Gaebler}}, \bibinfo {author} {\bibfnamefont {R.}~\bibnamefont {Bowler}}, \bibinfo {author} {\bibfnamefont {Y.}~\bibnamefont {Lin}}, \bibinfo {author} {\bibfnamefont {J.~D.}\ \bibnamefont {Jost}}, \bibinfo {author} {\bibfnamefont {D.}~\bibnamefont {Leibfried}}, \ and\ \bibinfo {author} {\bibfnamefont {D.~J.}\ \bibnamefont {Wineland}},\ }\bibfield  {title} {\emph {\bibinfo {title} {Demonstration of a dressed-state phase gate for trapped ions},\ }}\href {\doibase 10.1103/PhysRevLett.110.263002} {\bibfield  {journal} {\bibinfo  {journal} {Phys. Rev. Lett.}\ }\textbf {\bibinfo {volume} {110}},\ \bibinfo {pages} {263002} (\bibinfo {year} {2013})}\BibitemShut {NoStop}%
\bibitem [{\citenamefont {Harty}\ \emph {et~al.}(2016)\citenamefont {Harty}, \citenamefont {Sepiol}, \citenamefont {Allcock}, \citenamefont {Ballance}, \citenamefont {Tarlton},\ and\ \citenamefont {Lucas}}]{PhysRevLett.117.140501}%
  \BibitemOpen
  \bibfield  {author} {\bibinfo {author} {\bibfnamefont {T.~P.}\ \bibnamefont {Harty}}, \bibinfo {author} {\bibfnamefont {M.~A.}\ \bibnamefont {Sepiol}}, \bibinfo {author} {\bibfnamefont {D.~T.~C.}\ \bibnamefont {Allcock}}, \bibinfo {author} {\bibfnamefont {C.~J.}\ \bibnamefont {Ballance}}, \bibinfo {author} {\bibfnamefont {J.~E.}\ \bibnamefont {Tarlton}}, \ and\ \bibinfo {author} {\bibfnamefont {D.~M.}\ \bibnamefont {Lucas}},\ }\bibfield  {title} {\emph {\bibinfo {title} {High-fidelity trapped-ion quantum logic using near-field microwaves},\ }}\href {\doibase 10.1103/PhysRevLett.117.140501} {\bibfield  {journal} {\bibinfo  {journal} {Phys. Rev. Lett.}\ }\textbf {\bibinfo {volume} {117}},\ \bibinfo {pages} {140501} (\bibinfo {year} {2016})}\BibitemShut {NoStop}%
\bibitem [{\citenamefont {Leibfried}\ \emph {et~al.}(2003{\natexlab{a}})\citenamefont {Leibfried}, \citenamefont {DeMarco}, \citenamefont {Meyer}, \citenamefont {Lucas}, \citenamefont {Barrett}, \citenamefont {Britton}, \citenamefont {Itano}, \citenamefont {Jelenkovic}, \citenamefont {Langer}, \citenamefont {Rosenband},\ and\ \citenamefont {Wineland}}]{Leibfried2003}%
  \BibitemOpen
  \bibfield  {author} {\bibinfo {author} {\bibfnamefont {D.}~\bibnamefont {Leibfried}}, \bibinfo {author} {\bibfnamefont {B.}~\bibnamefont {DeMarco}}, \bibinfo {author} {\bibfnamefont {V.}~\bibnamefont {Meyer}}, \bibinfo {author} {\bibfnamefont {D.}~\bibnamefont {Lucas}}, \bibinfo {author} {\bibfnamefont {M.}~\bibnamefont {Barrett}}, \bibinfo {author} {\bibfnamefont {J.}~\bibnamefont {Britton}}, \bibinfo {author} {\bibfnamefont {W.~M.}\ \bibnamefont {Itano}}, \bibinfo {author} {\bibfnamefont {B.}~\bibnamefont {Jelenkovic}}, \bibinfo {author} {\bibfnamefont {C.}~\bibnamefont {Langer}}, \bibinfo {author} {\bibfnamefont {T.}~\bibnamefont {Rosenband}}, \ and\ \bibinfo {author} {\bibfnamefont {D.~J.}\ \bibnamefont {Wineland}},\ }\bibfield  {title} {\emph {\bibinfo {title} {Experimental demonstration of a robust, high-fidelity geometric two ion-qubit phase gate},\ }}\href {\doibase 10.1038/nature01492} {\bibfield  {journal} {\bibinfo  {journal} {Nature}\ }\textbf {\bibinfo {volume} {422}},\ \bibinfo {pages} {412}
  (\bibinfo {year} {2003}{\natexlab{a}})}\BibitemShut {NoStop}%
\bibitem [{\citenamefont {Ballance}\ \emph {et~al.}(2016)\citenamefont {Ballance}, \citenamefont {Harty}, \citenamefont {Linke}, \citenamefont {Sepiol},\ and\ \citenamefont {Lucas}}]{PhysRevLett.117.060504}%
  \BibitemOpen
  \bibfield  {author} {\bibinfo {author} {\bibfnamefont {C.~J.}\ \bibnamefont {Ballance}}, \bibinfo {author} {\bibfnamefont {T.~P.}\ \bibnamefont {Harty}}, \bibinfo {author} {\bibfnamefont {N.~M.}\ \bibnamefont {Linke}}, \bibinfo {author} {\bibfnamefont {M.~A.}\ \bibnamefont {Sepiol}}, \ and\ \bibinfo {author} {\bibfnamefont {D.~M.}\ \bibnamefont {Lucas}},\ }\bibfield  {title} {\emph {\bibinfo {title} {High-fidelity quantum logic gates using trapped-ion hyperfine qubits},\ }}\href {\doibase 10.1103/PhysRevLett.117.060504} {\bibfield  {journal} {\bibinfo  {journal} {Phys. Rev. Lett.}\ }\textbf {\bibinfo {volume} {117}},\ \bibinfo {pages} {060504} (\bibinfo {year} {2016})}\BibitemShut {NoStop}%
\bibitem [{\citenamefont {Baldwin}\ \emph {et~al.}(2021)\citenamefont {Baldwin}, \citenamefont {Bjork}, \citenamefont {Foss-Feig}, \citenamefont {Gaebler}, \citenamefont {Hayes}, \citenamefont {Kokish}, \citenamefont {Langer}, \citenamefont {Sedlacek}, \citenamefont {Stack},\ and\ \citenamefont {Vittorini}}]{PhysRevA.103.012603}%
  \BibitemOpen
  \bibfield  {author} {\bibinfo {author} {\bibfnamefont {C.~H.}\ \bibnamefont {Baldwin}}, \bibinfo {author} {\bibfnamefont {B.~J.}\ \bibnamefont {Bjork}}, \bibinfo {author} {\bibfnamefont {M.}~\bibnamefont {Foss-Feig}}, \bibinfo {author} {\bibfnamefont {J.~P.}\ \bibnamefont {Gaebler}}, \bibinfo {author} {\bibfnamefont {D.}~\bibnamefont {Hayes}}, \bibinfo {author} {\bibfnamefont {M.~G.}\ \bibnamefont {Kokish}}, \bibinfo {author} {\bibfnamefont {C.}~\bibnamefont {Langer}}, \bibinfo {author} {\bibfnamefont {J.~A.}\ \bibnamefont {Sedlacek}}, \bibinfo {author} {\bibfnamefont {D.}~\bibnamefont {Stack}}, \ and\ \bibinfo {author} {\bibfnamefont {G.}~\bibnamefont {Vittorini}},\ }\bibfield  {title} {\emph {\bibinfo {title} {High-fidelity light-shift gate for clock-state qubits},\ }}\href {\doibase 10.1103/PhysRevA.103.012603} {\bibfield  {journal} {\bibinfo  {journal} {Phys. Rev. A}\ }\textbf {\bibinfo {volume} {103}},\ \bibinfo {pages} {012603} (\bibinfo {year} {2021})}\BibitemShut {NoStop}%
\bibitem [{\citenamefont {Kielpinski}\ \emph {et~al.}(2002)\citenamefont {Kielpinski}, \citenamefont {Monroe},\ and\ \citenamefont {Wineland}}]{Kielpinski2002}%
  \BibitemOpen
  \bibfield  {author} {\bibinfo {author} {\bibfnamefont {D.}~\bibnamefont {Kielpinski}}, \bibinfo {author} {\bibfnamefont {C.}~\bibnamefont {Monroe}}, \ and\ \bibinfo {author} {\bibfnamefont {D.~J.}\ \bibnamefont {Wineland}},\ }\bibfield  {title} {\emph {\bibinfo {title} {Architecture for a large-scale ion-trap quantum computer},\ }}\href {\doibase 10.1038/nature00784} {\bibfield  {journal} {\bibinfo  {journal} {Nature}\ }\textbf {\bibinfo {volume} {417}},\ \bibinfo {pages} {709} (\bibinfo {year} {2002})}\BibitemShut {NoStop}%
\bibitem [{\citenamefont {Kaushal}\ \emph {et~al.}(2020)\citenamefont {Kaushal}, \citenamefont {Lekitsch}, \citenamefont {Stahl}, \citenamefont {Hilder}, \citenamefont {Pijn}, \citenamefont {Schmiegelow}, \citenamefont {Bermudez}, \citenamefont {Müller}, \citenamefont {Schmidt-Kaler},\ and\ \citenamefont {Poschinger}}]{doi:10.1116/1.5126186}%
  \BibitemOpen
  \bibfield  {author} {\bibinfo {author} {\bibfnamefont {V.}~\bibnamefont {Kaushal}}, \bibinfo {author} {\bibfnamefont {B.}~\bibnamefont {Lekitsch}}, \bibinfo {author} {\bibfnamefont {A.}~\bibnamefont {Stahl}}, \bibinfo {author} {\bibfnamefont {J.}~\bibnamefont {Hilder}}, \bibinfo {author} {\bibfnamefont {D.}~\bibnamefont {Pijn}}, \bibinfo {author} {\bibfnamefont {C.}~\bibnamefont {Schmiegelow}}, \bibinfo {author} {\bibfnamefont {A.}~\bibnamefont {Bermudez}}, \bibinfo {author} {\bibfnamefont {M.}~\bibnamefont {Müller}}, \bibinfo {author} {\bibfnamefont {F.}~\bibnamefont {Schmidt-Kaler}}, \ and\ \bibinfo {author} {\bibfnamefont {U.}~\bibnamefont {Poschinger}},\ }\bibfield  {title} {\emph {\bibinfo {title} {Shuttling-based trapped-ion quantum information processing},\ }}\href {\doibase 10.1116/1.5126186} {\bibfield  {journal} {\bibinfo  {journal} {AVS Quantum Science}\ }\textbf {\bibinfo {volume} {2}},\ \bibinfo {pages} {014101} (\bibinfo {year} {2020})}\BibitemShut {NoStop}%
\bibitem [{\citenamefont {Home}\ \emph {et~al.}(2009)\citenamefont {Home}, \citenamefont {Hanneke}, \citenamefont {Jost}, \citenamefont {Amini}, \citenamefont {Leibfried},\ and\ \citenamefont {Wineland}}]{doi:10.1126/science.1177077}%
  \BibitemOpen
  \bibfield  {author} {\bibinfo {author} {\bibfnamefont {J.~P.}\ \bibnamefont {Home}}, \bibinfo {author} {\bibfnamefont {D.}~\bibnamefont {Hanneke}}, \bibinfo {author} {\bibfnamefont {J.~D.}\ \bibnamefont {Jost}}, \bibinfo {author} {\bibfnamefont {J.~M.}\ \bibnamefont {Amini}}, \bibinfo {author} {\bibfnamefont {D.}~\bibnamefont {Leibfried}}, \ and\ \bibinfo {author} {\bibfnamefont {D.~J.}\ \bibnamefont {Wineland}},\ }\bibfield  {title} {\emph {\bibinfo {title} {Complete methods set for scalable ion trap quantum information processing},\ }}\href {\doibase 10.1126/science.1177077} {\bibfield  {journal} {\bibinfo  {journal} {Science}\ }\textbf {\bibinfo {volume} {325}},\ \bibinfo {pages} {1227} (\bibinfo {year} {2009})},\ \Eprint {http://arxiv.org/abs/https://www.science.org/doi/pdf/10.1126/science.1177077} {https://www.science.org/doi/pdf/10.1126/science.1177077} \BibitemShut {NoStop}%
\bibitem [{\citenamefont {Kaufmann}\ \emph {et~al.}(2017)\citenamefont {Kaufmann}, \citenamefont {Ruster}, \citenamefont {Schmiegelow}, \citenamefont {Luda}, \citenamefont {Kaushal}, \citenamefont {Schulz}, \citenamefont {von Lindenfels}, \citenamefont {Schmidt-Kaler},\ and\ \citenamefont {Poschinger}}]{PhysRevLett.119.150503}%
  \BibitemOpen
  \bibfield  {author} {\bibinfo {author} {\bibfnamefont {H.}~\bibnamefont {Kaufmann}}, \bibinfo {author} {\bibfnamefont {T.}~\bibnamefont {Ruster}}, \bibinfo {author} {\bibfnamefont {C.~T.}\ \bibnamefont {Schmiegelow}}, \bibinfo {author} {\bibfnamefont {M.~A.}\ \bibnamefont {Luda}}, \bibinfo {author} {\bibfnamefont {V.}~\bibnamefont {Kaushal}}, \bibinfo {author} {\bibfnamefont {J.}~\bibnamefont {Schulz}}, \bibinfo {author} {\bibfnamefont {D.}~\bibnamefont {von Lindenfels}}, \bibinfo {author} {\bibfnamefont {F.}~\bibnamefont {Schmidt-Kaler}}, \ and\ \bibinfo {author} {\bibfnamefont {U.~G.}\ \bibnamefont {Poschinger}},\ }\bibfield  {title} {\emph {\bibinfo {title} {Scalable creation of long-lived multipartite entanglement},\ }}\href {\doibase 10.1103/PhysRevLett.119.150503} {\bibfield  {journal} {\bibinfo  {journal} {Phys. Rev. Lett.}\ }\textbf {\bibinfo {volume} {119}},\ \bibinfo {pages} {150503} (\bibinfo {year} {2017})}\BibitemShut {NoStop}%
\bibitem [{\citenamefont {Pino}\ \emph {et~al.}(2021)\citenamefont {Pino} \emph {et~al.}}]{Pino2021}%
  \BibitemOpen
  \bibfield  {author} {\bibinfo {author} {\bibfnamefont {J.~M.}\ \bibnamefont {Pino}} \emph {et~al.},\ }\bibfield  {title} {\emph {\bibinfo {title} {Demonstration of the trapped-ion quantum {CCD} computer architecture},\ }}\href {\doibase 10.1038/s41586-021-03318-4} {\bibfield  {journal} {\bibinfo  {journal} {Nature}\ }\textbf {\bibinfo {volume} {592}},\ \bibinfo {pages} {209} (\bibinfo {year} {2021})}\BibitemShut {NoStop}%
\bibitem [{\citenamefont {Ryan-Anderson}\ \emph {et~al.}(2021)\citenamefont {Ryan-Anderson}, \citenamefont {Bohnet}, \citenamefont {Lee}, \citenamefont {Gresh}, \citenamefont {Hankin}, \citenamefont {Gaebler}, \citenamefont {Francois}, \citenamefont {Chernoguzov}, \citenamefont {Lucchetti}, \citenamefont {Brown}, \citenamefont {Gatterman}, \citenamefont {Halit}, \citenamefont {Gilmore}, \citenamefont {Gerber}, \citenamefont {Neyenhuis}, \citenamefont {Hayes},\ and\ \citenamefont {Stutz}}]{PhysRevX.11.041058}%
  \BibitemOpen
  \bibfield  {author} {\bibinfo {author} {\bibfnamefont {C.}~\bibnamefont {Ryan-Anderson}}, \bibinfo {author} {\bibfnamefont {J.~G.}\ \bibnamefont {Bohnet}}, \bibinfo {author} {\bibfnamefont {K.}~\bibnamefont {Lee}}, \bibinfo {author} {\bibfnamefont {D.}~\bibnamefont {Gresh}}, \bibinfo {author} {\bibfnamefont {A.}~\bibnamefont {Hankin}}, \bibinfo {author} {\bibfnamefont {J.~P.}\ \bibnamefont {Gaebler}}, \bibinfo {author} {\bibfnamefont {D.}~\bibnamefont {Francois}}, \bibinfo {author} {\bibfnamefont {A.}~\bibnamefont {Chernoguzov}}, \bibinfo {author} {\bibfnamefont {D.}~\bibnamefont {Lucchetti}}, \bibinfo {author} {\bibfnamefont {N.~C.}\ \bibnamefont {Brown}}, \bibinfo {author} {\bibfnamefont {T.~M.}\ \bibnamefont {Gatterman}}, \bibinfo {author} {\bibfnamefont {S.~K.}\ \bibnamefont {Halit}}, \bibinfo {author} {\bibfnamefont {K.}~\bibnamefont {Gilmore}}, \bibinfo {author} {\bibfnamefont {J.~A.}\ \bibnamefont {Gerber}}, \bibinfo {author} {\bibfnamefont {B.}~\bibnamefont {Neyenhuis}}, \bibinfo {author}
  {\bibfnamefont {D.}~\bibnamefont {Hayes}}, \ and\ \bibinfo {author} {\bibfnamefont {R.~P.}\ \bibnamefont {Stutz}},\ }\bibfield  {title} {\emph {\bibinfo {title} {Realization of real-time fault-tolerant quantum error correction},\ }}\href {\doibase 10.1103/PhysRevX.11.041058} {\bibfield  {journal} {\bibinfo  {journal} {Phys. Rev. X}\ }\textbf {\bibinfo {volume} {11}},\ \bibinfo {pages} {041058} (\bibinfo {year} {2021})}\BibitemShut {NoStop}%
\bibitem [{\citenamefont {Debnath}\ \emph {et~al.}(2016)\citenamefont {Debnath}, \citenamefont {Linke}, \citenamefont {Figgatt}, \citenamefont {Landsman}, \citenamefont {Wright},\ and\ \citenamefont {Monroe}}]{Debnath2016}%
  \BibitemOpen
  \bibfield  {author} {\bibinfo {author} {\bibfnamefont {S.}~\bibnamefont {Debnath}}, \bibinfo {author} {\bibfnamefont {N.~M.}\ \bibnamefont {Linke}}, \bibinfo {author} {\bibfnamefont {C.}~\bibnamefont {Figgatt}}, \bibinfo {author} {\bibfnamefont {K.~A.}\ \bibnamefont {Landsman}}, \bibinfo {author} {\bibfnamefont {K.}~\bibnamefont {Wright}}, \ and\ \bibinfo {author} {\bibfnamefont {C.}~\bibnamefont {Monroe}},\ }\bibfield  {title} {\emph {\bibinfo {title} {Demonstration of a small programmable quantum computer with atomic qubits},\ }}\href {\doibase 10.1038/nature18648} {\bibfield  {journal} {\bibinfo  {journal} {Nature}\ }\textbf {\bibinfo {volume} {536}},\ \bibinfo {pages} {63} (\bibinfo {year} {2016})}\BibitemShut {NoStop}%
\bibitem [{\citenamefont {Figgatt}\ \emph {et~al.}(2019)\citenamefont {Figgatt}, \citenamefont {Ostrander}, \citenamefont {Linke}, \citenamefont {Landsman}, \citenamefont {Zhu}, \citenamefont {Maslov},\ and\ \citenamefont {Monroe}}]{Figgatt2019}%
  \BibitemOpen
  \bibfield  {author} {\bibinfo {author} {\bibfnamefont {C.}~\bibnamefont {Figgatt}}, \bibinfo {author} {\bibfnamefont {A.}~\bibnamefont {Ostrander}}, \bibinfo {author} {\bibfnamefont {N.~M.}\ \bibnamefont {Linke}}, \bibinfo {author} {\bibfnamefont {K.~A.}\ \bibnamefont {Landsman}}, \bibinfo {author} {\bibfnamefont {D.}~\bibnamefont {Zhu}}, \bibinfo {author} {\bibfnamefont {D.}~\bibnamefont {Maslov}}, \ and\ \bibinfo {author} {\bibfnamefont {C.}~\bibnamefont {Monroe}},\ }\bibfield  {title} {\emph {\bibinfo {title} {Parallel entangling operations on a universal ion-trap quantum computer},\ }}\href {\doibase 10.1038/s41586-019-1427-5} {\bibfield  {journal} {\bibinfo  {journal} {Nature}\ }\textbf {\bibinfo {volume} {572}},\ \bibinfo {pages} {368} (\bibinfo {year} {2019})}\BibitemShut {NoStop}%
\bibitem [{\citenamefont {Pogorelov}\ \emph {et~al.}(2021)\citenamefont {Pogorelov}, \citenamefont {Feldker}, \citenamefont {Marciniak}, \citenamefont {Postler}, \citenamefont {Jacob}, \citenamefont {Krieglsteiner}, \citenamefont {Podlesnic}, \citenamefont {Meth}, \citenamefont {Negnevitsky}, \citenamefont {Stadler}, \citenamefont {H\"ofer}, \citenamefont {W\"achter}, \citenamefont {Lakhmanskiy}, \citenamefont {Blatt}, \citenamefont {Schindler},\ and\ \citenamefont {Monz}}]{PRXQuantum.2.020343}%
  \BibitemOpen
  \bibfield  {author} {\bibinfo {author} {\bibfnamefont {I.}~\bibnamefont {Pogorelov}}, \bibinfo {author} {\bibfnamefont {T.}~\bibnamefont {Feldker}}, \bibinfo {author} {\bibfnamefont {C.~D.}\ \bibnamefont {Marciniak}}, \bibinfo {author} {\bibfnamefont {L.}~\bibnamefont {Postler}}, \bibinfo {author} {\bibfnamefont {G.}~\bibnamefont {Jacob}}, \bibinfo {author} {\bibfnamefont {O.}~\bibnamefont {Krieglsteiner}}, \bibinfo {author} {\bibfnamefont {V.}~\bibnamefont {Podlesnic}}, \bibinfo {author} {\bibfnamefont {M.}~\bibnamefont {Meth}}, \bibinfo {author} {\bibfnamefont {V.}~\bibnamefont {Negnevitsky}}, \bibinfo {author} {\bibfnamefont {M.}~\bibnamefont {Stadler}}, \bibinfo {author} {\bibfnamefont {B.}~\bibnamefont {H\"ofer}}, \bibinfo {author} {\bibfnamefont {C.}~\bibnamefont {W\"achter}}, \bibinfo {author} {\bibfnamefont {K.}~\bibnamefont {Lakhmanskiy}}, \bibinfo {author} {\bibfnamefont {R.}~\bibnamefont {Blatt}}, \bibinfo {author} {\bibfnamefont {P.}~\bibnamefont {Schindler}}, \ and\ \bibinfo {author}
  {\bibfnamefont {T.}~\bibnamefont {Monz}},\ }\bibfield  {title} {\emph {\bibinfo {title} {Compact ion-trap quantum computing demonstrator},\ }}\href {\doibase 10.1103/PRXQuantum.2.020343} {\bibfield  {journal} {\bibinfo  {journal} {PRX Quantum}\ }\textbf {\bibinfo {volume} {2}},\ \bibinfo {pages} {020343} (\bibinfo {year} {2021})}\BibitemShut {NoStop}%
\bibitem [{\citenamefont {Bermudez}\ \emph {et~al.}(2017{\natexlab{a}})\citenamefont {Bermudez} \emph {et~al.}}]{PhysRevX.7.041061}%
  \BibitemOpen
  \bibfield  {author} {\bibinfo {author} {\bibfnamefont {A.}~\bibnamefont {Bermudez}} \emph {et~al.},\ }\bibfield  {title} {\emph {\bibinfo {title} {Assessing the progress of trapped-ion processors towards fault-tolerant quantum computation},\ }}\href {\doibase 10.1103/PhysRevX.7.041061} {\bibfield  {journal} {\bibinfo  {journal} {Phys. Rev. X}\ }\textbf {\bibinfo {volume} {7}},\ \bibinfo {pages} {041061} (\bibinfo {year} {2017}{\natexlab{a}})}\BibitemShut {NoStop}%
\bibitem [{\citenamefont {Guti\'errez}\ \emph {et~al.}(2019)\citenamefont {Guti\'errez}, \citenamefont {M\"uller},\ and\ \citenamefont {Berm\'udez}}]{PhysRevA.99.022330}%
  \BibitemOpen
  \bibfield  {author} {\bibinfo {author} {\bibfnamefont {M.}~\bibnamefont {Guti\'errez}}, \bibinfo {author} {\bibfnamefont {M.}~\bibnamefont {M\"uller}}, \ and\ \bibinfo {author} {\bibfnamefont {A.}~\bibnamefont {Berm\'udez}},\ }\bibfield  {title} {\emph {\bibinfo {title} {Transversality and lattice surgery: Exploring realistic routes toward coupled logical qubits with trapped-ion quantum processors},\ }}\href {\doibase 10.1103/PhysRevA.99.022330} {\bibfield  {journal} {\bibinfo  {journal} {Phys. Rev. A}\ }\textbf {\bibinfo {volume} {99}},\ \bibinfo {pages} {022330} (\bibinfo {year} {2019})}\BibitemShut {NoStop}%
\bibitem [{\citenamefont {Bermudez}\ \emph {et~al.}(2019)\citenamefont {Bermudez}, \citenamefont {Xu}, \citenamefont {Guti\'errez}, \citenamefont {Benjamin},\ and\ \citenamefont {M\"uller}}]{PhysRevA.100.062307}%
  \BibitemOpen
  \bibfield  {author} {\bibinfo {author} {\bibfnamefont {A.}~\bibnamefont {Bermudez}}, \bibinfo {author} {\bibfnamefont {X.}~\bibnamefont {Xu}}, \bibinfo {author} {\bibfnamefont {M.}~\bibnamefont {Guti\'errez}}, \bibinfo {author} {\bibfnamefont {S.~C.}\ \bibnamefont {Benjamin}}, \ and\ \bibinfo {author} {\bibfnamefont {M.}~\bibnamefont {M\"uller}},\ }\bibfield  {title} {\emph {\bibinfo {title} {Fault-tolerant protection of near-term trapped-ion topological qubits under realistic noise sources},\ }}\href {\doibase 10.1103/PhysRevA.100.062307} {\bibfield  {journal} {\bibinfo  {journal} {Phys. Rev. A}\ }\textbf {\bibinfo {volume} {100}},\ \bibinfo {pages} {062307} (\bibinfo {year} {2019})}\BibitemShut {NoStop}%
\bibitem [{\citenamefont {Parrado-Rodr{\'{i}}guez}\ \emph {et~al.}(2021)\citenamefont {Parrado-Rodr{\'{i}}guez}, \citenamefont {Ryan-Anderson}, \citenamefont {Bermudez},\ and\ \citenamefont {M{\"{u}}ller}}]{ParradoRodriguez2021crosstalk}%
  \BibitemOpen
  \bibfield  {author} {\bibinfo {author} {\bibfnamefont {P.}~\bibnamefont {Parrado-Rodr{\'{i}}guez}}, \bibinfo {author} {\bibfnamefont {C.}~\bibnamefont {Ryan-Anderson}}, \bibinfo {author} {\bibfnamefont {A.}~\bibnamefont {Bermudez}}, \ and\ \bibinfo {author} {\bibfnamefont {M.}~\bibnamefont {M{\"{u}}ller}},\ }\bibfield  {title} {\emph {\bibinfo {title} {Crosstalk {S}uppression for {F}ault-tolerant {Q}uantum {E}rror {C}orrection with {T}rapped {I}ons},\ }}\href {\doibase 10.22331/q-2021-06-29-487} {\bibfield  {journal} {\bibinfo  {journal} {{Quantum}}\ }\textbf {\bibinfo {volume} {5}},\ \bibinfo {pages} {487} (\bibinfo {year} {2021})}\BibitemShut {NoStop}%
\bibitem [{\citenamefont {Trout}\ \emph {et~al.}(2018)\citenamefont {Trout}, \citenamefont {Li}, \citenamefont {Guti{\'{e}}rrez}, \citenamefont {Wu}, \citenamefont {Wang}, \citenamefont {Duan},\ and\ \citenamefont {Brown}}]{Trout_2018}%
  \BibitemOpen
  \bibfield  {author} {\bibinfo {author} {\bibfnamefont {C.~J.}\ \bibnamefont {Trout}}, \bibinfo {author} {\bibfnamefont {M.}~\bibnamefont {Li}}, \bibinfo {author} {\bibfnamefont {M.}~\bibnamefont {Guti{\'{e}}rrez}}, \bibinfo {author} {\bibfnamefont {Y.}~\bibnamefont {Wu}}, \bibinfo {author} {\bibfnamefont {S.-T.}\ \bibnamefont {Wang}}, \bibinfo {author} {\bibfnamefont {L.}~\bibnamefont {Duan}}, \ and\ \bibinfo {author} {\bibfnamefont {K.~R.}\ \bibnamefont {Brown}},\ }\bibfield  {title} {\emph {\bibinfo {title} {Simulating the performance of a distance-3 surface code in a linear ion trap},\ }}\href {\doibase 10.1088/1367-2630/aab341} {\bibfield  {journal} {\bibinfo  {journal} {New Journal of Physics}\ }\textbf {\bibinfo {volume} {20}},\ \bibinfo {pages} {043038} (\bibinfo {year} {2018})}\BibitemShut {NoStop}%
\bibitem [{\citenamefont {Debroy}\ \emph {et~al.}(2020)\citenamefont {Debroy}, \citenamefont {Li}, \citenamefont {Huang},\ and\ \citenamefont {Brown}}]{Debroy_2020}%
  \BibitemOpen
  \bibfield  {author} {\bibinfo {author} {\bibfnamefont {D.~M.}\ \bibnamefont {Debroy}}, \bibinfo {author} {\bibfnamefont {M.}~\bibnamefont {Li}}, \bibinfo {author} {\bibfnamefont {S.}~\bibnamefont {Huang}}, \ and\ \bibinfo {author} {\bibfnamefont {K.~R.}\ \bibnamefont {Brown}},\ }\bibfield  {title} {\emph {\bibinfo {title} {Logical performance of 9 qubit compass codes in ion traps with crosstalk errors},\ }}\href {\doibase 10.1088/2058-9565/ab7e80} {\bibfield  {journal} {\bibinfo  {journal} {Quantum Science and Technology}\ }\textbf {\bibinfo {volume} {5}},\ \bibinfo {pages} {034002} (\bibinfo {year} {2020})}\BibitemShut {NoStop}%
\bibitem [{\citenamefont {Murali}\ \emph {et~al.}(2020)\citenamefont {Murali}, \citenamefont {Debroy}, \citenamefont {Brown},\ and\ \citenamefont {Martonosi}}]{https://doi.org/10.48550/arxiv.2004.04706}%
  \BibitemOpen
  \bibfield  {author} {\bibinfo {author} {\bibfnamefont {P.}~\bibnamefont {Murali}}, \bibinfo {author} {\bibfnamefont {D.~M.}\ \bibnamefont {Debroy}}, \bibinfo {author} {\bibfnamefont {K.~R.}\ \bibnamefont {Brown}}, \ and\ \bibinfo {author} {\bibfnamefont {M.}~\bibnamefont {Martonosi}},\ }\href {\doibase 10.48550/ARXIV.2004.04706} {\bibinfo {title} {Architecting noisy intermediate-scale trapped ion quantum computers},\ } (\bibinfo {year} {2020})\BibitemShut {NoStop}%
\bibitem [{\citenamefont {Wu}\ \emph {et~al.}(2021)\citenamefont {Wu}, \citenamefont {Debroy}, \citenamefont {Ding}, \citenamefont {Baker}, \citenamefont {Alexeev}, \citenamefont {Brown},\ and\ \citenamefont {Chong}}]{9407237}%
  \BibitemOpen
  \bibfield  {author} {\bibinfo {author} {\bibfnamefont {X.-C.}\ \bibnamefont {Wu}}, \bibinfo {author} {\bibfnamefont {D.~M.}\ \bibnamefont {Debroy}}, \bibinfo {author} {\bibfnamefont {Y.}~\bibnamefont {Ding}}, \bibinfo {author} {\bibfnamefont {J.~M.}\ \bibnamefont {Baker}}, \bibinfo {author} {\bibfnamefont {Y.}~\bibnamefont {Alexeev}}, \bibinfo {author} {\bibfnamefont {K.~R.}\ \bibnamefont {Brown}}, \ and\ \bibinfo {author} {\bibfnamefont {F.~T.}\ \bibnamefont {Chong}},\ }in\ \href {\doibase 10.1109/HPCA51647.2021.00023} {\emph {\bibinfo {booktitle} {2021 IEEE International Symposium on High-Performance Computer Architecture (HPCA)}}}\ (\bibinfo {year} {2021})\ pp.\ \bibinfo {pages} {153--166}\BibitemShut {NoStop}%
\bibitem [{\citenamefont {Tinkey}\ \emph {et~al.}(2021)\citenamefont {Tinkey}, \citenamefont {Meier}, \citenamefont {Clark}, \citenamefont {Seck},\ and\ \citenamefont {Brown}}]{Tinkey_2021}%
  \BibitemOpen
  \bibfield  {author} {\bibinfo {author} {\bibfnamefont {H.~N.}\ \bibnamefont {Tinkey}}, \bibinfo {author} {\bibfnamefont {A.~M.}\ \bibnamefont {Meier}}, \bibinfo {author} {\bibfnamefont {C.~R.}\ \bibnamefont {Clark}}, \bibinfo {author} {\bibfnamefont {C.~M.}\ \bibnamefont {Seck}}, \ and\ \bibinfo {author} {\bibfnamefont {K.~R.}\ \bibnamefont {Brown}},\ }\bibfield  {title} {\emph {\bibinfo {title} {Quantum process tomography of a m{\o}lmer-s{\o}rensen gate via a global beam},\ }}\href {\doibase 10.1088/2058-9565/ac0543} {\bibfield  {journal} {\bibinfo  {journal} {Quantum Science and Technology}\ }\textbf {\bibinfo {volume} {6}},\ \bibinfo {pages} {034013} (\bibinfo {year} {2021})}\BibitemShut {NoStop}%
\bibitem [{\citenamefont {Schindler}\ \emph {et~al.}(2013)\citenamefont {Schindler} \emph {et~al.}}]{Schindler2013}%
  \BibitemOpen
  \bibfield  {author} {\bibinfo {author} {\bibfnamefont {P.}~\bibnamefont {Schindler}} \emph {et~al.},\ }\bibfield  {title} {\emph {\bibinfo {title} {A quantum information processor with trapped ions},\ }}\href {\doibase 10.1088/1367-2630/15/12/123012} {\bibfield  {journal} {\bibinfo  {journal} {New Journal of Physics}\ }\textbf {\bibinfo {volume} {15}},\ \bibinfo {pages} {123012} (\bibinfo {year} {2013})}\BibitemShut {NoStop}%
\bibitem [{\citenamefont {Ballance}()}]{ballance_thesis}%
  \BibitemOpen
  \bibfield  {author} {\bibinfo {author} {\bibfnamefont {C.}~\bibnamefont {Ballance}},\ }\bibfield  {title} {\emph {\bibinfo {title} {High-fidelity quantum logic in ca+},\ }}\href {https://link.springer.com/book/10.1007/978-3-319-68216-7} {\ }\BibitemShut {NoStop}%
\bibitem [{\citenamefont {Sutherland}\ \emph {et~al.}(2022)\citenamefont {Sutherland}, \citenamefont {Yu}, \citenamefont {Beck},\ and\ \citenamefont {H\"affner}}]{PhysRevA.105.022437}%
  \BibitemOpen
  \bibfield  {author} {\bibinfo {author} {\bibfnamefont {R.~T.}\ \bibnamefont {Sutherland}}, \bibinfo {author} {\bibfnamefont {Q.}~\bibnamefont {Yu}}, \bibinfo {author} {\bibfnamefont {K.~M.}\ \bibnamefont {Beck}}, \ and\ \bibinfo {author} {\bibfnamefont {H.}~\bibnamefont {H\"affner}},\ }\bibfield  {title} {\emph {\bibinfo {title} {One- and two-qubit gate infidelities due to motional errors in trapped ions and electrons},\ }}\href {\doibase 10.1103/PhysRevA.105.022437} {\bibfield  {journal} {\bibinfo  {journal} {Phys. Rev. A}\ }\textbf {\bibinfo {volume} {105}},\ \bibinfo {pages} {022437} (\bibinfo {year} {2022})}\BibitemShut {NoStop}%
\bibitem [{\citenamefont {Preskill}(1998)}]{preskill_notes}%
  \BibitemOpen
  \bibfield  {author} {\bibinfo {author} {\bibfnamefont {J.}~\bibnamefont {Preskill}},\ }\href {http://theory.caltech.edu/~preskill/ph229/notes/chap4_01.pdf} {\emph {\bibinfo {title} {Lecture notes for Physics 219: Quantum Computation}}}\ (\bibinfo  {publisher} {California Institute of Technology},\ \bibinfo {address} {Pasadena, CA},\ \bibinfo {year} {1998})\BibitemShut {NoStop}%
\bibitem [{\citenamefont {Bennett}\ \emph {et~al.}(1996)\citenamefont {Bennett}, \citenamefont {DiVincenzo}, \citenamefont {Smolin},\ and\ \citenamefont {Wootters}}]{PhysRevA.54.3824}%
  \BibitemOpen
  \bibfield  {author} {\bibinfo {author} {\bibfnamefont {C.~H.}\ \bibnamefont {Bennett}}, \bibinfo {author} {\bibfnamefont {D.~P.}\ \bibnamefont {DiVincenzo}}, \bibinfo {author} {\bibfnamefont {J.~A.}\ \bibnamefont {Smolin}}, \ and\ \bibinfo {author} {\bibfnamefont {W.~K.}\ \bibnamefont {Wootters}},\ }\bibfield  {title} {\emph {\bibinfo {title} {Mixed-state entanglement and quantum error correction},\ }}\href {\doibase 10.1103/PhysRevA.54.3824} {\bibfield  {journal} {\bibinfo  {journal} {Phys. Rev. A}\ }\textbf {\bibinfo {volume} {54}},\ \bibinfo {pages} {3824} (\bibinfo {year} {1996})}\BibitemShut {NoStop}%
\bibitem [{\citenamefont {Lidar}\ and\ \citenamefont {Brun}(2013)}]{entanglementassistQECC}%
  \BibitemOpen
  \bibfield  {author} {\bibinfo {author} {\bibfnamefont {D.~A.}\ \bibnamefont {Lidar}}\ and\ \bibinfo {author} {\bibfnamefont {T.~A.}\ \bibnamefont {Brun}},\ }\href@noop {} {\emph {\bibinfo {title} {Quantum Error Correction}}}\ (\bibinfo  {publisher} {Cambridge University Press},\ \bibinfo {address} {United Kingdom},\ \bibinfo {year} {2013})\BibitemShut {NoStop}%
\bibitem [{\citenamefont {Almheiri}\ \emph {et~al.}(2015)\citenamefont {Almheiri}, \citenamefont {Dong},\ and\ \citenamefont {Harlow}}]{Almheiri2015}%
  \BibitemOpen
  \bibfield  {author} {\bibinfo {author} {\bibfnamefont {A.}~\bibnamefont {Almheiri}}, \bibinfo {author} {\bibfnamefont {X.}~\bibnamefont {Dong}}, \ and\ \bibinfo {author} {\bibfnamefont {D.}~\bibnamefont {Harlow}},\ }\bibfield  {title} {\emph {\bibinfo {title} {Bulk locality and quantum error correction in ads/cft},\ }}\href {\doibase 10.1007/JHEP04(2015)163} {\bibfield  {journal} {\bibinfo  {journal} {Journal of High Energy Physics}\ }\textbf {\bibinfo {volume} {2015}},\ \bibinfo {pages} {163} (\bibinfo {year} {2015})}\BibitemShut {NoStop}%
\bibitem [{\citenamefont {Gottesman}(1997)}]{stabilisers}%
  \BibitemOpen
  \bibfield  {author} {\bibinfo {author} {\bibfnamefont {D.}~\bibnamefont {Gottesman}},\ }\bibfield  {title} {\emph {\bibinfo {title} {Stabilizer codes and quantum error correc- tion},\ }}\href {https://arxiv.org/abs/quant-ph/9705052} {\bibfield  {journal} {\bibinfo  {journal} {Ph.D. thesis, arXiv:9705052(1997)}\ } (\bibinfo {year} {1997})}\BibitemShut {NoStop}%
\bibitem [{\citenamefont {Chiaverini}\ \emph {et~al.}(2004)\citenamefont {Chiaverini}, \citenamefont {Leibfried}, \citenamefont {Schaetz}, \citenamefont {Barrett}, \citenamefont {Blakestad}, \citenamefont {Britton}, \citenamefont {Itano}, \citenamefont {Jost}, \citenamefont {Knill}, \citenamefont {Langer}, \citenamefont {Ozeri},\ and\ \citenamefont {Wineland}}]{Chiaverini2004}%
  \BibitemOpen
  \bibfield  {author} {\bibinfo {author} {\bibfnamefont {J.}~\bibnamefont {Chiaverini}}, \bibinfo {author} {\bibfnamefont {D.}~\bibnamefont {Leibfried}}, \bibinfo {author} {\bibfnamefont {T.}~\bibnamefont {Schaetz}}, \bibinfo {author} {\bibfnamefont {M.~D.}\ \bibnamefont {Barrett}}, \bibinfo {author} {\bibfnamefont {R.~B.}\ \bibnamefont {Blakestad}}, \bibinfo {author} {\bibfnamefont {J.}~\bibnamefont {Britton}}, \bibinfo {author} {\bibfnamefont {W.~M.}\ \bibnamefont {Itano}}, \bibinfo {author} {\bibfnamefont {J.~D.}\ \bibnamefont {Jost}}, \bibinfo {author} {\bibfnamefont {E.}~\bibnamefont {Knill}}, \bibinfo {author} {\bibfnamefont {C.}~\bibnamefont {Langer}}, \bibinfo {author} {\bibfnamefont {R.}~\bibnamefont {Ozeri}}, \ and\ \bibinfo {author} {\bibfnamefont {D.~J.}\ \bibnamefont {Wineland}},\ }\bibfield  {title} {\emph {\bibinfo {title} {Realization of quantum error correction},\ }}\href {\doibase 10.1038/nature03074} {\bibfield  {journal} {\bibinfo  {journal} {Nature}\ }\textbf {\bibinfo {volume} {432}},\
  \bibinfo {pages} {602} (\bibinfo {year} {2004})}\BibitemShut {NoStop}%
\bibitem [{\citenamefont {Schindler}\ \emph {et~al.}(2011)\citenamefont {Schindler}, \citenamefont {Barreiro}, \citenamefont {Monz}, \citenamefont {Nebendahl}, \citenamefont {Nigg}, \citenamefont {Chwalla}, \citenamefont {Hennrich},\ and\ \citenamefont {Blatt}}]{Schindler1059}%
  \BibitemOpen
  \bibfield  {author} {\bibinfo {author} {\bibfnamefont {P.}~\bibnamefont {Schindler}}, \bibinfo {author} {\bibfnamefont {J.~T.}\ \bibnamefont {Barreiro}}, \bibinfo {author} {\bibfnamefont {T.}~\bibnamefont {Monz}}, \bibinfo {author} {\bibfnamefont {V.}~\bibnamefont {Nebendahl}}, \bibinfo {author} {\bibfnamefont {D.}~\bibnamefont {Nigg}}, \bibinfo {author} {\bibfnamefont {M.}~\bibnamefont {Chwalla}}, \bibinfo {author} {\bibfnamefont {M.}~\bibnamefont {Hennrich}}, \ and\ \bibinfo {author} {\bibfnamefont {R.}~\bibnamefont {Blatt}},\ }\bibfield  {title} {\emph {\bibinfo {title} {Experimental repetitive quantum error correction},\ }}\href {\doibase 10.1126/science.1203329} {\bibfield  {journal} {\bibinfo  {journal} {Science}\ }\textbf {\bibinfo {volume} {332}},\ \bibinfo {pages} {1059} (\bibinfo {year} {2011})}\BibitemShut {NoStop}%
\bibitem [{\citenamefont {Nigg}\ \emph {et~al.}(2014)\citenamefont {Nigg}, \citenamefont {M{\"u}ller}, \citenamefont {Martinez}, \citenamefont {Schindler}, \citenamefont {Hennrich}, \citenamefont {Monz}, \citenamefont {Martin-Delgado},\ and\ \citenamefont {Blatt}}]{Nigg302}%
  \BibitemOpen
  \bibfield  {author} {\bibinfo {author} {\bibfnamefont {D.}~\bibnamefont {Nigg}}, \bibinfo {author} {\bibfnamefont {M.}~\bibnamefont {M{\"u}ller}}, \bibinfo {author} {\bibfnamefont {E.~A.}\ \bibnamefont {Martinez}}, \bibinfo {author} {\bibfnamefont {P.}~\bibnamefont {Schindler}}, \bibinfo {author} {\bibfnamefont {M.}~\bibnamefont {Hennrich}}, \bibinfo {author} {\bibfnamefont {T.}~\bibnamefont {Monz}}, \bibinfo {author} {\bibfnamefont {M.~A.}\ \bibnamefont {Martin-Delgado}}, \ and\ \bibinfo {author} {\bibfnamefont {R.}~\bibnamefont {Blatt}},\ }\bibfield  {title} {\emph {\bibinfo {title} {Quantum computations on a topologically encoded qubit},\ }}\href {\doibase 10.1126/science.1253742} {\bibfield  {journal} {\bibinfo  {journal} {Science}\ }\textbf {\bibinfo {volume} {345}},\ \bibinfo {pages} {302} (\bibinfo {year} {2014})}\BibitemShut {NoStop}%
\bibitem [{\citenamefont {Linke}\ \emph {et~al.}(2017)\citenamefont {Linke} \emph {et~al.}}]{Linkee1701074}%
  \BibitemOpen
  \bibfield  {author} {\bibinfo {author} {\bibfnamefont {N.~M.}\ \bibnamefont {Linke}} \emph {et~al.},\ }\bibfield  {title} {\emph {\bibinfo {title} {Fault-tolerant quantum error detection},\ }}\href {https://advances.sciencemag.org/content/3/10/e1701074} {\bibfield  {journal} {\bibinfo  {journal} {Science Advances}\ }\textbf {\bibinfo {volume} {3}} (\bibinfo {year} {2017})}\BibitemShut {NoStop}%
\bibitem [{\citenamefont {Stricker}\ \emph {et~al.}(2020)\citenamefont {Stricker}, \citenamefont {Vodola}, \citenamefont {Erhard}, \citenamefont {Postler}, \citenamefont {Meth}, \citenamefont {Ringbauer}, \citenamefont {Schindler}, \citenamefont {Monz}, \citenamefont {M{\"u}ller},\ and\ \citenamefont {Blatt}}]{Stricker2020}%
  \BibitemOpen
  \bibfield  {author} {\bibinfo {author} {\bibfnamefont {R.}~\bibnamefont {Stricker}}, \bibinfo {author} {\bibfnamefont {D.}~\bibnamefont {Vodola}}, \bibinfo {author} {\bibfnamefont {A.}~\bibnamefont {Erhard}}, \bibinfo {author} {\bibfnamefont {L.}~\bibnamefont {Postler}}, \bibinfo {author} {\bibfnamefont {M.}~\bibnamefont {Meth}}, \bibinfo {author} {\bibfnamefont {M.}~\bibnamefont {Ringbauer}}, \bibinfo {author} {\bibfnamefont {P.}~\bibnamefont {Schindler}}, \bibinfo {author} {\bibfnamefont {T.}~\bibnamefont {Monz}}, \bibinfo {author} {\bibfnamefont {M.}~\bibnamefont {M{\"u}ller}}, \ and\ \bibinfo {author} {\bibfnamefont {R.}~\bibnamefont {Blatt}},\ }\bibfield  {title} {\emph {\bibinfo {title} {Experimental deterministic correction of qubit loss},\ }}\href {\doibase 10.1038/s41586-020-2667-0} {\bibfield  {journal} {\bibinfo  {journal} {Nature}\ }\textbf {\bibinfo {volume} {585}},\ \bibinfo {pages} {207} (\bibinfo {year} {2020})}\BibitemShut {NoStop}%
\bibitem [{\citenamefont {Erhard}\ \emph {et~al.}(2021)\citenamefont {Erhard} \emph {et~al.}}]{Erhard2021}%
  \BibitemOpen
  \bibfield  {author} {\bibinfo {author} {\bibfnamefont {A.}~\bibnamefont {Erhard}} \emph {et~al.},\ }\bibfield  {title} {\emph {\bibinfo {title} {Entangling logical qubits with lattice surgery},\ }}\href {\doibase 10.1038/s41586-020-03079-6} {\bibfield  {journal} {\bibinfo  {journal} {Nature}\ }\textbf {\bibinfo {volume} {589}},\ \bibinfo {pages} {220} (\bibinfo {year} {2021})}\BibitemShut {NoStop}%
\bibitem [{\citenamefont {Egan}\ \emph {et~al.}(2021)\citenamefont {Egan}, \citenamefont {Debroy}, \citenamefont {Noel}, \citenamefont {Risinger}, \citenamefont {Zhu}, \citenamefont {Biswas}, \citenamefont {Newman}, \citenamefont {Li}, \citenamefont {Brown}, \citenamefont {Cetina},\ and\ \citenamefont {Monroe}}]{Egan2021}%
  \BibitemOpen
  \bibfield  {author} {\bibinfo {author} {\bibfnamefont {L.}~\bibnamefont {Egan}}, \bibinfo {author} {\bibfnamefont {D.~M.}\ \bibnamefont {Debroy}}, \bibinfo {author} {\bibfnamefont {C.}~\bibnamefont {Noel}}, \bibinfo {author} {\bibfnamefont {A.}~\bibnamefont {Risinger}}, \bibinfo {author} {\bibfnamefont {D.}~\bibnamefont {Zhu}}, \bibinfo {author} {\bibfnamefont {D.}~\bibnamefont {Biswas}}, \bibinfo {author} {\bibfnamefont {M.}~\bibnamefont {Newman}}, \bibinfo {author} {\bibfnamefont {M.}~\bibnamefont {Li}}, \bibinfo {author} {\bibfnamefont {K.~R.}\ \bibnamefont {Brown}}, \bibinfo {author} {\bibfnamefont {M.}~\bibnamefont {Cetina}}, \ and\ \bibinfo {author} {\bibfnamefont {C.}~\bibnamefont {Monroe}},\ }\bibfield  {title} {\emph {\bibinfo {title} {Fault-tolerant control of an error-corrected qubit},\ }}\href {\doibase 10.1038/s41586-021-03928-y} {\bibfield  {journal} {\bibinfo  {journal} {Nature}\ }\textbf {\bibinfo {volume} {598}},\ \bibinfo {pages} {281} (\bibinfo {year} {2021})}\BibitemShut {NoStop}%
\bibitem [{\citenamefont {Debroy}\ \emph {et~al.}(2021)\citenamefont {Debroy}, \citenamefont {Egan}, \citenamefont {Noel}, \citenamefont {Risinger}, \citenamefont {Zhu}, \citenamefont {Biswas}, \citenamefont {Cetina}, \citenamefont {Monroe},\ and\ \citenamefont {Brown}}]{PhysRevLett.127.240501}%
  \BibitemOpen
  \bibfield  {author} {\bibinfo {author} {\bibfnamefont {D.~M.}\ \bibnamefont {Debroy}}, \bibinfo {author} {\bibfnamefont {L.}~\bibnamefont {Egan}}, \bibinfo {author} {\bibfnamefont {C.}~\bibnamefont {Noel}}, \bibinfo {author} {\bibfnamefont {A.}~\bibnamefont {Risinger}}, \bibinfo {author} {\bibfnamefont {D.}~\bibnamefont {Zhu}}, \bibinfo {author} {\bibfnamefont {D.}~\bibnamefont {Biswas}}, \bibinfo {author} {\bibfnamefont {M.}~\bibnamefont {Cetina}}, \bibinfo {author} {\bibfnamefont {C.}~\bibnamefont {Monroe}}, \ and\ \bibinfo {author} {\bibfnamefont {K.~R.}\ \bibnamefont {Brown}},\ }\bibfield  {title} {\emph {\bibinfo {title} {Optimizing stabilizer parities for improved logical qubit memories},\ }}\href {\doibase 10.1103/PhysRevLett.127.240501} {\bibfield  {journal} {\bibinfo  {journal} {Phys. Rev. Lett.}\ }\textbf {\bibinfo {volume} {127}},\ \bibinfo {pages} {240501} (\bibinfo {year} {2021})}\BibitemShut {NoStop}%
\bibitem [{\citenamefont {Postler}\ \emph {et~al.}(2022)\citenamefont {Postler}, \citenamefont {Heu$\beta$en}, \citenamefont {Pogorelov}, \citenamefont {Rispler}, \citenamefont {Feldker}, \citenamefont {Meth}, \citenamefont {Marciniak}, \citenamefont {Stricker}, \citenamefont {Ringbauer}, \citenamefont {Blatt}, \citenamefont {Schindler}, \citenamefont {M{\"u}ller},\ and\ \citenamefont {Monz}}]{Postler2022}%
  \BibitemOpen
  \bibfield  {author} {\bibinfo {author} {\bibfnamefont {L.}~\bibnamefont {Postler}}, \bibinfo {author} {\bibfnamefont {S.}~\bibnamefont {Heu$\beta$en}}, \bibinfo {author} {\bibfnamefont {I.}~\bibnamefont {Pogorelov}}, \bibinfo {author} {\bibfnamefont {M.}~\bibnamefont {Rispler}}, \bibinfo {author} {\bibfnamefont {T.}~\bibnamefont {Feldker}}, \bibinfo {author} {\bibfnamefont {M.}~\bibnamefont {Meth}}, \bibinfo {author} {\bibfnamefont {C.~D.}\ \bibnamefont {Marciniak}}, \bibinfo {author} {\bibfnamefont {R.}~\bibnamefont {Stricker}}, \bibinfo {author} {\bibfnamefont {M.}~\bibnamefont {Ringbauer}}, \bibinfo {author} {\bibfnamefont {R.}~\bibnamefont {Blatt}}, \bibinfo {author} {\bibfnamefont {P.}~\bibnamefont {Schindler}}, \bibinfo {author} {\bibfnamefont {M.}~\bibnamefont {M{\"u}ller}}, \ and\ \bibinfo {author} {\bibfnamefont {T.}~\bibnamefont {Monz}},\ }\bibfield  {title} {\emph {\bibinfo {title} {Demonstration of fault-tolerant universal quantum gate operations},\ }}\href {\doibase 10.1038/s41586-022-04721-1}
  {\bibfield  {journal} {\bibinfo  {journal} {Nature}\ }\textbf {\bibinfo {volume} {605}},\ \bibinfo {pages} {675} (\bibinfo {year} {2022})}\BibitemShut {NoStop}%
\bibitem [{\citenamefont {Ryan-Anderson}\ \emph {et~al.}(2022)\citenamefont {Ryan-Anderson}, \citenamefont {Brown}, \citenamefont {Allman}, \citenamefont {Arkin}, \citenamefont {Asa-Attuah}, \citenamefont {Baldwin}, \citenamefont {Berg}, \citenamefont {Bohnet}, \citenamefont {Braxton}, \citenamefont {Burdick}, \citenamefont {Campora}, \citenamefont {Chernoguzov}, \citenamefont {Esposito}, \citenamefont {Evans}, \citenamefont {Francois}, \citenamefont {Gaebler}, \citenamefont {Gatterman}, \citenamefont {Gerber}, \citenamefont {Gilmore}, \citenamefont {Gresh}, \citenamefont {Hall}, \citenamefont {Hankin}, \citenamefont {Hostetter}, \citenamefont {Lucchetti}, \citenamefont {Mayer}, \citenamefont {Myers}, \citenamefont {Neyenhuis}, \citenamefont {Santiago}, \citenamefont {Sedlacek}, \citenamefont {Skripka}, \citenamefont {Slattery}, \citenamefont {Stutz}, \citenamefont {Tait}, \citenamefont {Tobey}, \citenamefont {Vittorini}, \citenamefont {Walker},\ and\ \citenamefont
  {Hayes}}]{https://doi.org/10.48550/arxiv.2208.01863}%
  \BibitemOpen
  \bibfield  {author} {\bibinfo {author} {\bibfnamefont {C.}~\bibnamefont {Ryan-Anderson}}, \bibinfo {author} {\bibfnamefont {N.~C.}\ \bibnamefont {Brown}}, \bibinfo {author} {\bibfnamefont {M.~S.}\ \bibnamefont {Allman}}, \bibinfo {author} {\bibfnamefont {B.}~\bibnamefont {Arkin}}, \bibinfo {author} {\bibfnamefont {G.}~\bibnamefont {Asa-Attuah}}, \bibinfo {author} {\bibfnamefont {C.}~\bibnamefont {Baldwin}}, \bibinfo {author} {\bibfnamefont {J.}~\bibnamefont {Berg}}, \bibinfo {author} {\bibfnamefont {J.~G.}\ \bibnamefont {Bohnet}}, \bibinfo {author} {\bibfnamefont {S.}~\bibnamefont {Braxton}}, \bibinfo {author} {\bibfnamefont {N.}~\bibnamefont {Burdick}}, \bibinfo {author} {\bibfnamefont {J.~P.}\ \bibnamefont {Campora}}, \bibinfo {author} {\bibfnamefont {A.}~\bibnamefont {Chernoguzov}}, \bibinfo {author} {\bibfnamefont {J.}~\bibnamefont {Esposito}}, \bibinfo {author} {\bibfnamefont {B.}~\bibnamefont {Evans}}, \bibinfo {author} {\bibfnamefont {D.}~\bibnamefont {Francois}}, \bibinfo {author} {\bibfnamefont
  {J.~P.}\ \bibnamefont {Gaebler}}, \bibinfo {author} {\bibfnamefont {T.~M.}\ \bibnamefont {Gatterman}}, \bibinfo {author} {\bibfnamefont {J.}~\bibnamefont {Gerber}}, \bibinfo {author} {\bibfnamefont {K.}~\bibnamefont {Gilmore}}, \bibinfo {author} {\bibfnamefont {D.}~\bibnamefont {Gresh}}, \bibinfo {author} {\bibfnamefont {A.}~\bibnamefont {Hall}}, \bibinfo {author} {\bibfnamefont {A.}~\bibnamefont {Hankin}}, \bibinfo {author} {\bibfnamefont {J.}~\bibnamefont {Hostetter}}, \bibinfo {author} {\bibfnamefont {D.}~\bibnamefont {Lucchetti}}, \bibinfo {author} {\bibfnamefont {K.}~\bibnamefont {Mayer}}, \bibinfo {author} {\bibfnamefont {J.}~\bibnamefont {Myers}}, \bibinfo {author} {\bibfnamefont {B.}~\bibnamefont {Neyenhuis}}, \bibinfo {author} {\bibfnamefont {J.}~\bibnamefont {Santiago}}, \bibinfo {author} {\bibfnamefont {J.}~\bibnamefont {Sedlacek}}, \bibinfo {author} {\bibfnamefont {T.}~\bibnamefont {Skripka}}, \bibinfo {author} {\bibfnamefont {A.}~\bibnamefont {Slattery}}, \bibinfo {author} {\bibfnamefont
  {R.~P.}\ \bibnamefont {Stutz}}, \bibinfo {author} {\bibfnamefont {J.}~\bibnamefont {Tait}}, \bibinfo {author} {\bibfnamefont {R.}~\bibnamefont {Tobey}}, \bibinfo {author} {\bibfnamefont {G.}~\bibnamefont {Vittorini}}, \bibinfo {author} {\bibfnamefont {J.}~\bibnamefont {Walker}}, \ and\ \bibinfo {author} {\bibfnamefont {D.}~\bibnamefont {Hayes}},\ }\href {\doibase 10.48550/ARXIV.2208.01863} {\bibinfo {title} {Implementing fault-tolerant entangling gates on the five-qubit code and the color code},\ } (\bibinfo {year} {2022})\BibitemShut {NoStop}%
\bibitem [{\citenamefont {Bombin}\ and\ \citenamefont {Martin-Delgado}(2006)}]{PhysRevLett.97.180501}%
  \BibitemOpen
  \bibfield  {author} {\bibinfo {author} {\bibfnamefont {H.}~\bibnamefont {Bombin}}\ and\ \bibinfo {author} {\bibfnamefont {M.~A.}\ \bibnamefont {Martin-Delgado}},\ }\bibfield  {title} {\emph {\bibinfo {title} {Topological quantum distillation},\ }}\href {\doibase 10.1103/PhysRevLett.97.180501} {\bibfield  {journal} {\bibinfo  {journal} {Phys. Rev. Lett.}\ }\textbf {\bibinfo {volume} {97}},\ \bibinfo {pages} {180501} (\bibinfo {year} {2006})}\BibitemShut {NoStop}%
\bibitem [{\citenamefont {Chao}\ and\ \citenamefont {Reichardt}(2018)}]{PhysRevLett.121.050502}%
  \BibitemOpen
  \bibfield  {author} {\bibinfo {author} {\bibfnamefont {R.}~\bibnamefont {Chao}}\ and\ \bibinfo {author} {\bibfnamefont {B.~W.}\ \bibnamefont {Reichardt}},\ }\bibfield  {title} {\emph {\bibinfo {title} {Quantum error correction with only two extra qubits},\ }}\href {\doibase 10.1103/PhysRevLett.121.050502} {\bibfield  {journal} {\bibinfo  {journal} {Phys. Rev. Lett.}\ }\textbf {\bibinfo {volume} {121}},\ \bibinfo {pages} {050502} (\bibinfo {year} {2018})}\BibitemShut {NoStop}%
\bibitem [{\citenamefont {Chamberland}\ and\ \citenamefont {Beverland}(2018)}]{Chamberland2018flagfaulttolerant}%
  \BibitemOpen
  \bibfield  {author} {\bibinfo {author} {\bibfnamefont {C.}~\bibnamefont {Chamberland}}\ and\ \bibinfo {author} {\bibfnamefont {M.~E.}\ \bibnamefont {Beverland}},\ }\bibfield  {title} {\emph {\bibinfo {title} {Flag fault-tolerant error correction with arbitrary distance codes},\ }}\href {\doibase 10.22331/q-2018-02-08-53} {\bibfield  {journal} {\bibinfo  {journal} {{Quantum}}\ }\textbf {\bibinfo {volume} {2}},\ \bibinfo {pages} {53} (\bibinfo {year} {2018})}\BibitemShut {NoStop}%
\bibitem [{\citenamefont {Chao}\ and\ \citenamefont {Reichardt}(2020{\natexlab{a}})}]{Chao2020}%
  \BibitemOpen
  \bibfield  {author} {\bibinfo {author} {\bibfnamefont {R.}~\bibnamefont {Chao}}\ and\ \bibinfo {author} {\bibfnamefont {B.~W.}\ \bibnamefont {Reichardt}},\ }\bibfield  {title} {\emph {\bibinfo {title} {Flag fault-tolerant error correction for any stabilizer code},\ }}\href {\doibase 10.1103/PRXQuantum.1.010302} {\bibfield  {journal} {\bibinfo  {journal} {PRX Quantum}\ }\textbf {\bibinfo {volume} {1}},\ \bibinfo {pages} {010302} (\bibinfo {year} {2020}{\natexlab{a}})}\BibitemShut {NoStop}%
\bibitem [{\citenamefont {Steane}(1996{\natexlab{b}})}]{doi:10.1098/rspa.1996.0136}%
  \BibitemOpen
  \bibfield  {author} {\bibinfo {author} {\bibfnamefont {A.}~\bibnamefont {Steane}},\ }\bibfield  {title} {\emph {\bibinfo {title} {Multiple-particle interference and quantum error correction},\ }}\href {\doibase 10.1098/rspa.1996.0136} {\bibfield  {journal} {\bibinfo  {journal} {Proceedings of the Royal Society of London. Series A: Mathematical, Physical and Engineering Sciences}\ }\textbf {\bibinfo {volume} {452}},\ \bibinfo {pages} {2551} (\bibinfo {year} {1996}{\natexlab{b}})},\ \Eprint {http://arxiv.org/abs/https://royalsocietypublishing.org/doi/pdf/10.1098/rspa.1996.0136} {https://royalsocietypublishing.org/doi/pdf/10.1098/rspa.1996.0136} \BibitemShut {NoStop}%
\bibitem [{\citenamefont {Sackett}\ \emph {et~al.}(2000)\citenamefont {Sackett}, \citenamefont {Kielpinski}, \citenamefont {King}, \citenamefont {Langer}, \citenamefont {Meyer}, \citenamefont {Myatt}, \citenamefont {Rowe}, \citenamefont {Turchette}, \citenamefont {Itano}, \citenamefont {Wineland},\ and\ \citenamefont {Monroe}}]{Sackett2000}%
  \BibitemOpen
  \bibfield  {author} {\bibinfo {author} {\bibfnamefont {C.~A.}\ \bibnamefont {Sackett}}, \bibinfo {author} {\bibfnamefont {D.}~\bibnamefont {Kielpinski}}, \bibinfo {author} {\bibfnamefont {B.~E.}\ \bibnamefont {King}}, \bibinfo {author} {\bibfnamefont {C.}~\bibnamefont {Langer}}, \bibinfo {author} {\bibfnamefont {V.}~\bibnamefont {Meyer}}, \bibinfo {author} {\bibfnamefont {C.~J.}\ \bibnamefont {Myatt}}, \bibinfo {author} {\bibfnamefont {M.}~\bibnamefont {Rowe}}, \bibinfo {author} {\bibfnamefont {Q.~A.}\ \bibnamefont {Turchette}}, \bibinfo {author} {\bibfnamefont {W.~M.}\ \bibnamefont {Itano}}, \bibinfo {author} {\bibfnamefont {D.~J.}\ \bibnamefont {Wineland}}, \ and\ \bibinfo {author} {\bibfnamefont {C.}~\bibnamefont {Monroe}},\ }\bibfield  {title} {\emph {\bibinfo {title} {Experimental entanglement of four particles},\ }}\href {\doibase 10.1038/35005011} {\bibfield  {journal} {\bibinfo  {journal} {Nature}\ }\textbf {\bibinfo {volume} {404}},\ \bibinfo {pages} {256} (\bibinfo {year} {2000})}\BibitemShut
  {NoStop}%
\bibitem [{\citenamefont {Benhelm}\ \emph {et~al.}(2008)\citenamefont {Benhelm}, \citenamefont {Kirchmair}, \citenamefont {Roos},\ and\ \citenamefont {Blatt}}]{Benhelm2008}%
  \BibitemOpen
  \bibfield  {author} {\bibinfo {author} {\bibfnamefont {J.}~\bibnamefont {Benhelm}}, \bibinfo {author} {\bibfnamefont {G.}~\bibnamefont {Kirchmair}}, \bibinfo {author} {\bibfnamefont {C.~F.}\ \bibnamefont {Roos}}, \ and\ \bibinfo {author} {\bibfnamefont {R.}~\bibnamefont {Blatt}},\ }\bibfield  {title} {\emph {\bibinfo {title} {Towards fault-tolerant quantum computing with trapped ions},\ }}\href {\doibase 10.1038/nphys961} {\bibfield  {journal} {\bibinfo  {journal} {Nature Physics}\ }\textbf {\bibinfo {volume} {4}},\ \bibinfo {pages} {463} (\bibinfo {year} {2008})}\BibitemShut {NoStop}%
\bibitem [{\citenamefont {Poschinger}\ \emph {et~al.}(2009)\citenamefont {Poschinger}, \citenamefont {Huber}, \citenamefont {Ziesel}, \citenamefont {Dei{\ss}}, \citenamefont {Hettrich}, \citenamefont {Schulz}, \citenamefont {Singer}, \citenamefont {Poulsen}, \citenamefont {Drewsen}, \citenamefont {Hendricks},\ and\ \citenamefont {Schmidt-Kaler}}]{Poschinger_2009}%
  \BibitemOpen
  \bibfield  {author} {\bibinfo {author} {\bibfnamefont {U.~G.}\ \bibnamefont {Poschinger}}, \bibinfo {author} {\bibfnamefont {G.}~\bibnamefont {Huber}}, \bibinfo {author} {\bibfnamefont {F.}~\bibnamefont {Ziesel}}, \bibinfo {author} {\bibfnamefont {M.}~\bibnamefont {Dei{\ss}}}, \bibinfo {author} {\bibfnamefont {M.}~\bibnamefont {Hettrich}}, \bibinfo {author} {\bibfnamefont {S.~A.}\ \bibnamefont {Schulz}}, \bibinfo {author} {\bibfnamefont {K.}~\bibnamefont {Singer}}, \bibinfo {author} {\bibfnamefont {G.}~\bibnamefont {Poulsen}}, \bibinfo {author} {\bibfnamefont {M.}~\bibnamefont {Drewsen}}, \bibinfo {author} {\bibfnamefont {R.~J.}\ \bibnamefont {Hendricks}}, \ and\ \bibinfo {author} {\bibfnamefont {F.}~\bibnamefont {Schmidt-Kaler}},\ }\bibfield  {title} {\emph {\bibinfo {title} {Coherent manipulation of a \ce{^{40}Ca+} spin qubit in a micro ion trap},\ }}\href {\doibase 10.1088/0953-4075/42/15/154013} {\bibfield  {journal} {\bibinfo  {journal} {Journal of Physics B: Atomic, Molecular and Optical Physics}\
  }\textbf {\bibinfo {volume} {42}},\ \bibinfo {pages} {154013} (\bibinfo {year} {2009})}\BibitemShut {NoStop}%
\bibitem [{\citenamefont {Leibfried}\ \emph {et~al.}(2003{\natexlab{b}})\citenamefont {Leibfried}, \citenamefont {Blatt}, \citenamefont {Monroe},\ and\ \citenamefont {Wineland}}]{RevModPhys.75.281}%
  \BibitemOpen
  \bibfield  {author} {\bibinfo {author} {\bibfnamefont {D.}~\bibnamefont {Leibfried}}, \bibinfo {author} {\bibfnamefont {R.}~\bibnamefont {Blatt}}, \bibinfo {author} {\bibfnamefont {C.}~\bibnamefont {Monroe}}, \ and\ \bibinfo {author} {\bibfnamefont {D.}~\bibnamefont {Wineland}},\ }\bibfield  {title} {\emph {\bibinfo {title} {Quantum dynamics of single trapped ions},\ }}\href {\doibase 10.1103/RevModPhys.75.281} {\bibfield  {journal} {\bibinfo  {journal} {Rev. Mod. Phys.}\ }\textbf {\bibinfo {volume} {75}},\ \bibinfo {pages} {281} (\bibinfo {year} {2003}{\natexlab{b}})}\BibitemShut {NoStop}%
\bibitem [{\citenamefont {Nebendahl}\ \emph {et~al.}(2009)\citenamefont {Nebendahl}, \citenamefont {H\"affner},\ and\ \citenamefont {Roos}}]{Nebendahl2009}%
  \BibitemOpen
  \bibfield  {author} {\bibinfo {author} {\bibfnamefont {V.}~\bibnamefont {Nebendahl}}, \bibinfo {author} {\bibfnamefont {H.}~\bibnamefont {H\"affner}}, \ and\ \bibinfo {author} {\bibfnamefont {C.~F.}\ \bibnamefont {Roos}},\ }\bibfield  {title} {\emph {\bibinfo {title} {Optimal control of entangling operations for trapped-ion quantum computing},\ }}\href {\doibase 10.1103/PhysRevA.79.012312} {\bibfield  {journal} {\bibinfo  {journal} {Phys. Rev. A}\ }\textbf {\bibinfo {volume} {79}},\ \bibinfo {pages} {012312} (\bibinfo {year} {2009})}\BibitemShut {NoStop}%
\bibitem [{\citenamefont {Schmidt-Kaler}\ \emph {et~al.}(2003)\citenamefont {Schmidt-Kaler}, \citenamefont {H{\"{a}}ffner}, \citenamefont {Riebe}, \citenamefont {Gulde}, \citenamefont {Lancaster}, \citenamefont {Deuschle}, \citenamefont {Becher}, \citenamefont {Roos}, \citenamefont {Eschner},\ and\ \citenamefont {Blatt}}]{SchmidtKaler2003}%
  \BibitemOpen
  \bibfield  {author} {\bibinfo {author} {\bibfnamefont {F.}~\bibnamefont {Schmidt-Kaler}}, \bibinfo {author} {\bibfnamefont {H.}~\bibnamefont {H{\"{a}}ffner}}, \bibinfo {author} {\bibfnamefont {M.}~\bibnamefont {Riebe}}, \bibinfo {author} {\bibfnamefont {S.}~\bibnamefont {Gulde}}, \bibinfo {author} {\bibfnamefont {G.~P.~T.}\ \bibnamefont {Lancaster}}, \bibinfo {author} {\bibfnamefont {T.}~\bibnamefont {Deuschle}}, \bibinfo {author} {\bibfnamefont {C.}~\bibnamefont {Becher}}, \bibinfo {author} {\bibfnamefont {C.~F.}\ \bibnamefont {Roos}}, \bibinfo {author} {\bibfnamefont {J.}~\bibnamefont {Eschner}}, \ and\ \bibinfo {author} {\bibfnamefont {R.}~\bibnamefont {Blatt}},\ }\bibfield  {title} {\emph {\bibinfo {title} {{Realization of the Cirac–Zoller controlled-NOT quantum gate}},\ }}\href {\doibase 10.1038/nature01494} {\bibfield  {journal} {\bibinfo  {journal} {Nature}\ }\textbf {\bibinfo {volume} {422}},\ \bibinfo {pages} {408} (\bibinfo {year} {2003})}\BibitemShut {NoStop}%
\bibitem [{\citenamefont {DiVincenzo}\ and\ \citenamefont {Shor}(1996)}]{PhysRevLett.77.3260}%
  \BibitemOpen
  \bibfield  {author} {\bibinfo {author} {\bibfnamefont {D.~P.}\ \bibnamefont {DiVincenzo}}\ and\ \bibinfo {author} {\bibfnamefont {P.~W.}\ \bibnamefont {Shor}},\ }\bibfield  {title} {\emph {\bibinfo {title} {Fault-tolerant error correction with efficient quantum codes},\ }}\href {\doibase 10.1103/PhysRevLett.77.3260} {\bibfield  {journal} {\bibinfo  {journal} {Phys. Rev. Lett.}\ }\textbf {\bibinfo {volume} {77}},\ \bibinfo {pages} {3260} (\bibinfo {year} {1996})}\BibitemShut {NoStop}%
\bibitem [{\citenamefont {DiVincenzo}\ and\ \citenamefont {Aliferis}(2007)}]{PhysRevLett.98.020501}%
  \BibitemOpen
  \bibfield  {author} {\bibinfo {author} {\bibfnamefont {D.~P.}\ \bibnamefont {DiVincenzo}}\ and\ \bibinfo {author} {\bibfnamefont {P.}~\bibnamefont {Aliferis}},\ }\bibfield  {title} {\emph {\bibinfo {title} {Effective fault-tolerant quantum computation with slow measurements},\ }}\href {\doibase 10.1103/PhysRevLett.98.020501} {\bibfield  {journal} {\bibinfo  {journal} {Phys. Rev. Lett.}\ }\textbf {\bibinfo {volume} {98}},\ \bibinfo {pages} {020501} (\bibinfo {year} {2007})}\BibitemShut {NoStop}%
\bibitem [{\citenamefont {Chamberland}\ and\ \citenamefont {Cross}(2019)}]{Chamberland2019faulttolerantmagic}%
  \BibitemOpen
  \bibfield  {author} {\bibinfo {author} {\bibfnamefont {C.}~\bibnamefont {Chamberland}}\ and\ \bibinfo {author} {\bibfnamefont {A.~W.}\ \bibnamefont {Cross}},\ }\bibfield  {title} {\emph {\bibinfo {title} {Fault-tolerant magic state preparation with flag qubits},\ }}\href {\doibase 10.22331/q-2019-05-20-143} {\bibfield  {journal} {\bibinfo  {journal} {{Quantum}}\ }\textbf {\bibinfo {volume} {3}},\ \bibinfo {pages} {143} (\bibinfo {year} {2019})}\BibitemShut {NoStop}%
\bibitem [{\citenamefont {Reichardt}(2020)}]{Reichardt_2020}%
  \BibitemOpen
  \bibfield  {author} {\bibinfo {author} {\bibfnamefont {B.~W.}\ \bibnamefont {Reichardt}},\ }\bibfield  {title} {\emph {\bibinfo {title} {Fault-tolerant quantum error correction for steane's seven-qubit color code with few or no extra qubits},\ }}\href {\doibase 10.1088/2058-9565/abc6f4} {\bibfield  {journal} {\bibinfo  {journal} {Quantum Science and Technology}\ }\textbf {\bibinfo {volume} {6}},\ \bibinfo {pages} {015007} (\bibinfo {year} {2020})}\BibitemShut {NoStop}%
\bibitem [{\citenamefont {Tansuwannont}\ \emph {et~al.}(2020)\citenamefont {Tansuwannont}, \citenamefont {Chamberland},\ and\ \citenamefont {Leung}}]{PhysRevA.101.012342}%
  \BibitemOpen
  \bibfield  {author} {\bibinfo {author} {\bibfnamefont {T.}~\bibnamefont {Tansuwannont}}, \bibinfo {author} {\bibfnamefont {C.}~\bibnamefont {Chamberland}}, \ and\ \bibinfo {author} {\bibfnamefont {D.}~\bibnamefont {Leung}},\ }\bibfield  {title} {\emph {\bibinfo {title} {Flag fault-tolerant error correction, measurement, and quantum computation for cyclic calderbank-shor-steane codes},\ }}\href {\doibase 10.1103/PhysRevA.101.012342} {\bibfield  {journal} {\bibinfo  {journal} {Phys. Rev. A}\ }\textbf {\bibinfo {volume} {101}},\ \bibinfo {pages} {012342} (\bibinfo {year} {2020})}\BibitemShut {NoStop}%
\bibitem [{\citenamefont {Chamberland}\ \emph {et~al.}(2020)\citenamefont {Chamberland}, \citenamefont {Kubica}, \citenamefont {Yoder},\ and\ \citenamefont {Zhu}}]{Chamberland_2020}%
  \BibitemOpen
  \bibfield  {author} {\bibinfo {author} {\bibfnamefont {C.}~\bibnamefont {Chamberland}}, \bibinfo {author} {\bibfnamefont {A.}~\bibnamefont {Kubica}}, \bibinfo {author} {\bibfnamefont {T.~J.}\ \bibnamefont {Yoder}}, \ and\ \bibinfo {author} {\bibfnamefont {G.}~\bibnamefont {Zhu}},\ }\bibfield  {title} {\emph {\bibinfo {title} {Triangular color codes on trivalent graphs with flag qubits},\ }}\href {\doibase 10.1088/1367-2630/ab68fd} {\bibfield  {journal} {\bibinfo  {journal} {New Journal of Physics}\ }\textbf {\bibinfo {volume} {22}},\ \bibinfo {pages} {023019} (\bibinfo {year} {2020})}\BibitemShut {NoStop}%
\bibitem [{\citenamefont {Chao}\ and\ \citenamefont {Reichardt}(2020{\natexlab{b}})}]{PRXQuantum.1.010302}%
  \BibitemOpen
  \bibfield  {author} {\bibinfo {author} {\bibfnamefont {R.}~\bibnamefont {Chao}}\ and\ \bibinfo {author} {\bibfnamefont {B.~W.}\ \bibnamefont {Reichardt}},\ }\bibfield  {title} {\emph {\bibinfo {title} {Flag fault-tolerant error correction for any stabilizer code},\ }}\href {\doibase 10.1103/PRXQuantum.1.010302} {\bibfield  {journal} {\bibinfo  {journal} {PRX Quantum}\ }\textbf {\bibinfo {volume} {1}},\ \bibinfo {pages} {010302} (\bibinfo {year} {2020}{\natexlab{b}})}\BibitemShut {NoStop}%
\bibitem [{\citenamefont {Horodecki}\ \emph {et~al.}(1996)\citenamefont {Horodecki}, \citenamefont {Horodecki},\ and\ \citenamefont {Horodecki}}]{HORODECKI19961}%
  \BibitemOpen
  \bibfield  {author} {\bibinfo {author} {\bibfnamefont {M.}~\bibnamefont {Horodecki}}, \bibinfo {author} {\bibfnamefont {P.}~\bibnamefont {Horodecki}}, \ and\ \bibinfo {author} {\bibfnamefont {R.}~\bibnamefont {Horodecki}},\ }\bibfield  {title} {\emph {\bibinfo {title} {Separability of mixed states: necessary and sufficient conditions},\ }}\href {\doibase https://doi.org/10.1016/S0375-9601(96)00706-2} {\bibfield  {journal} {\bibinfo  {journal} {Physics Letters A}\ }\textbf {\bibinfo {volume} {223}},\ \bibinfo {pages} {1} (\bibinfo {year} {1996})}\BibitemShut {NoStop}%
\bibitem [{\citenamefont {Terhal}(2000)}]{Terhal2000}%
  \BibitemOpen
  \bibfield  {author} {\bibinfo {author} {\bibfnamefont {B.~M.}\ \bibnamefont {Terhal}},\ }\bibfield  {title} {\emph {\bibinfo {title} {{Bell inequalities and the separability criterion}},\ }}\href {\doibase 10.1016/S0375-9601(00)00401-1} {\bibfield  {journal} {\bibinfo  {journal} {Physics Letters A}\ }\textbf {\bibinfo {volume} {271}},\ \bibinfo {pages} {319} (\bibinfo {year} {2000})}\BibitemShut {NoStop}%
\bibitem [{\citenamefont {Lewenstein}\ \emph {et~al.}(2000)\citenamefont {Lewenstein}, \citenamefont {Kraus}, \citenamefont {Cirac},\ and\ \citenamefont {Horodecki}}]{Lewenstein2000}%
  \BibitemOpen
  \bibfield  {author} {\bibinfo {author} {\bibfnamefont {M.}~\bibnamefont {Lewenstein}}, \bibinfo {author} {\bibfnamefont {B.}~\bibnamefont {Kraus}}, \bibinfo {author} {\bibfnamefont {J.~I.}\ \bibnamefont {Cirac}}, \ and\ \bibinfo {author} {\bibfnamefont {P.}~\bibnamefont {Horodecki}},\ }\bibfield  {title} {\emph {\bibinfo {title} {{Optimization of entanglement witnesses}},\ }}\href {\doibase 10.1103/PhysRevA.62.052310} {\bibfield  {journal} {\bibinfo  {journal} {Phys. Rev. A}\ }\textbf {\bibinfo {volume} {62}},\ \bibinfo {pages} {052310} (\bibinfo {year} {2000})}\BibitemShut {NoStop}%
\bibitem [{\citenamefont {Sperling}\ and\ \citenamefont {Vogel}(2013)}]{Sperling2013}%
  \BibitemOpen
  \bibfield  {author} {\bibinfo {author} {\bibfnamefont {J.}~\bibnamefont {Sperling}}\ and\ \bibinfo {author} {\bibfnamefont {W.}~\bibnamefont {Vogel}},\ }\bibfield  {title} {\emph {\bibinfo {title} {{Multipartite Entanglement Witnesses}},\ }}\href {\doibase 10.1103/PhysRevLett.111.110503} {\bibfield  {journal} {\bibinfo  {journal} {Phys. Rev. Lett.}\ }\textbf {\bibinfo {volume} {111}},\ \bibinfo {pages} {110503} (\bibinfo {year} {2013})}\BibitemShut {NoStop}%
\bibitem [{\citenamefont {Horodecki}\ \emph {et~al.}(2009)\citenamefont {Horodecki}, \citenamefont {Horodecki}, \citenamefont {Horodecki},\ and\ \citenamefont {Horodecki}}]{Horodecki2009}%
  \BibitemOpen
  \bibfield  {author} {\bibinfo {author} {\bibfnamefont {R.}~\bibnamefont {Horodecki}}, \bibinfo {author} {\bibfnamefont {P.}~\bibnamefont {Horodecki}}, \bibinfo {author} {\bibfnamefont {M.}~\bibnamefont {Horodecki}}, \ and\ \bibinfo {author} {\bibfnamefont {K.}~\bibnamefont {Horodecki}},\ }\bibfield  {title} {\emph {\bibinfo {title} {Quantum entanglement},\ }}\href {\doibase 10.1103/RevModPhys.81.865} {\bibfield  {journal} {\bibinfo  {journal} {Rev. Mod. Phys.}\ }\textbf {\bibinfo {volume} {81}},\ \bibinfo {pages} {865} (\bibinfo {year} {2009})}\BibitemShut {NoStop}%
\bibitem [{\citenamefont {T\'oth}\ and\ \citenamefont {G\"uhne}(2005{\natexlab{a}})}]{Toth2005}%
  \BibitemOpen
  \bibfield  {author} {\bibinfo {author} {\bibfnamefont {G.}~\bibnamefont {T\'oth}}\ and\ \bibinfo {author} {\bibfnamefont {O.}~\bibnamefont {G\"uhne}},\ }\bibfield  {title} {\emph {\bibinfo {title} {Entanglement detection in the stabilizer formalism},\ }}\href {\doibase 10.1103/PhysRevA.72.022340} {\bibfield  {journal} {\bibinfo  {journal} {Phys. Rev. A}\ }\textbf {\bibinfo {volume} {72}},\ \bibinfo {pages} {022340} (\bibinfo {year} {2005}{\natexlab{a}})}\BibitemShut {NoStop}%
\bibitem [{\citenamefont {T\'oth}\ and\ \citenamefont {G\"uhne}(2005{\natexlab{b}})}]{PhysRevLett.94.060501}%
  \BibitemOpen
  \bibfield  {author} {\bibinfo {author} {\bibfnamefont {G.}~\bibnamefont {T\'oth}}\ and\ \bibinfo {author} {\bibfnamefont {O.}~\bibnamefont {G\"uhne}},\ }\bibfield  {title} {\emph {\bibinfo {title} {Detecting genuine multipartite entanglement with two local measurements},\ }}\href {\doibase 10.1103/PhysRevLett.94.060501} {\bibfield  {journal} {\bibinfo  {journal} {Phys. Rev. Lett.}\ }\textbf {\bibinfo {volume} {94}},\ \bibinfo {pages} {060501} (\bibinfo {year} {2005}{\natexlab{b}})}\BibitemShut {NoStop}%
\bibitem [{\citenamefont {Werner}(1989)}]{PhysRevA.40.4277}%
  \BibitemOpen
  \bibfield  {author} {\bibinfo {author} {\bibfnamefont {R.~F.}\ \bibnamefont {Werner}},\ }\bibfield  {title} {\emph {\bibinfo {title} {Quantum states with einstein-podolsky-rosen correlations admitting a hidden-variable model},\ }}\href {\doibase 10.1103/PhysRevA.40.4277} {\bibfield  {journal} {\bibinfo  {journal} {Phys. Rev. A}\ }\textbf {\bibinfo {volume} {40}},\ \bibinfo {pages} {4277} (\bibinfo {year} {1989})}\BibitemShut {NoStop}%
\bibitem [{\citenamefont {Devitt}\ \emph {et~al.}(2013)\citenamefont {Devitt}, \citenamefont {Munro},\ and\ \citenamefont {Nemoto}}]{Devitt_2013}%
  \BibitemOpen
  \bibfield  {author} {\bibinfo {author} {\bibfnamefont {S.~J.}\ \bibnamefont {Devitt}}, \bibinfo {author} {\bibfnamefont {W.~J.}\ \bibnamefont {Munro}}, \ and\ \bibinfo {author} {\bibfnamefont {K.}~\bibnamefont {Nemoto}},\ }\bibfield  {title} {\emph {\bibinfo {title} {Quantum error correction for beginners},\ }}\href {\doibase 10.1088/0034-4885/76/7/076001} {\bibfield  {journal} {\bibinfo  {journal} {Reports on Progress in Physics}\ }\textbf {\bibinfo {volume} {76}},\ \bibinfo {pages} {076001} (\bibinfo {year} {2013})}\BibitemShut {NoStop}%
\bibitem [{\citenamefont {G{\"{u}}hne}\ and\ \citenamefont {T{\'{o}}th}(2009)}]{Guhne2009}%
  \BibitemOpen
  \bibfield  {author} {\bibinfo {author} {\bibfnamefont {O.}~\bibnamefont {G{\"{u}}hne}}\ and\ \bibinfo {author} {\bibfnamefont {G.}~\bibnamefont {T{\'{o}}th}},\ }\bibfield  {title} {\emph {\bibinfo {title} {{Entanglement detection}},\ }}\href {\doibase 10.1016/j.physrep.2009.02.004} {\bibfield  {journal} {\bibinfo  {journal} {Physics Reports}\ }\textbf {\bibinfo {volume} {474}},\ \bibinfo {pages} {1} (\bibinfo {year} {2009})}\BibitemShut {NoStop}%
\bibitem [{\citenamefont {Huber}\ \emph {et~al.}(2010)\citenamefont {Huber}, \citenamefont {Mintert}, \citenamefont {Gabriel},\ and\ \citenamefont {Hiesmayr}}]{Huber2010}%
  \BibitemOpen
  \bibfield  {author} {\bibinfo {author} {\bibfnamefont {M.}~\bibnamefont {Huber}}, \bibinfo {author} {\bibfnamefont {F.}~\bibnamefont {Mintert}}, \bibinfo {author} {\bibfnamefont {A.}~\bibnamefont {Gabriel}}, \ and\ \bibinfo {author} {\bibfnamefont {B.~C.}\ \bibnamefont {Hiesmayr}},\ }\bibfield  {title} {\emph {\bibinfo {title} {Detection of high-dimensional genuine multipartite entanglement of mixed states},\ }}\href {\doibase 10.1103/PhysRevLett.104.210501} {\bibfield  {journal} {\bibinfo  {journal} {Phys. Rev. Lett.}\ }\textbf {\bibinfo {volume} {104}},\ \bibinfo {pages} {210501} (\bibinfo {year} {2010})}\BibitemShut {NoStop}%
\bibitem [{\citenamefont {Verstraete}\ \emph {et~al.}(2004)\citenamefont {Verstraete}, \citenamefont {Popp},\ and\ \citenamefont {Cirac}}]{Verstraete2004}%
  \BibitemOpen
  \bibfield  {author} {\bibinfo {author} {\bibfnamefont {F.}~\bibnamefont {Verstraete}}, \bibinfo {author} {\bibfnamefont {M.}~\bibnamefont {Popp}}, \ and\ \bibinfo {author} {\bibfnamefont {J.~I.}\ \bibnamefont {Cirac}},\ }\bibfield  {title} {\emph {\bibinfo {title} {{Entanglement versus Correlations in Spin Systems}},\ }}\href {\doibase 10.1103/PhysRevLett.92.027901} {\bibfield  {journal} {\bibinfo  {journal} {Phys. Rev. Lett.}\ }\textbf {\bibinfo {volume} {92}},\ \bibinfo {pages} {027901} (\bibinfo {year} {2004})}\BibitemShut {NoStop}%
\bibitem [{\citenamefont {Popp}\ \emph {et~al.}(2005)\citenamefont {Popp}, \citenamefont {Verstraete}, \citenamefont {Mart{\'{i}}n-Delgado},\ and\ \citenamefont {Cirac}}]{Popp2005}%
  \BibitemOpen
  \bibfield  {author} {\bibinfo {author} {\bibfnamefont {M.}~\bibnamefont {Popp}}, \bibinfo {author} {\bibfnamefont {F.}~\bibnamefont {Verstraete}}, \bibinfo {author} {\bibfnamefont {M.~A.}\ \bibnamefont {Mart{\'{i}}n-Delgado}}, \ and\ \bibinfo {author} {\bibfnamefont {J.~I.}\ \bibnamefont {Cirac}},\ }\bibfield  {title} {\emph {\bibinfo {title} {{Localizable entanglement}},\ }}\href {\doibase 10.1103/PhysRevA.71.042306} {\bibfield  {journal} {\bibinfo  {journal} {Phys. Rev. A}\ }\textbf {\bibinfo {volume} {71}},\ \bibinfo {pages} {042306} (\bibinfo {year} {2005})}\BibitemShut {NoStop}%
\bibitem [{\citenamefont {James}(1998)}]{James1998}%
  \BibitemOpen
  \bibfield  {author} {\bibinfo {author} {\bibfnamefont {D.~F.~V.}\ \bibnamefont {James}},\ }\bibfield  {title} {\emph {\bibinfo {title} {Quantum dynamics of cold trapped ions with application to quantum computation},\ }}\href {\doibase 10.1007/s003400050373} {\bibfield  {journal} {\bibinfo  {journal} {Applied Physics B}\ }\textbf {\bibinfo {volume} {66}},\ \bibinfo {pages} {181} (\bibinfo {year} {1998})}\BibitemShut {NoStop}%
\bibitem [{\citenamefont {Bermudez}\ \emph {et~al.}(2017{\natexlab{b}})\citenamefont {Bermudez}, \citenamefont {Schindler}, \citenamefont {Monz}, \citenamefont {Blatt},\ and\ \citenamefont {Müller}}]{Bermudez_2017}%
  \BibitemOpen
  \bibfield  {author} {\bibinfo {author} {\bibfnamefont {A.}~\bibnamefont {Bermudez}}, \bibinfo {author} {\bibfnamefont {P.}~\bibnamefont {Schindler}}, \bibinfo {author} {\bibfnamefont {T.}~\bibnamefont {Monz}}, \bibinfo {author} {\bibfnamefont {R.}~\bibnamefont {Blatt}}, \ and\ \bibinfo {author} {\bibfnamefont {M.}~\bibnamefont {Müller}},\ }\bibfield  {title} {\emph {\bibinfo {title} {Micromotion-enabled improvement of quantum logic gates with trapped ions},\ }}\href {\doibase 10.1088/1367-2630/aa86eb} {\bibfield  {journal} {\bibinfo  {journal} {New Journal of Physics}\ }\textbf {\bibinfo {volume} {19}},\ \bibinfo {pages} {113038} (\bibinfo {year} {2017}{\natexlab{b}})}\BibitemShut {NoStop}%
\bibitem [{\citenamefont {Magnus}(1954)}]{https://doi.org/10.1002/cpa.3160070404}%
  \BibitemOpen
  \bibfield  {author} {\bibinfo {author} {\bibfnamefont {W.}~\bibnamefont {Magnus}},\ }\bibfield  {title} {\emph {\bibinfo {title} {On the exponential solution of differential equations for a linear operator},\ }}\href {\doibase https://doi.org/10.1002/cpa.3160070404} {\bibfield  {journal} {\bibinfo  {journal} {Communications on Pure and Applied Mathematics}\ }\textbf {\bibinfo {volume} {7}},\ \bibinfo {pages} {649} (\bibinfo {year} {1954})},\ \Eprint {http://arxiv.org/abs/https://onlinelibrary.wiley.com/doi/pdf/10.1002/cpa.3160070404} {https://onlinelibrary.wiley.com/doi/pdf/10.1002/cpa.3160070404} \BibitemShut {NoStop}%
\bibitem [{\citenamefont {Blanes}\ \emph {et~al.}(2010)\citenamefont {Blanes}, \citenamefont {Casas}, \citenamefont {Oteo},\ and\ \citenamefont {Ros}}]{Blanes_2010}%
  \BibitemOpen
  \bibfield  {author} {\bibinfo {author} {\bibfnamefont {S.}~\bibnamefont {Blanes}}, \bibinfo {author} {\bibfnamefont {F.}~\bibnamefont {Casas}}, \bibinfo {author} {\bibfnamefont {J.~A.}\ \bibnamefont {Oteo}}, \ and\ \bibinfo {author} {\bibfnamefont {J.}~\bibnamefont {Ros}},\ }\bibfield  {title} {\emph {\bibinfo {title} {A pedagogical approach to the magnus expansion},\ }}\href {\doibase 10.1088/0143-0807/31/4/020} {\bibfield  {journal} {\bibinfo  {journal} {European Journal of Physics}\ }\textbf {\bibinfo {volume} {31}},\ \bibinfo {pages} {907} (\bibinfo {year} {2010})}\BibitemShut {NoStop}%
\bibitem [{\citenamefont {Porras}\ and\ \citenamefont {Cirac}(2004)}]{PhysRevLett.92.207901}%
  \BibitemOpen
  \bibfield  {author} {\bibinfo {author} {\bibfnamefont {D.}~\bibnamefont {Porras}}\ and\ \bibinfo {author} {\bibfnamefont {J.~I.}\ \bibnamefont {Cirac}},\ }\bibfield  {title} {\emph {\bibinfo {title} {Effective quantum spin systems with trapped ions},\ }}\href {\doibase 10.1103/PhysRevLett.92.207901} {\bibfield  {journal} {\bibinfo  {journal} {Phys. Rev. Lett.}\ }\textbf {\bibinfo {volume} {92}},\ \bibinfo {pages} {207901} (\bibinfo {year} {2004})}\BibitemShut {NoStop}%
\bibitem [{\citenamefont {Zhu}\ \emph {et~al.}(2006)\citenamefont {Zhu}, \citenamefont {Monroe},\ and\ \citenamefont {Duan}}]{PhysRevLett.97.050505}%
  \BibitemOpen
  \bibfield  {author} {\bibinfo {author} {\bibfnamefont {S.-L.}\ \bibnamefont {Zhu}}, \bibinfo {author} {\bibfnamefont {C.}~\bibnamefont {Monroe}}, \ and\ \bibinfo {author} {\bibfnamefont {L.-M.}\ \bibnamefont {Duan}},\ }\bibfield  {title} {\emph {\bibinfo {title} {Trapped ion quantum computation with transverse phonon modes},\ }}\href {\doibase 10.1103/PhysRevLett.97.050505} {\bibfield  {journal} {\bibinfo  {journal} {Phys. Rev. Lett.}\ }\textbf {\bibinfo {volume} {97}},\ \bibinfo {pages} {050505} (\bibinfo {year} {2006})}\BibitemShut {NoStop}%
\bibitem [{\citenamefont {Lin}\ \emph {et~al.}(2009)\citenamefont {Lin}, \citenamefont {Zhu}, \citenamefont {Islam}, \citenamefont {Kim}, \citenamefont {Chang}, \citenamefont {Korenblit}, \citenamefont {Monroe},\ and\ \citenamefont {Duan}}]{Lin_2009}%
  \BibitemOpen
  \bibfield  {author} {\bibinfo {author} {\bibfnamefont {G.-D.}\ \bibnamefont {Lin}}, \bibinfo {author} {\bibfnamefont {S.-L.}\ \bibnamefont {Zhu}}, \bibinfo {author} {\bibfnamefont {R.}~\bibnamefont {Islam}}, \bibinfo {author} {\bibfnamefont {K.}~\bibnamefont {Kim}}, \bibinfo {author} {\bibfnamefont {M.-S.}\ \bibnamefont {Chang}}, \bibinfo {author} {\bibfnamefont {S.}~\bibnamefont {Korenblit}}, \bibinfo {author} {\bibfnamefont {C.}~\bibnamefont {Monroe}}, \ and\ \bibinfo {author} {\bibfnamefont {L.-M.}\ \bibnamefont {Duan}},\ }\bibfield  {title} {\emph {\bibinfo {title} {Large-scale quantum computation in an anharmonic linear ion trap},\ }}\href {\doibase 10.1209/0295-5075/86/60004} {\bibfield  {journal} {\bibinfo  {journal} {Europhysics Letters}\ }\textbf {\bibinfo {volume} {86}},\ \bibinfo {pages} {60004} (\bibinfo {year} {2009})}\BibitemShut {NoStop}%
\bibitem [{\citenamefont {{Brickman Soderberg}}\ and\ \citenamefont {{Monroe}}(2010)}]{2010RPPh...73c6401B}%
  \BibitemOpen
  \bibfield  {author} {\bibinfo {author} {\bibfnamefont {K.~A.}\ \bibnamefont {{Brickman Soderberg}}}\ and\ \bibinfo {author} {\bibfnamefont {C.}~\bibnamefont {{Monroe}}},\ }\bibfield  {title} {\emph {\bibinfo {title} {{Phonon-mediated entanglement for trapped ion quantum computing}},\ }}\href {\doibase 10.1088/0034-4885/73/3/036401} {\bibfield  {journal} {\bibinfo  {journal} {Reports on Progress in Physics}\ }\textbf {\bibinfo {volume} {73}},\ \bibinfo {eid} {036401} (\bibinfo {year} {2010})}\BibitemShut {NoStop}%
\bibitem [{\citenamefont {Schumacher}(1996)}]{PhysRevA.54.2614}%
  \BibitemOpen
  \bibfield  {author} {\bibinfo {author} {\bibfnamefont {B.}~\bibnamefont {Schumacher}},\ }\bibfield  {title} {\emph {\bibinfo {title} {Sending entanglement through noisy quantum channels},\ }}\href {\doibase 10.1103/PhysRevA.54.2614} {\bibfield  {journal} {\bibinfo  {journal} {Phys. Rev. A}\ }\textbf {\bibinfo {volume} {54}},\ \bibinfo {pages} {2614} (\bibinfo {year} {1996})}\BibitemShut {NoStop}%
\bibitem [{\citenamefont {Horodecki}\ \emph {et~al.}(1999)\citenamefont {Horodecki}, \citenamefont {Horodecki},\ and\ \citenamefont {Horodecki}}]{PhysRevA.60.1888}%
  \BibitemOpen
  \bibfield  {author} {\bibinfo {author} {\bibfnamefont {M.}~\bibnamefont {Horodecki}}, \bibinfo {author} {\bibfnamefont {P.}~\bibnamefont {Horodecki}}, \ and\ \bibinfo {author} {\bibfnamefont {R.}~\bibnamefont {Horodecki}},\ }\bibfield  {title} {\emph {\bibinfo {title} {General teleportation channel, singlet fraction, and quasidistillation},\ }}\href {\doibase 10.1103/PhysRevA.60.1888} {\bibfield  {journal} {\bibinfo  {journal} {Phys. Rev. A}\ }\textbf {\bibinfo {volume} {60}},\ \bibinfo {pages} {1888} (\bibinfo {year} {1999})}\BibitemShut {NoStop}%
\bibitem [{\citenamefont {Nielsen}(2002)}]{NIELSEN2002249}%
  \BibitemOpen
  \bibfield  {author} {\bibinfo {author} {\bibfnamefont {M.~A.}\ \bibnamefont {Nielsen}},\ }\bibfield  {title} {\emph {\bibinfo {title} {A simple formula for the average gate fidelity of a quantum dynamical operation},\ }}\href {\doibase https://doi.org/10.1016/S0375-9601(02)01272-0} {\bibfield  {journal} {\bibinfo  {journal} {Physics Letters A}\ }\textbf {\bibinfo {volume} {303}},\ \bibinfo {pages} {249} (\bibinfo {year} {2002})}\BibitemShut {NoStop}%
\bibitem [{\citenamefont {Kim}\ \emph {et~al.}(2009)\citenamefont {Kim}, \citenamefont {Chang}, \citenamefont {Islam}, \citenamefont {Korenblit}, \citenamefont {Duan},\ and\ \citenamefont {Monroe}}]{PhysRevLett.103.120502}%
  \BibitemOpen
  \bibfield  {author} {\bibinfo {author} {\bibfnamefont {K.}~\bibnamefont {Kim}}, \bibinfo {author} {\bibfnamefont {M.-S.}\ \bibnamefont {Chang}}, \bibinfo {author} {\bibfnamefont {R.}~\bibnamefont {Islam}}, \bibinfo {author} {\bibfnamefont {S.}~\bibnamefont {Korenblit}}, \bibinfo {author} {\bibfnamefont {L.-M.}\ \bibnamefont {Duan}}, \ and\ \bibinfo {author} {\bibfnamefont {C.}~\bibnamefont {Monroe}},\ }\bibfield  {title} {\emph {\bibinfo {title} {Entanglement and tunable spin-spin couplings between trapped ions using multiple transverse modes},\ }}\href {\doibase 10.1103/PhysRevLett.103.120502} {\bibfield  {journal} {\bibinfo  {journal} {Phys. Rev. Lett.}\ }\textbf {\bibinfo {volume} {103}},\ \bibinfo {pages} {120502} (\bibinfo {year} {2009})}\BibitemShut {NoStop}%
\bibitem [{\citenamefont {Reiter}\ and\ \citenamefont {S\o{}rensen}(2012)}]{PhysRevA.85.032111}%
  \BibitemOpen
  \bibfield  {author} {\bibinfo {author} {\bibfnamefont {F.}~\bibnamefont {Reiter}}\ and\ \bibinfo {author} {\bibfnamefont {A.~S.}\ \bibnamefont {S\o{}rensen}},\ }\bibfield  {title} {\emph {\bibinfo {title} {Effective operator formalism for open quantum systems},\ }}\href {\doibase 10.1103/PhysRevA.85.032111} {\bibfield  {journal} {\bibinfo  {journal} {Phys. Rev. A}\ }\textbf {\bibinfo {volume} {85}},\ \bibinfo {pages} {032111} (\bibinfo {year} {2012})}\BibitemShut {NoStop}%
\bibitem [{\citenamefont {Bermudez}\ \emph {et~al.}(2012{\natexlab{b}})\citenamefont {Bermudez}, \citenamefont {Almeida}, \citenamefont {Ott}, \citenamefont {Kaufmann}, \citenamefont {Ulm}, \citenamefont {Poschinger}, \citenamefont {Schmidt-Kaler}, \citenamefont {Retzker},\ and\ \citenamefont {Plenio}}]{Bermudez_2012}%
  \BibitemOpen
  \bibfield  {author} {\bibinfo {author} {\bibfnamefont {A.}~\bibnamefont {Bermudez}}, \bibinfo {author} {\bibfnamefont {J.}~\bibnamefont {Almeida}}, \bibinfo {author} {\bibfnamefont {K.}~\bibnamefont {Ott}}, \bibinfo {author} {\bibfnamefont {H.}~\bibnamefont {Kaufmann}}, \bibinfo {author} {\bibfnamefont {S.}~\bibnamefont {Ulm}}, \bibinfo {author} {\bibfnamefont {U.}~\bibnamefont {Poschinger}}, \bibinfo {author} {\bibfnamefont {F.}~\bibnamefont {Schmidt-Kaler}}, \bibinfo {author} {\bibfnamefont {A.}~\bibnamefont {Retzker}}, \ and\ \bibinfo {author} {\bibfnamefont {M.~B.}\ \bibnamefont {Plenio}},\ }\bibfield  {title} {\emph {\bibinfo {title} {Quantum magnetism of spin-ladder compounds with trapped-ion crystals},\ }}\href {\doibase 10.1088/1367-2630/14/9/093042} {\bibfield  {journal} {\bibinfo  {journal} {New Journal of Physics}\ }\textbf {\bibinfo {volume} {14}},\ \bibinfo {pages} {093042} (\bibinfo {year} {2012}{\natexlab{b}})}\BibitemShut {NoStop}%
\bibitem [{\citenamefont {Ozeri}\ \emph {et~al.}(2005)\citenamefont {Ozeri}, \citenamefont {Langer}, \citenamefont {Jost}, \citenamefont {DeMarco}, \citenamefont {Ben-Kish}, \citenamefont {Blakestad}, \citenamefont {Britton}, \citenamefont {Chiaverini}, \citenamefont {Itano}, \citenamefont {Hume}, \citenamefont {Leibfried}, \citenamefont {Rosenband}, \citenamefont {Schmidt},\ and\ \citenamefont {Wineland}}]{PhysRevLett.95.030403}%
  \BibitemOpen
  \bibfield  {author} {\bibinfo {author} {\bibfnamefont {R.}~\bibnamefont {Ozeri}}, \bibinfo {author} {\bibfnamefont {C.}~\bibnamefont {Langer}}, \bibinfo {author} {\bibfnamefont {J.~D.}\ \bibnamefont {Jost}}, \bibinfo {author} {\bibfnamefont {B.}~\bibnamefont {DeMarco}}, \bibinfo {author} {\bibfnamefont {A.}~\bibnamefont {Ben-Kish}}, \bibinfo {author} {\bibfnamefont {B.~R.}\ \bibnamefont {Blakestad}}, \bibinfo {author} {\bibfnamefont {J.}~\bibnamefont {Britton}}, \bibinfo {author} {\bibfnamefont {J.}~\bibnamefont {Chiaverini}}, \bibinfo {author} {\bibfnamefont {W.~M.}\ \bibnamefont {Itano}}, \bibinfo {author} {\bibfnamefont {D.~B.}\ \bibnamefont {Hume}}, \bibinfo {author} {\bibfnamefont {D.}~\bibnamefont {Leibfried}}, \bibinfo {author} {\bibfnamefont {T.}~\bibnamefont {Rosenband}}, \bibinfo {author} {\bibfnamefont {P.~O.}\ \bibnamefont {Schmidt}}, \ and\ \bibinfo {author} {\bibfnamefont {D.~J.}\ \bibnamefont {Wineland}},\ }\bibfield  {title} {\emph {\bibinfo {title} {Hyperfine coherence in the presence of
  spontaneous photon scattering},\ }}\href {\doibase 10.1103/PhysRevLett.95.030403} {\bibfield  {journal} {\bibinfo  {journal} {Phys. Rev. Lett.}\ }\textbf {\bibinfo {volume} {95}},\ \bibinfo {pages} {030403} (\bibinfo {year} {2005})}\BibitemShut {NoStop}%
\bibitem [{\citenamefont {Uys}\ \emph {et~al.}(2010)\citenamefont {Uys}, \citenamefont {Biercuk}, \citenamefont {VanDevender}, \citenamefont {Ospelkaus}, \citenamefont {Meiser}, \citenamefont {Ozeri},\ and\ \citenamefont {Bollinger}}]{PhysRevLett.105.200401}%
  \BibitemOpen
  \bibfield  {author} {\bibinfo {author} {\bibfnamefont {H.}~\bibnamefont {Uys}}, \bibinfo {author} {\bibfnamefont {M.~J.}\ \bibnamefont {Biercuk}}, \bibinfo {author} {\bibfnamefont {A.~P.}\ \bibnamefont {VanDevender}}, \bibinfo {author} {\bibfnamefont {C.}~\bibnamefont {Ospelkaus}}, \bibinfo {author} {\bibfnamefont {D.}~\bibnamefont {Meiser}}, \bibinfo {author} {\bibfnamefont {R.}~\bibnamefont {Ozeri}}, \ and\ \bibinfo {author} {\bibfnamefont {J.~J.}\ \bibnamefont {Bollinger}},\ }\bibfield  {title} {\emph {\bibinfo {title} {Decoherence due to elastic rayleigh scattering},\ }}\href {\doibase 10.1103/PhysRevLett.105.200401} {\bibfield  {journal} {\bibinfo  {journal} {Phys. Rev. Lett.}\ }\textbf {\bibinfo {volume} {105}},\ \bibinfo {pages} {200401} (\bibinfo {year} {2010})}\BibitemShut {NoStop}%
\end{thebibliography}%
\bibliographystyle{apsrev4-1new}

\end{document}